\documentclass[preprint2]{emulateapj}

\newcommand{\bzcat}{Roma-BZCAT}

\newcommand{\chn}{{\it Chandra}}

\newcommand{\fer}{{\it Fermi}}

\newcommand{\swf}{{\it Swift}}
\newcommand{\suz}{{\it Suzaku}}
\newcommand{\xmm}{{\it XMM-Newton}}
\newcommand{\wse}{{\it WISE}}

\usepackage{graphicx}
\usepackage{longtable}
\usepackage{mdwlist}

\usepackage{natbib}
\usepackage{multirow}
\usepackage{upgreek}
\usepackage[colorlinks=true, pdfstartview=FitV, linkcolor=blue,citecolor=blue, urlcolor=blue]{hyperref}
\usepackage{wasysym}
\usepackage{txfonts}
\usepackage{gensymb}
\usepackage{morefloats}
\usepackage{threeparttable}
\usepackage{rotating}
\usepackage{graphicx}

\slugcomment{version \today: fm}
\shorttitle{The refined associations in both the \fer\ catalogs}
\shortauthors{F. Massaro et al. 2014}

\begin{document}
\title{Refining the associations of the \fer\ Large Area Telescope Source Catalogs}
\author{
F. Massaro\altaffilmark{1,2},
R. D'Abrusco\altaffilmark{3},
M. Landoni\altaffilmark{4},
A. Paggi\altaffilmark{3},
N. Masetti\altaffilmark{5},
M. Giroletti\altaffilmark{6},\\
H. Ot\'i-Floranes\altaffilmark{7,8},
V. Chavushyan\altaffilmark{9},
E. Jim\'enez-Bail\'on\altaffilmark{7},
V. Pati\~no-\'Alvarez\altaffilmark{9},
S. W. Digel\altaffilmark{10},
Howard A. Smith\altaffilmark{3}
\&
G. Tosti\altaffilmark{11}
}

\altaffiltext{1}{Dipartimento di Fisica, Universit\`a degli Studi di Torino, via Pietro Giuria 1, I-10125 Torino, Italy}
\altaffiltext{2}{Yale Center for Astronomy and Astrophysics, Physics Department, Yale University, PO Box 208120, New Haven, CT 06520-8120, USA}
\altaffiltext{3}{Harvard - Smithsonian Astrophysical Observatory, 60 Garden Street, Cambridge, MA 02138, USA}
\altaffiltext{4}{INAF-Osservatorio Astronomico di Brera, Via Emilio Bianchi 46, I-23807 Merate, Italy}
\altaffiltext{5}{INAF - Istituto di Astrofisica Spaziale e Fisica Cosmica di Bologna, via Gobetti 101, 40129, Bologna, Italy}
\altaffiltext{6}{INAF Istituto di Radioastronomia, via Gobetti 101, 40129, Bologna, Italy}
\altaffiltext{7}{Instituto de Astronom\'{\i}a, Universidad Nacional Aut\'onoma de M\'exico, Apdo. Postal 877, Ensenada, 22800 Baja California, M\'exico}
\altaffiltext{8}{Centro de Radioastronom\'{i}a and Astrof\'{i}sica, UNAM, Campus Morelia, M\'exico}
\altaffiltext{9}{Instituto Nacional de Astrof\'{i}sica, \'Optica y Electr\'onica, Apartado Postal 51-216, 72000 Puebla, M\'exico}
\altaffiltext{10}{SLAC National Accelerator Laboratory and Kavli Institute for Particle Astrophysics and Cosmology, 2575 Sand Hill Road, Menlo Park, CA 94025, USA}
\altaffiltext{11}{Dipartimento di Fisica, Universit\`a degli Studi di Perugia, 06123 Perugia, Italy}

\begin{abstract}
The \fer-Large Area Telescope (LAT) First Source Catalog (1FGL) was released in February 2010 and
the \fer-LAT 2-Year Source Catalog (2FGL) appeared in April 2012, based on data from 24 months of operation. 
Since their releases, many follow up observations of unidentified gamma-ray sources (UGSs) 
were performed and new procedures to associate gamma-ray sources with potential counterparts at other wavelengths were developed. 
Here we review and characterize all the associations as published in the 1FGL and 2FGL catalog on the basis of multifrequency archival observations. 
In particular we located 177 spectra for the low-energy counterparts that were not listed in the previous \fer\ catalogs, 
and in addition we present new spectroscopic observations of 8 $\gamma$-ray blazar candidates.  
Based on our investigations, we introduce a new counterpart category of ``candidate associations'' 
and propose a refined classification for the candidate low-energy counterparts of the \fer\ sources.  
We compare the 1FGL-assigned counterparts with those listed in the 2FGL
to determine which unassociated sources became associated in later releases of the \fer\ catalogs. 
We also search for potential counterparts to all the remaining unassociated \fer sources. 
Finally, we prepare a refined and merged list of all the associations of the 1FGL plus 2FGL catalogs 
that includes 2219 unique \fer\ objects. This is the most comprehensive 
and systematic study of all the associations collected for the $\gamma$-ray sources available to date. 
We conclude that 80\% of the \fer\ sources have at least one known plausible gamma-ray emitter within their positional uncertainty regions.
\end{abstract}


\section{Introduction}
\label{sec:intro}
Despite the large progress in $\gamma$-ray source localization made by the \fer\ Large Area Telescope (LAT) \citep{atwood09}
coupled with the improvements in our knowledge of the diffuse Galactic $\gamma$-ray emission \citep[e.g.,][]{moskalenko07,abdo09},
the positional uncertainties of the \fer\ sources are still large with respect to typical localization precisions at X-ray and lower energies,
making a significant fraction of the sources in the $\gamma$-ray sky yet unknown \citep{abdo10a,nolan12}.
About 43\% of the \fer\ sources detected in the \fer-Large Area Telescope (LAT) First Source Catalog \citep[1FGL][]{abdo10a} 
were listed as UGSs while there were $\sim$33\% $\gamma$-ray objects unassociated in the \fer-LAT 2-Year Source Catalog \citep[2FGL][]{nolan12}.

To decrease the number of unidentified gamma-ray sources (UGSs) many methods based on 
multifrequency approaches or statistical analyses have been recently adopted.
Radio follow up observations of the \fer\ UGSs have already been performed \citep[e.g.,][]{kovalev09,hovatta12,petrov13,hovatta14,schinzel14}
and the \swf\ X-ray survey for all the UGSs listed in the 
\fer\ source catalogs is still on going\footnote{\underline{http://www.swift.psu.edu/unassociated/}} 
\citep[e.g.,][]{mirabal09,paggi13,takeuchi13,stroh13,acero13}.
Additional X-ray observations performed with \chn\ and \suz\ improved our knowledge 
on the UGSs \citep[e.g.,][]{maeda11,cheung12,kataoka12,takahashi12}.
Statistical studies based on the $\gamma$-ray source properties also allowed us
to recognize the nature of the potential counterparts for the UGSs \citep[e.g.,][]{ackermann12,mirabal12,hassan13,doert14}.
Moreover a tight connection between the infrared (IR) sky seen by Wide-Field Infrared Survey Explorer \citep[\wse][]{wright10}
and the \fer\ one has been recently discovered for blazars, the rarest class of active galaxies \citep{paper1,paper2,ugs1}.
These works greatly decreased the fraction of UGSs 
with no assigned counterpart at low energies \citep{paper3,paper4,ugs2}.
More recently also low frequency radio observations (i.e., below $\sim$1~GHz) 
revealed a new spectral behavior that allowed us to search for blazar-like sources 
lying within the positional uncertainty regions of the UGSs \citep{ugs3,ugs6}.
Additional studies have been also carried out with near-infrared observations \citep{raitieri14}
as well as in the sub millimeter range \citep[e.g.,][]{giommi12,lopez13}.
Optical spectroscopic campaigns also have been crucial to
disentangle and/or confirm the natures of the low-energy counterparts selected 
with different methods \citep[e.g.,][]{masetti13,shaw13a,shaw13b,paggi14,sdss,landoni14}.

The main aim of the analysis presented here is to confirm
the current status of the associations 
for the 1FGL and the 2FGL $\gamma$-ray sources. Thus we prepare a single list, 
created by merging the unique sources in the 1FGL and in the 2FGL catalogs. 
On the basis of the information reported in the First \fer-LAT Catalog of Sources Above 10 GeV 
\citep[1FHL][]{ackermann13} and in both the First and the Second Catalog of Active Galactic Nuclei Detected by \fer\ 
\citep[1LAC and 2LAC, respectively][]{abdo10b,ackermann11a}, together 
with an extensive literature search in multifrequency archives, we verify the current status of the $\gamma$-ray associations.
In addition, we also present new optical spectroscopic observations
of eight $\gamma$-ray blazar candidates selected during our multifrequency study.
The result is a refined and merged list of all the associations for the 1FGL-plus-2FGL catalog 
with updated information on the low-energy counterparts including specific multifrequency notes.

The paper is organized as follows: in Section~\ref{sec:cat} we discuss the \fer\ procedures
to assign identifications and associations with potential counterparts at low energies,
and the categories of $\gamma$-ray source associations.
Section~\ref{sec:original} describes the 
content and the main properties of both the 1FGL and the 2FGL catalogs.
In Section~\ref{sec:assoc} we introduce an updated $\gamma$-ray source classification 
for the counterparts of the \fer\ detected objects.
Details on the correlation with multifrequency databases and catalogs
are in Section~\ref{sec:crossmatches}.
In Section~\ref{sec:refined} we present the refined version of the associations listed in the 
\fer\ catalogs (hereinafter designated 1FGLR and 2FGLR, respectively, for brevity).
The analysis of the infrared colors for the sources associated in the 1FGL is given in Section~\ref{sec:kde}
and new optical spectroscopic observations of $\gamma$-ray blazar candidates are presented in Section~\ref{sec:optical}.
In Section~\ref{sec:statistical} we compare our results with those achieved from statistical analyses performed 
on the UGSs listed in both the 1FGL and the 2FGL.
In Section~\ref{sec:connections} we speculate
on both the radio-$\gamma$-ray and the IR-$\gamma$-ray connections for the \fer\ blazars
to check the consistency of sources classified as blazar candidates with these multifrequency behaviors. 
The summary and conclusions are given in Section~\ref{sec:conclusions}.
We use cgs units unless stated otherwise. Spectral indices, $\alpha$, 
are defined by flux density, S$_{\nu}\propto\nu^{-\alpha}$ and
\wse\ magnitudes at [3.4], [4.6], [12], [22] $\mu$m (i.e., the nominal bands)
are in the Vega system. For numerical results a flat cosmology was assumed 
with $H_0=67.3$ km s$^{-1}$ Mpc$^{-1}$, 
$\Omega_{M}=0.315$ and $\Omega_{\Lambda}=0.685$ \citep{planck14}.
{\bf The astronomical survey and source class acronyms used in the paper are listed in Table~\ref{tab:acronyms}}.

\begin{table}
\tiny
\caption{List of acronyms used in the paper.}
\begin{center}
\begin{tabular}{|ll|}
\hline
Acronym  &  Term \\
\hline
\noalign{\smallskip}
1FGL   & {\it Fermi}-Large Area Telescope First Source Catalog\\
1FGLR  & 1FGL Refined association catalog\\
1FHL   & First {\it Fermi}-Large Area Telescope Catalog of Sources Above 10 GeV\\
1LAC   & First Catalog of Active Galactic Nuclei Detected by {\it Fermi}\\
1XSPS  & Deep Swift X-Ray Telescope Point Source Catalog\\
2FGL   & {\it Fermi}-Large Area Telescope 2-Year Source Catalog\\
2FGLR  & 2FGL Refined association catalog\\
2LAC   & Second Catalog of Active Galactic Nuclei Detected by {\it Fermi}\\
2MASS  & Two Micron All-Sky Survey\\
3C     & Third Cambridge Catalogue of Radio Sources\\
6dFGS  & Six-degree-Field Galaxy Redshift Survey\\
\noalign{\smallskip}
\hline
\noalign{\smallskip}
AGN    & Active Galactic Nucleus\\
AGU    & AGN of Uncertain type\\
AT20G  & Australia Telescope 20 GHz Survey\\
ATNF   & Australia Telescope National Facility\\
BZB    & BL Lac object\\
BZQ    & flat spectrum radio quasar\\
CRATES & Combined Radio All-Sky Targeted Eight-GHz Survey radio catalog\\
CSC    & Chandra Source Catalog\\
CT     & Classification Tree method\\
FIRST  & VLA Faint Images of the Radio Sky at Twenty-Centimeters\\
GB6    & Green Bank 6-cm Radio Source Catalog\\
GLC    & Globular Cluster\\
HMB    & High Mass X-ray Binary\\
KDE    & Kernel Density Estimation\\
IRAF   & Image Reduction and Analysis Facility\\
IRAS   & InfraRed Astronomical Satellite\\
LAT    & Large Area Telescope\\
LBA    & Australian Long Baseline Array\\
LCS1   & LBA Calibrator Survey\\
LORCAT & Low-frequency Radio Catalog of flat-spectrum Sources\\
LR     & Logistic Regression method\\
MGPS   & Molonglo Galactic Plane Survey\\
MSP    & Millisecond Pulsar\\
NED    & NASA/IPAC Extragalactic Database\\
NOV    & Nova\\
NRAO   & National Radio Astronomy Observatory\\
NVSS   & NRAO VLA Sky Survey Catalog\\
OAGH   & Observatorio Astrofisico Guillermo Haro\\\
OAN    & Observatorio Astron\'omico Nacional\\
PKS    & Parkes Southern Radio Source catalog\\
PMN    & Parkes-MIT-NRAO Surveys\\
PSR    & Pulsar\\
PWN    & Pulsar wind nebula\\
RBSC   & ROSAT Bright Source Catalog\\
RFSC   & ROSAT Faint Source Catalog\\
SDSS   & Sloan Digitized Sky Survey\\
SED    & Spectral Energy Distribution\\
SFR    & Star Forming - HII regions\\
SNR    & Supernova Remnant\\
SUMSS  & Sydney University Molonglo Sky Survey\\
TEXAS  & Texas Survey of Radio Sources at 365 MHz\\
UGS    & Unidentified Gamma-ray Source\\
VLA    & Very Large Array\\
VLSS   & VLA Low-Frequency Sky Survey Discrete Source Catalog\\
USNO & United States Naval Observatory\\
XMMSL  & {\it XMM-Newton} Slew Survey\\
WENSS  & Westerbork Northern Sky Survey\\
WISE   & Wide-Field Infrared Survey Explorer\\
WISH   & Westerbork in the Southern Hemisphere Source Catalog\\
WSRT   & Westerbork Synthesis Radio Telescope\\
WSRTGP & WSRT Galactic Plane Compact 327-MHz Source Catalog\\
\noalign{\smallskip}
\hline
\end{tabular}
\end{center}
\label{tab:acronyms}
\end{table}

\section{Categories of gamma-ray source associations}
\label{sec:cat}
In all the \fer\ catalogs there is an important distinction between {\it identification} of low-energy counterparts 
for the \fer\ sources and {\it association}.
{\bf Identification is based on i) spin or orbital periodicity (e.g., pulsars, binary systems) or on 
ii) correlated variability at other wavelengths (e.g., blazars, active galaxies)
or on iii) the consistency between the measured angular sizes in $\gamma$-ray and at lower energies (e.g., supernova remnants).}
On the other hand, the {\it association} designation indeed depends on the results of different procedures adopted 
in both \fer\ catalogs \citep{abdo10a,nolan12}. These procedures are: 

\begin{enumerate}

\item {\it The Bayesian Association Method}: Initially applied
to associate EGRET sources with flat-spectrum radio sources \citep[e.g.,][]{mattox97,mattox01,abdo10a},
this method assesses the probability of association between a $\gamma$-ray source
and a candidate counterpart taking into account their local densities.
Local density is estimated simply by counting candidates in a nearby region of the sky.

\item {\it The Likelihood Ratio method}:  Used to search for possible counterparts in uniform surveys in the
radio and in X-ray bands,
this procedure was originally proposed by Richter (1975) and subsequently 
applied by and modified by de Ruiter, Willis \& Arp (1977),
Prestage \& Peacock (1983), Wolstencroft et al. (1986) 
and by Sutherland \& Saunders (1992).

\item {\it The logN-logS association method}: This is a modified version of the Bayesian method for blazars
taking into account their $\log N$-$\log S$ \citep[see][for more details]{abdo10b,ackermann11a}.
\end{enumerate}

There are several differences between the source associations presented between the 1FGL and the 2FGL catalogs
mostly related to the $\gamma$-ray analysis and to the improvements achieved in the 
models of Galactic $\gamma$-ray diffuse emission as well as in evaluation of the LAT response functions. 
Improving the \fer\ source localization not only makes 
easier the search for low-energy counterparts 
but allows more accurate estimates of the association probability
since they depend on the $\gamma$-ray positional uncertainty \citep[see e.g.,][]{abdo10b,ackermann11a}. 
These differences also relate to updates
of comparison surveys and catalogs used to search for the low-energy counterparts and
on multifrequency observations performed on UGS samples.

The use of catalogs/surveys that were previously not considered for the \fer\ catalog preparation,
as for example the infrared all-sky survey performed by \wse\ and the low-frequency radio catalogs,
revealing new connections between the low and the high-energy skies \citep[e.g.,][]{paper1,ugs1,ugs3}
has also decreased the unknown fraction of the $\gamma$-ray sky \citep[e.g.,][]{paper4,ugs2}.
This affects the \fer\ associations because 
the methods adopted to assign low-energy counterparts depend 
on the source densities of the catalogs \citep{nolan12,ackermann11a}.
A better estimate of the counterpart density leads to a better estimate of the 
false positive associations and of the association probability
and this implies that a previously unassociated source could have an assigned counterpart in a new release of the \fer\ catalog.

Motivated by the changes occurring between the 1FGL and the 2FGL associations
here we propose to introduce, together with identified and associated sources, 
the category of {\it ``candidate associations''}. These are $\gamma$-ray sources that have a potential 
low-energy counterpart in a specific class of well-known $\gamma$-ray emitters 
lying within the \fer\ positional uncertainty region at 95\%
level of confidence and/or having angular separations between the \fer\ and the counterpart position
smaller than the maximum one for all the associated sources of the same class.
Candidate associations have the potential to be promoted to associations in the future (see Section~\ref{sec:assoc}).

We considered as counterpart classes for the candidate associations only a few kinds of sources:
i) blazar-like sources, as defined in the following, ii) known pulsars (PSRs), 
iii) pulsar wind nebulae (PWNe), supernova remnants (SNRs) and star forming and HII regions (hereinafter simply indicated as SFRs).
The first two source classes 
constitute the two largest populations of known $\gamma$-ray emitters
while the other choices were considered because 15 out of the 44 SNRs and 
10 out of the 63 PSRs and PWNe associated in the 1FGL lie within SFRs
and a similar situation occurs for the 2FGL.

Our choice concerning the SFRs, especially for the most massive cases, 
is indeed motivated by the possibility that they could host type O and/or B stars, 
potentially associated with open star clusters, and this being relatively
likely to have some members that have progressed to become a PSR, PWN, and/or SNR, 
all known classes of $\gamma$-ray sources \citep[see also][]{montmerle79,montmerle09}.

In addition, considering SFRs as candidate associations is also motivated by the recent 
idea, partially supported by the \fer\ discovery of $\gamma$-ray emission arising from Eta Carinae \citep{abdo10c},
that shells detected in the radio, infrared, and X-rays
could mark termination shocks of stellar winds. These winds interacting with the stars'
natal molecular clouds can be considered as plausible sites to accelerate particles that 
emit $\gamma$-rays \citep[e.g.,][]{araudo07,boschramon10,rowell10}.
Bowshocks are associated with many source classes 
such as PSRs, cataclysmic variables, colliding wind binaries, cometary H II regions
and particles accelerated therein can radiate up to the MeV-GeV energy range \citep[see e.g.][]{delvalle12}.
However we remark that evidence of $\gamma$-ray 
emission from colliding wind binaries associated with Wolf-Rayet stars 
was not found by \fer, for a selected sample of 7 sources \citep{werner13};
although, as highlighted by the same authors, 
searching for such $\gamma$-ray detections is extremely complicated due 
to the uncertainties of the diffuse emission in the Galactic plane.

Some candidate associations lack multifrequency information as well as
correct spectroscopic identifications but {\bf exhibit some characteristic}
features, typical of known $\gamma$-ray emitters, mainly blazars and PSRs.
They were not associated in the previous releases of the \fer\ catalogs mostly because they were not listed 
in one of the appropriate comparison catalogs of potential counterparts used to associated \fer\ sources.
However once follow up observations {\bf were} performed 
\citep[see e.g.][for blazar-like candidates]{paggi13,masetti13,petrov13,sdss,landoni14}, then 
they could be promoted to associated sources in new releases of the \fer\ catalogs. 
In addition, within this category we also considered \fer\ sources having SFRs 
within their $\gamma$-ray positional uncertainty regions
for which multifrequency observations could reveal the presence of unknown PSRs, PWNe and SNRs,
already counterparts of $\sim$5\% of the $\gamma$-ray sources.

We emphasize three advantages obtained from introducing the concept of candidate associations:
\begin{enumerate}
\item Designating candidate association category allows us to flag potential counterparts
that might be associated or identified in future releases of the \fer\ catalogs,
as previously occurred to 171 1FGL sources then recognized, classified and associated in the 2FGL. 
This also facilitates planning follow up multifrequency observations, in particular spectroscopic ones,
necessary to determine and/or confirm their nature.

\item The category of candidate associations permits us to establish which is the number of \fer\ sources 
not yet having potential counterparts. The precise knowledge 
of this number is extremely relevant to set better constraints on dark matter scenarios 
\citep[e.g.,][]{zechlin12,mirabal12,mirabal13a,mirabal13b,berlin14}.  

\item For all the methods adopted to associate sources in the \fer\ catalogs,
the resulting associations depend on the comparison catalog/survey used. The same low-energy counterpart
may be associated with a \fer\ source using a certain catalog but not with a different survey,
even when the positions in the two catalogs are the same. This occurs because the source density 
of the comparison catalogs, which is used to determine the probability of false positive associations, 
affects the association probability. Thus observing these candidate associations with follow up campaigns 
to determine their natures will permit us to add blazars and pulsars to the correct comparison catalogs 
and associate them in future releases of the \fer\ surveys. 
\end{enumerate}

To reiterate the importance of having candidate associations we highlight the recent work of Schinzel et al. (2014) 
on radio very-long-baseline interferometry (VLBI) observations of \fer\ unassociated sources in the 2FGL \citep[see also][]{petrov13}.
In the final list of high-confidence associations presented in Schinzel et al. (2014)
more than 90\% of the sources are also detected in the major radio surveys such {\bf as the 
National Radio Astronomy Observatory (NRAO) Very Large Array (VLA) Sky Survey Catalog \citep[NVSS;][]{condon98} 
and the Sydney University Molonglo Sky Survey \citep[SUMSS;][]{mauch03}.}
These counterparts were not associated using the methods adopted in \fer\ catalogs
due to the high density of sources at low flux level in the above surveys.
However thanks to the high resolution radio follow up observations, which reveled the presence of radio compact cores, 
it has been possible to use a different selected sample of potential counterparts 
to find new associations via the likelihood procedure \citep{schinzel14,petrov13}. 
Thus it is extremely important to have targets for follow up observations that, once performed, 
will allow us to insert the potential counterparts in the correct catalogs used to search $\gamma$-ray associations.

Identifications, associations and candidate associations' of $\gamma$-ray sources
are labelled in both the refined version of the \fer\ catalogs as the {\it category of the $\gamma$-ray source association}.
These are defined as: {\it identified} (I), {\it associated} (A), {\it candidate association} (C)
in the tables while sources for which no potential counterpart 
was found are assigned to the UGS category (U).

\section{The \fer\ catalogs}
\label{sec:original}

\subsection{The \fer-Large Area Telescope First Source Catalog}
\label{sec:1fgl}
The 1FGL catalog\footnote{\underline{http://fermi.gsfc.nasa.gov/ssc/data/access/lat/1yr\_catalog/}}
contains 1451 sources detected and characterized in the 100 MeV to 100 GeV energy range. 
For each \fer\ source listed therein, positional uncertainty regions at 68\% and 95\% level of confidence, 
results from spectral fits as well as flux measurements in five energy bands are reported \citep{abdo10a}. 
Firm identification or plausible association with a low-energy counterpart was provided for each 1FGL source
on the basis of the comparison with other astronomical catalogs and on the basis of the association procedures previously mentioned 
(see Section~\ref{sec:cat} and Abdo et al. 2010a for details). 

The published 1FGL catalog lists several classes of counterparts: 
63 pulsars (PSRs), 7 associated plus 56 identified (24 of them discovered thanks to the \fer-LAT observations), 
2 pulsars wind nebulae (PWNe)
44 $\gamma$-ray sources corresponding to 38 SNRs, 41 associated and 3 identified, 
8 globular clusters (GLC), 
2 high mass X-ray binaries (HMB), all identified,
295 BL Lac objects (BZBs)
278 flat spectrum radio quasars (BZQs),
28 non blazar-like active galactic nuclei (AGNs)
92 AGN of uncertain type (AGUs)
6 normal galaxies and 2 starburst galaxies.
The remaining 630 \fer\ sources are all UGSs.

A detailed comparison between the 1FGL and the 2FGL catalog is reported in Nolan et al. (2012). 
There are 1099 sources included in both \fer\ catalogs with only 381 1FGL sources being listed in the 1FHL \citep{ackermann13}.
It is worth noting that 267 out of 1099 sources in common between 1FGL and 2FGL changed classification from 1FGL to 2FGL
and among them 172, previously unassociated in 1FGL, obtained an assigned low-energy counterpart in 2FGL.
This sample of modified associations includes: 16 sources indicated as generic AGNs 
that have been classified plus one that became a UGS;
54 AGUs that have been recognized as blazars (either BZBs or BZQs) and 2 as AGNs; 
then 17 blazars actually changed their classification
from 1FGL to 2FGL thanks to new spectroscopic data available and, among them, 6 actually disappeared 
since they were listed in the 2FGL as unassociated as occurred for other 5 sources of Galactic origin (i.e., PSRs, PWNe and SNRs).
On the other hand, there are 172 UGSs of the 1FGL that appear to be associated in the 2FGL: 20 associated with PSRs, 11 with SNRs and 1 with a GLC while 
all the others with various AGN classes (i.e. blazars, radio galaxies, Seyfert galaxies etc.).

The 1FGL catalog includes 216 sources with $\gamma$-ray analysis flags
and their detection has to be considered carefully since they could have been artifacts of the analysis \citep{abdo10a}.
In particular, 103 of these 216 1FGL flagged sources have the ``c'' flag following the 1FGL name;
this indicates that the source is found in a region with bright and/or possibly incorrectly modeled diffuse emission. 
On the other hand any non-zero entry in the Flags column indicates inconsistencies found
during the analysis \citep[see][for details]{abdo10a}. 
However we note that 36 out of 103 ``c''-flagged 1FGL sources are also detected in the 2FGL and only 
25 of them were still indicated with a ``c'' flag therein.  Thus we decided to keep all the c-flagged sources
in our analysis since they could be useful for  reference in future releases of \fer\ catalogs.

\subsection{The \fer-Large Area Telescope Second Source Catalog}
\label{sec:2fgl}
The 2FGL catalog\footnote{\underline{http://fermi.gsfc.nasa.gov/ssc/data/access/lat/2yr\_catalog/}}
based on 24 months of \fer\ operation lists 1873 $\gamma$-ray sources \citep{nolan12}.
It includes:
\begin{itemize}
\item 108 PSRs, 25 associated plus 83 identified;
\item 3 PWNe, all 3 identified;
\item 61 $\gamma$-ray sources corresponding to 50 SNRs, 55 associated plus 6 identified;
\item 11 associated globular clusters (GLC), 4 high mass X-ray binaries (HMB), all identified and 1 nova (NOV),
\end{itemize} 
among the Galactic sources, while in the extragalactic sky:
\begin{itemize}
\item 435 BL Lac objects (BZBs), only 7 identified;
\item 370 flat spectrum radio quasars (BZQs), 17 identified;
\item 11 non blazar-like active galactic nuclei (AGNs), 1 identified;
\item 257 AGNs of uncertain type (AGUs);
\item 6 normal galaxies and 4 starburst galaxies .
\end{itemize} 
The remaining 577 \fer\ sources were are all UGSs.

The 2FGL catalog {\bf lists} 472 sources with $\gamma$-ray analysis flags
and 155 of them also have the ``c'' flag following the 2FGL name.
It is also worth noting that with respect to the 1FGL catalog the fraction of UGSs is decreased 
from 630 out of 1451 sources ($\sim$43\%) 
to 577 out of 1873  ($\sim$31\%) in the 2FGL.

\section{Introducing the classification for the low-energy counterparts of the \fer\ sources}
\label{sec:assoc}
We adopt the following class definitions for Galactic counterparts of the \fer\ sources,
summarized in Table~\ref{tab:classes}:
high mass X-ray binary (hmb), globular cluster (glc), nova (nov),
millisecond pulsar (msp), pulsar (psr), pulsar wind nebula (pwn), supernova remnant (snr),
star forming region or H\,{\sc ii} region (sfr) and binary star (bin).
The classes that we assign for extragalactic counterparts are
BL Lac object (bzb), quasar radio loud of flat spectrum type (bzq), blazar of uncertain type (bzu),
radio galaxy (rdg), normal galaxy (gal), starburst galaxy (sbg) and Seyfert galaxy (sey).
{\bf We note that we used the three-letter designators to indicate the class for each counterpart; 
these are shown in lower case to differentiate them from the acronyms used throughout the paper (see Table~\ref{tab:acronyms}).}
We emphasize that the labels used for the extragalactic blazar subclasses are used only if the 
counterpart corresponds to a source classified in the \bzcat\ \citep{massaro09,massaro11}.
{\bf We used version 4.1 of the \bzcat\ in our analysis.}
We also emphasize that the proposed classification depends on the availability 
of the optical spectroscopic information for the low-energy counterparts.

We also considered as a counterpart class blazar candidates (bcn) to indicate sources having at least one of the following requirements:
1) being classified as BL Lac candidate in the \bzcat;
2) sources that belong to the Combined Radio All-Sky Targeted Eight-GHz Survey radio catalog \citep[CRATES;][]{healey07} 
with a BL Lac-like or a quasar-like {\bf optical} spectrum available in literature;
3) radio sources with a blazar-like optical spectrum available in literature 
(i.e., a published spectrum or a description of the features found in the near infrared--optical band) and with a \wse\ counterpart.

Finally, we labelled as unclassified counterparts (unc) those sources satisfying at least one of the following criteria:
1) \wse\ Infrared colors similar to those of $\gamma$-ray blazars selected
according to the association methods developed by D'Abrusco et al. (2013)
and/or with the kernel density estimator analysis \citep[][see also Section~\ref{sec:kde} for a refined analysis]{paper3,paper5};
2) a flat radio spectrum at frequencies below $\sim$1 GHz (i.e., $\alpha<$0.5);
3) a radio source with or without an X-ray counterpart associated with the methods adopted in the 1LAC and the 2LAC;
4) a radio and/or an X-ray source that lies at angular separation between the \fer\ and the counterpart positions
smaller than the maximum one for the blazar class (i.e., 0.4525 \degr\ corresponding to 2FGL J1746.0+2316). 
All the remaining sources are still unidentified (ugs).

The proposed classification scheme has three main advantages. 
\begin{enumerate}
\item To have a homogeneous classification of the blazars listed in our analysis and to avoid misclassifications
we adopted the following criterion.
We indicate as blazars only sources that were previously recognized and classified in the \bzcat. 
There are indeed several counterparts classified as BZB or BZQ listed in the 1LAC and in the 2LAC catalogs 
that do not have the same classification according to the \bzcat\ 
or for which no optical spectra were found in literature to confirm their classification.

\item We do not classify a source with a radio and/or X-ray counterpart as an active galaxy of uncertain type (AGU)
since these requirements are not strictly sufficient to claim a source as an AGN or as blazar-like
and the lack of multifrequency information does not allow us to determine precisely its nature. 
Consequently, we introduced the bcn and the unc classes. Once optical spectra and
multifrequency observations become available a refined classification will be provided.

\item We introduced the sfr class for cases where \fer\ position is consistent with a known massive SFR for three reasons.
First, to highlight the possibility that an unknown SNR, PWN and/or PSR is embedded therein, 
for which multifrequency observations are necessary to confirm their presence. 
Second, because the use of this information could be adopted to 
refine $\gamma$-ray diffuse models in future releases of the \fer\ catalogs.
In particular, the largest fraction of these SFR potential associations were determined using the 
recent \wse\ Catalog of Galactic H\,{\sc ii} Regions\footnote{\underline{http://astro.phys.wvu.edu/wise/}}
\citep{anderson14}, which was not available during the preparation of the currently-published \fer\ catalogs.
This could also reveal new correspondences between the uncertainties of the diffuse $\gamma$-ray background models
and regions of interstellar medium in the Galactic plane.
Third, this could be useful for future investigations of the environments of the Galactic $\gamma$-ray sources.
\end{enumerate}

Finally, we remark that for the sources indicated as candidate associations and classified as SFRs, SNRs, PWNs and PSRs 
we investigated the radio and infrared images available in the NVSS, the Westerbork Synthesis Radio Telescope (WSRT) 
Galactic Plane Compact 327-MHz Source Catalog \citep[WSRTGP;][]{taylor96}, 
the Molonglo Galactic Plane Survey\footnote{\underline{http://www.astrop.physics.usyd.edu.au/MGPS/}} 
\citep[MGPS;][]{green99} catalogs and in the \wse\ archives searching for signatures of potential $\gamma$-ray emitters. 

\begin{figure}[] 
          \includegraphics[height=7.2cm,width=8.4cm,angle=0]{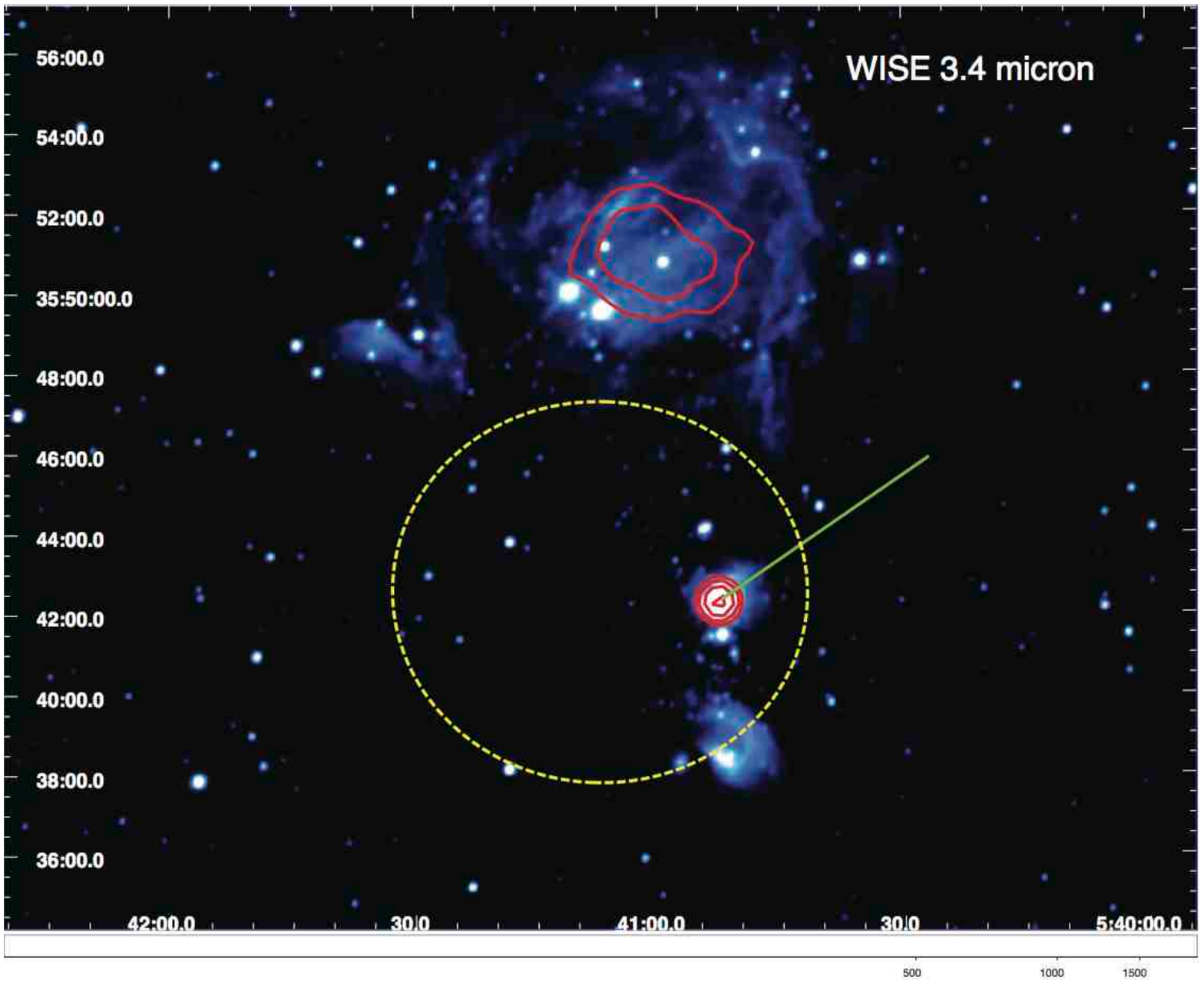}
          \includegraphics[height=7.2cm,width=8.4cm,angle=0]{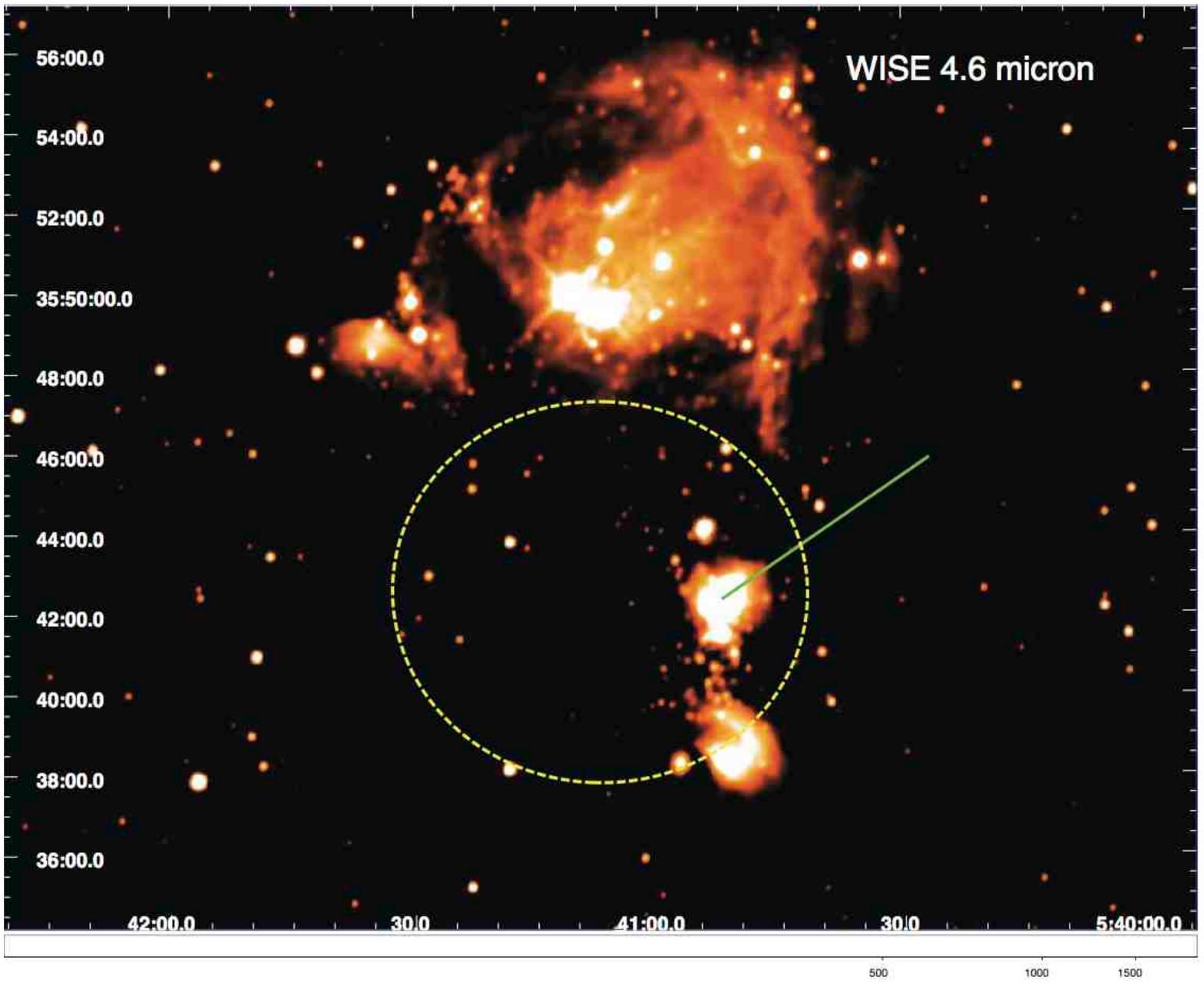}
           \caption{The \wse\ images of the \fer\ source 1FGLJ0541.1+3542c in two different bands at 3.4$\mu$m (top), 4.6$\mu$m (bottom).
                          In each image the yellow ellipse corresponds to the \fer\ positional uncertainty 
                          region at 68\% level of confidence. The red contours on the \wse\ image at 3.4$\mu$m 
                          indicate the radio brightness at 1.4 GHz from the NVSS. 
                          The radio-infrared extended structure, marked by the green arrow, suggests that 
                          a shock occurred in the interstellar medium that could be responsible for the $\gamma$-ray emission.
                          The radio source is also detected in the Westerbork Northern Sky Survey \citep[WENSS;][]{rengelink97}
                          and it is associated with an open star cluster.}
           \label{fig:1FGLJ0541.1+3542c}
          \end{figure}

\begin{figure}[] 
          \includegraphics[height=7.2cm,width=8.4cm,angle=0]{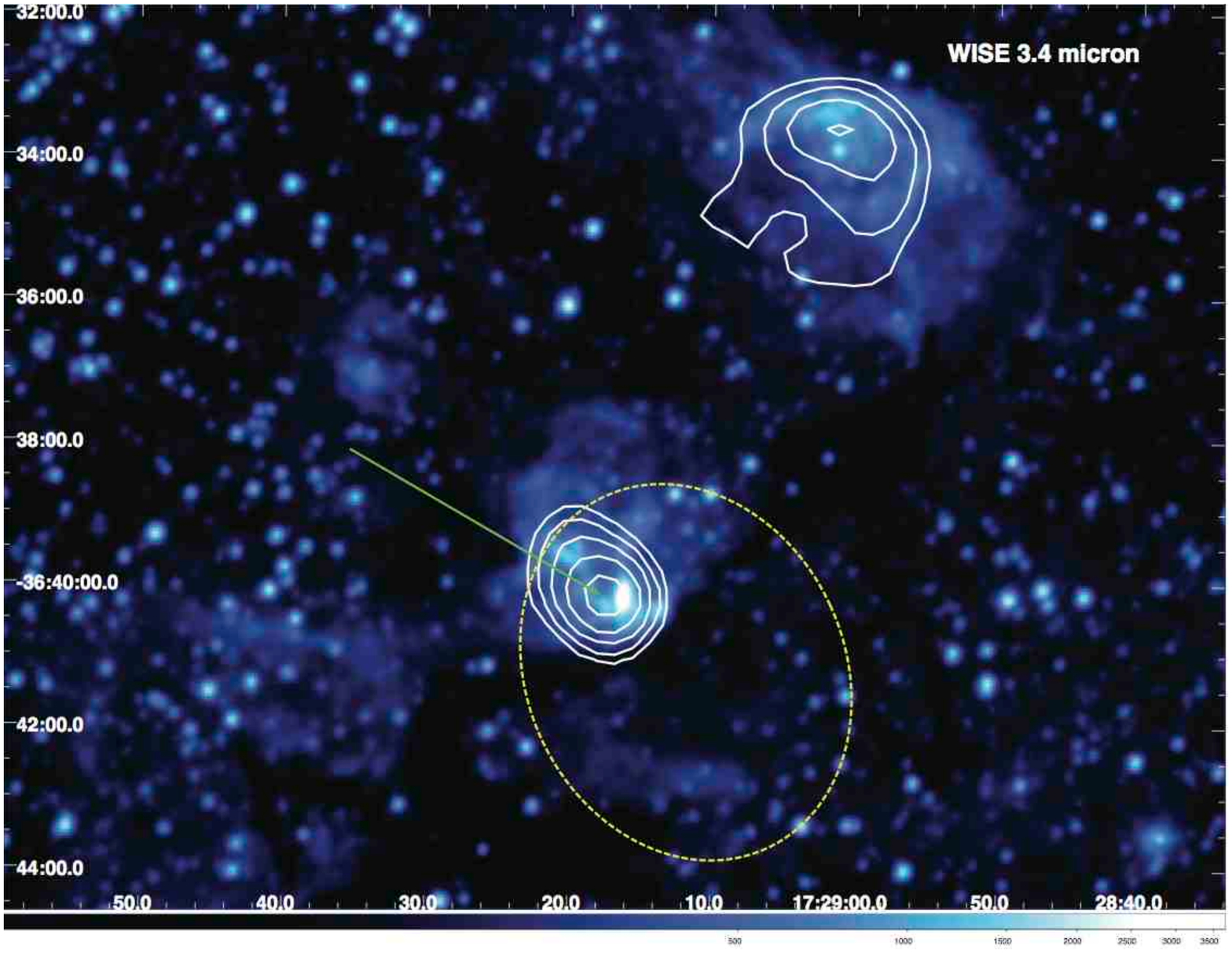}
          \includegraphics[height=7.2cm,width=8.4cm,angle=0]{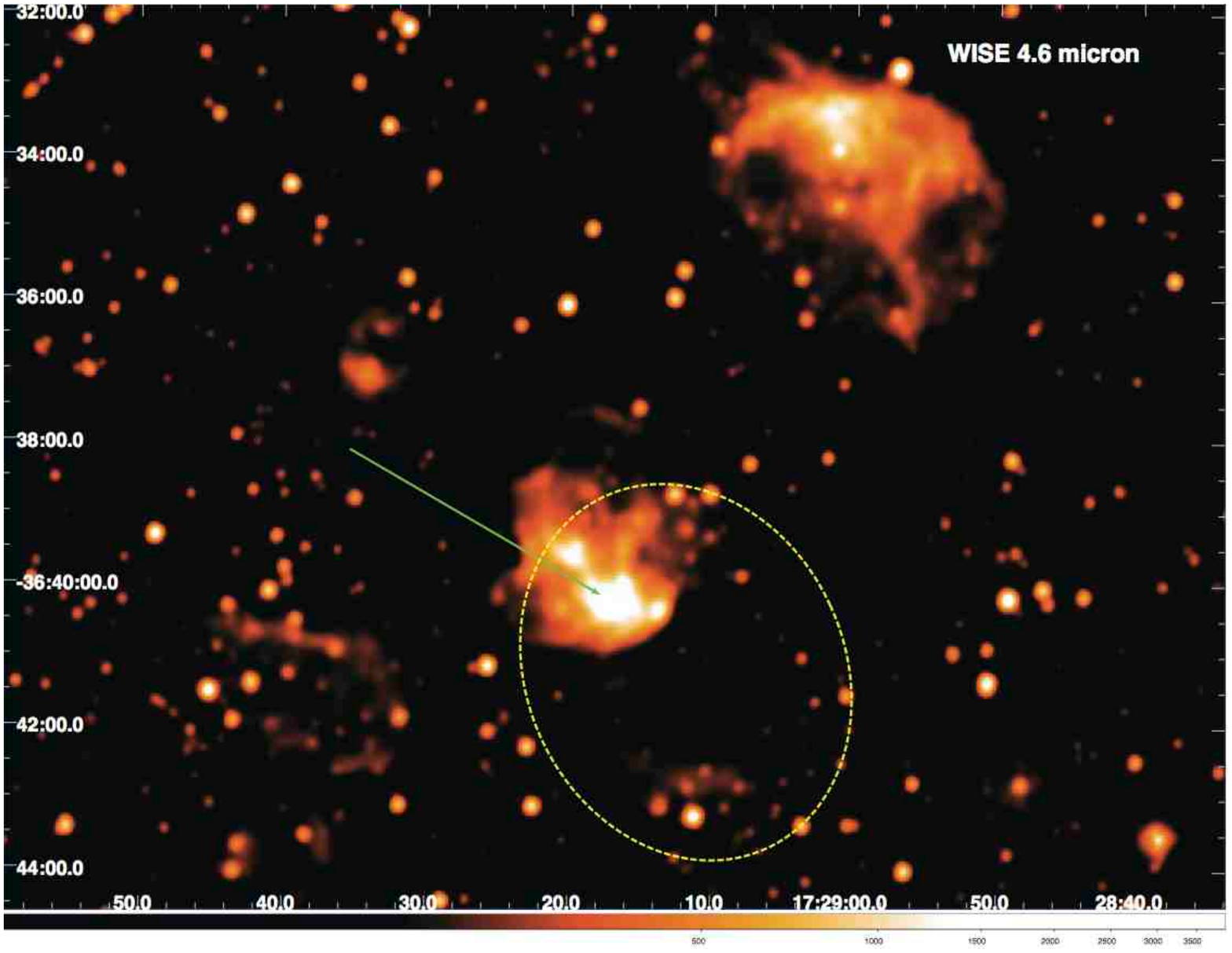}
           \caption{The \wse\ images of the \fer\ source 1FGLJ1729.1$-$3641c in two different bands at 3.4$\mu$m (top), 4.6$\mu$m (bottom).
                          In each image the yellow ellipse corresponds to the \fer\ positional uncertainty 
                          region at 68\% level of confidence. 
                          The white radio contours drawn from the NVSS image are overlaid on the \wse\ image at 3.4$\mu$m.
                          This is an example of a candidate association with a SFR that includes the open star cluster 165 from the
                          Northern and Equatorial Milky Way catalog built with the Two Micron All-Sky Survey (2MASS) observations \citep{bica03}.}
           \label{fig:1FGLJ1729.1-3641c}
          \end{figure}

In Figures~\ref{fig:1FGLJ0541.1+3542c} and \ref{fig:1FGLJ1729.1-3641c} we show two cases of $\gamma$-ray sources 
having extended structures associated with an open star cluster in a SFR 
found within their \fer\ positional uncertainty regions. 
These are two examples of star clusters where some members could have been progressed becoming 
a PSR, PWN, and/or a SNR, and so being a plausible counterpart for the $\gamma$-ray emission.

\begin{figure}[] 
          \includegraphics[height=7.2cm,width=8.4cm,angle=0]{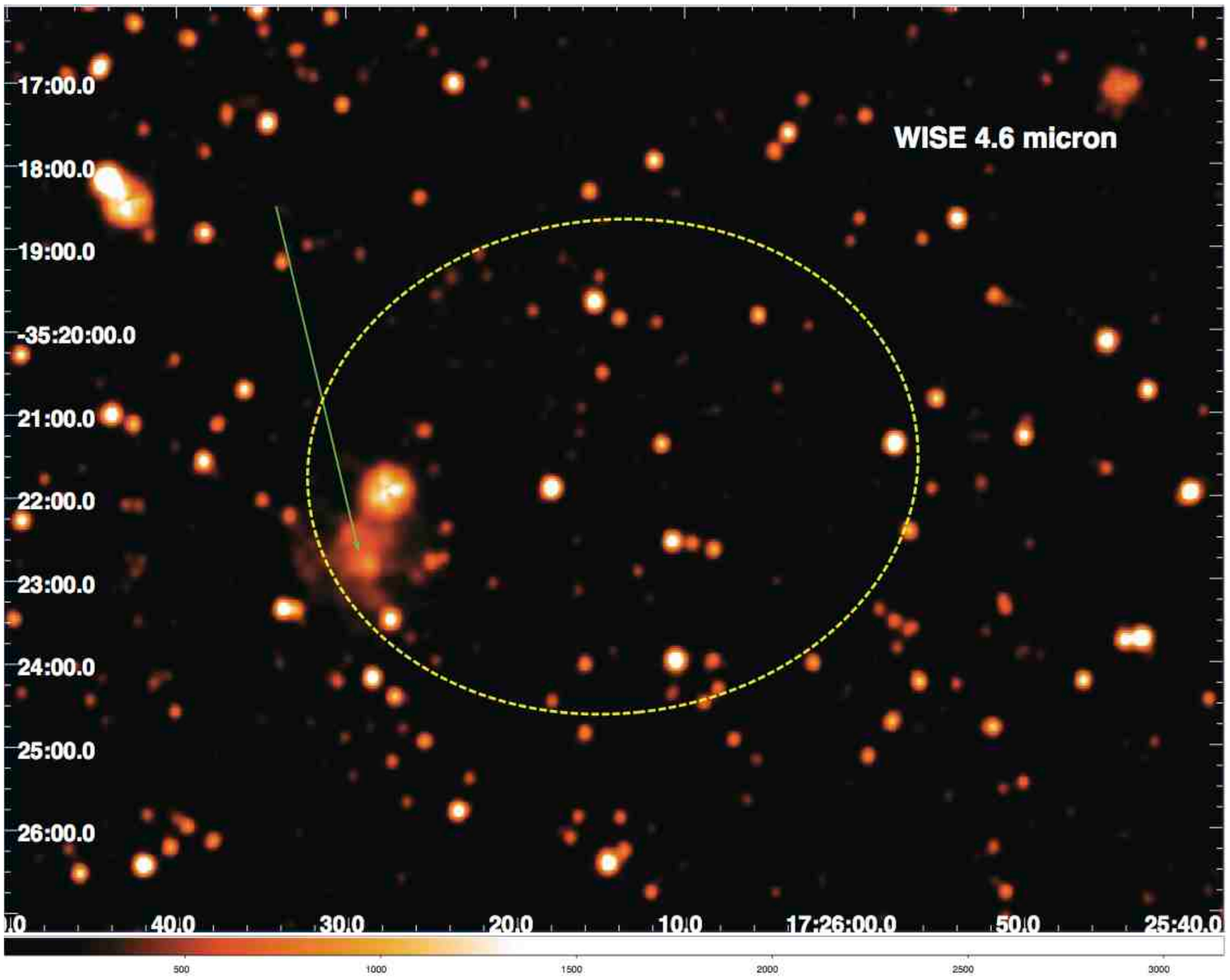}
          \includegraphics[height=7.2cm,width=8.4cm,angle=0]{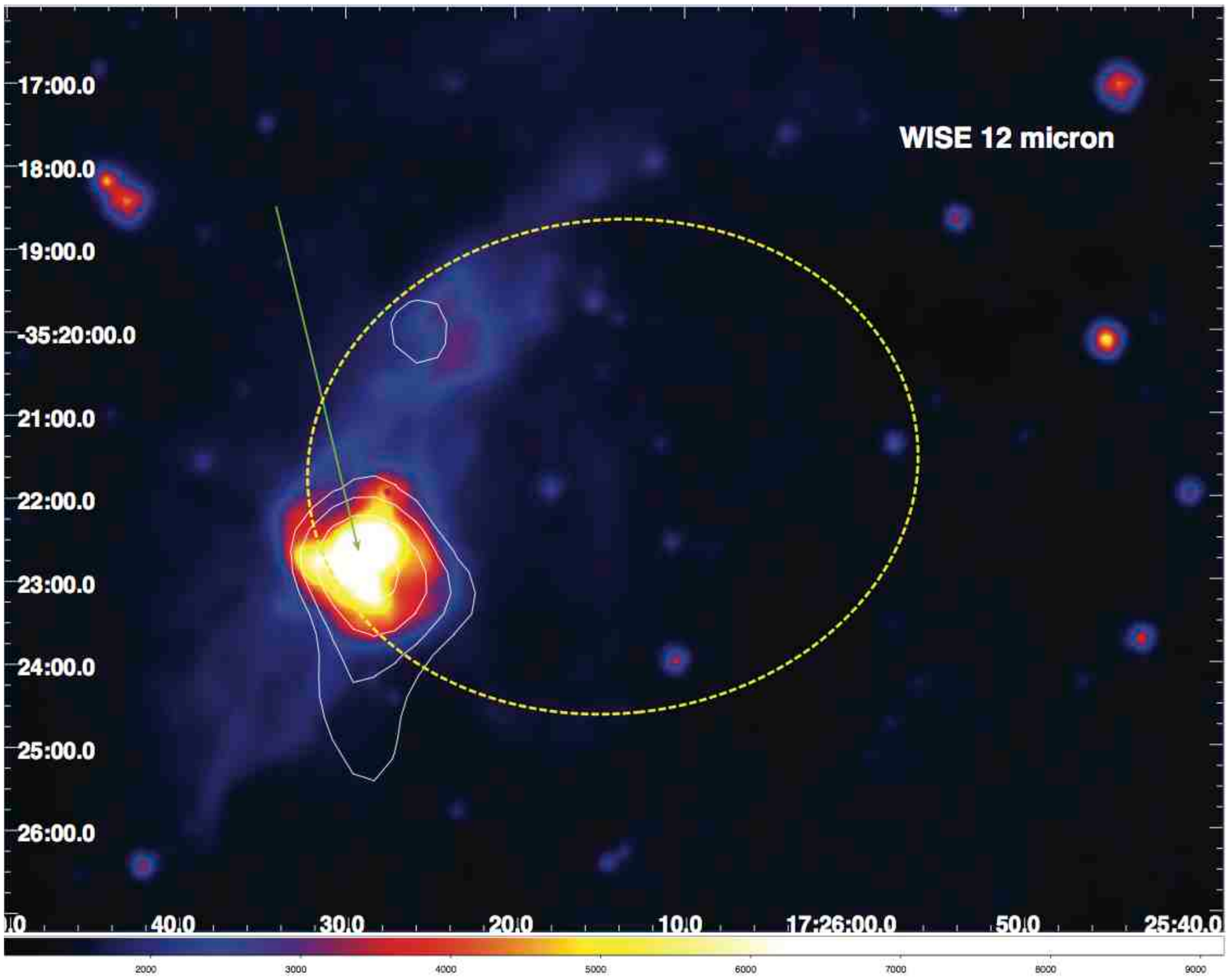}
           \caption{The \wse\ images of the \fer\ source 1FGLJ1726.2-3521c in two different bands at 4.6$\mu$m (top), 12$\mu$m (bottom).
                          In each image the yellow ellipse corresponds to the \fer\ positional uncertainty 
                          region at 68\% level of confidence. The white contours are overlaid on the \wse\ image at 12$\mu$m 
                          are the 1.4 GHz radio emission from the NVSS. 
                          The radio-infrared extended structure, marked by the green arrow, clearly indicates that a shock occurred in the interstellar medium
                          within the \fer\ positional uncertainty region. This could be responsible for the $\gamma$-ray emission.}
           \label{fig:1FGLJ1726.2-3521c}
          \end{figure}

\begin{figure}[] 
          \includegraphics[height=7.2cm,width=8.4cm,angle=0]{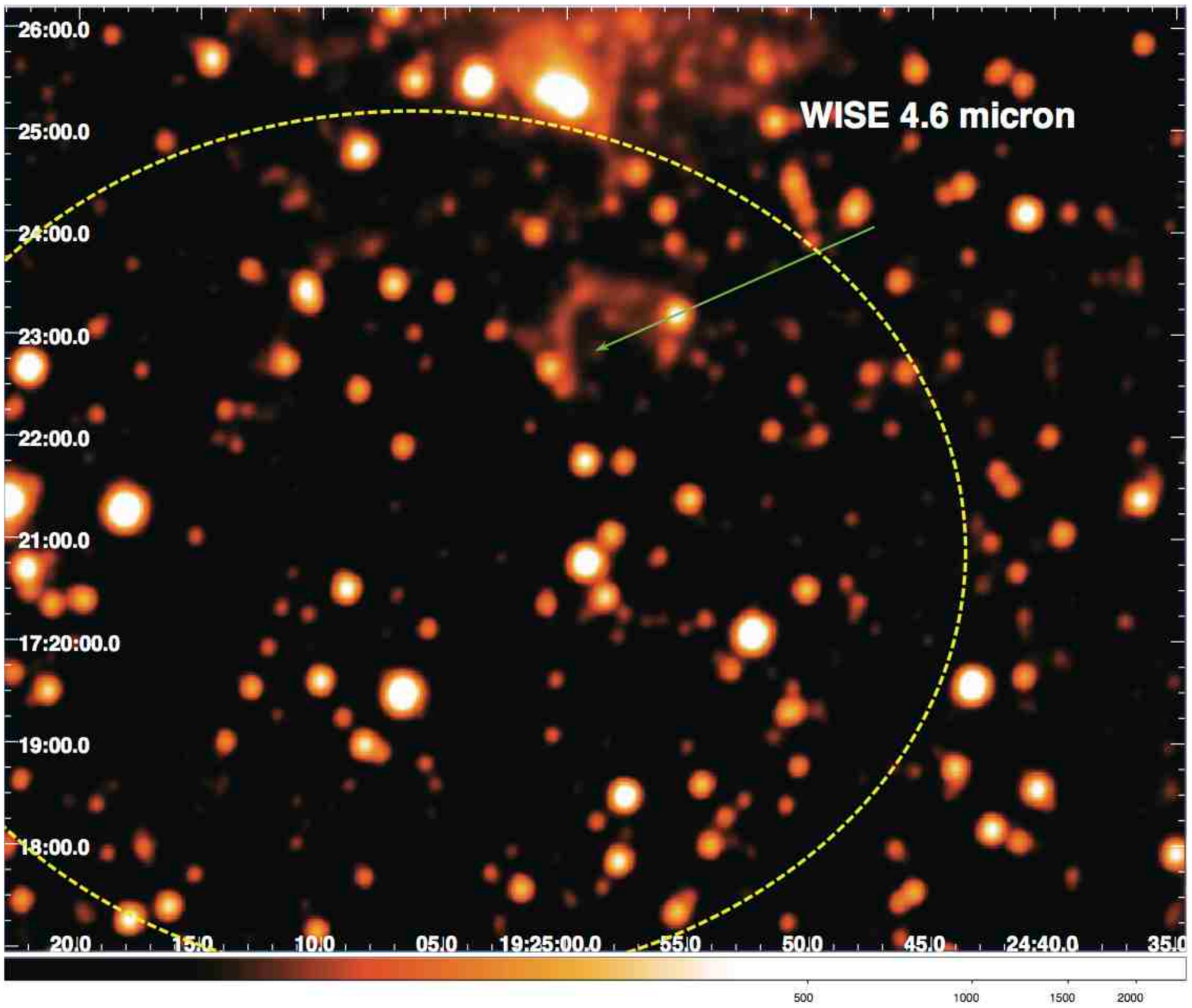}
          \includegraphics[height=7.2cm,width=8.4cm,angle=0]{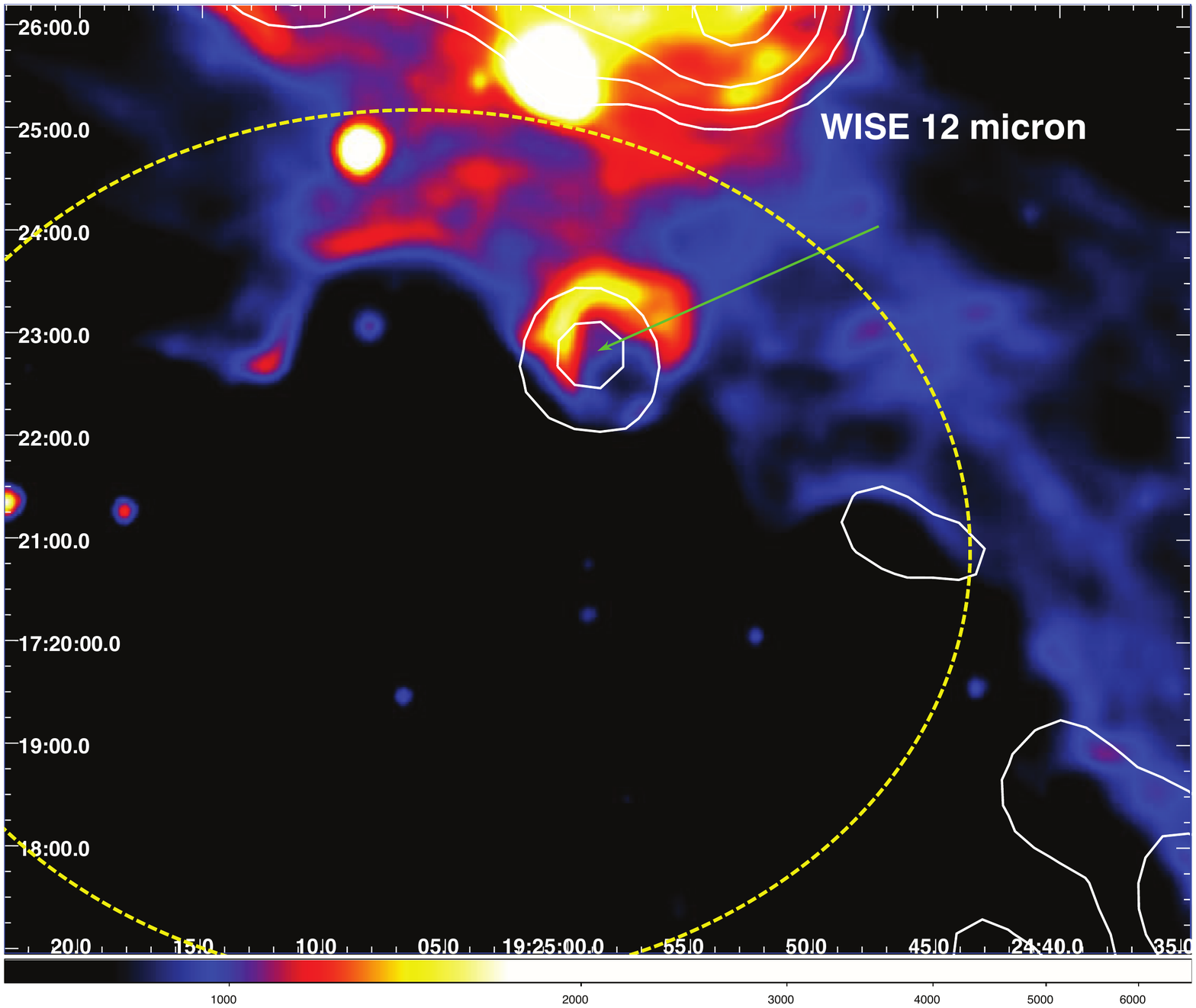}
           \caption{The \wse\ images of the \fer\ source 1FGLJ1925.0+1720c in two different bands at 4.6$\mu$m (top), 12$\mu$m (bottom).
                          In each image the yellow ellipse corresponds to the \fer\ positional uncertainty 
                          region at 68\% level of confidence. The white contours are overlaid on the \wse\ image at 12$\mu$m 
                          are the 1.4 GHz radio emission from the NVSS. The ring-shaped structure, indicated by the green arrow,
                          is extended for about 87\arcsec\ and resembles an unknown SNR or a PWN. It
                          clearly lies within the \fer\ positional uncertainty region and is also emitting in the radio band.
                          This is a clear example of a candidate association with a SFR including a potential SNR/PWN for which 
                          additional multifrequency observations are necessary to confirm its origin.}
           \label{fig:1FGLJ1925.0+1720c}
          \end{figure}

Then in Figures~\ref{fig:1FGLJ1726.2-3521c} and \ref{fig:1FGLJ1925.0+1720c} 
we report the best two cases of SFRs for which a signature of a shell-like or ring-like extended structures
was found within the \fer\ positional uncertainty region that could be ascribed as a shock signature 
potentially responsible for the $\gamma$-ray emission.

\begin{table}
\tiny
\caption{Source classification for the $\gamma$-ray counterparts.}
\begin{center}
\begin{tabular}{|lc|}
\hline
Class  &  code \\
\hline
\noalign{\smallskip}
GALACTIC SOURCES & \\
\noalign{\smallskip}
\hline
\noalign{\smallskip}
High mass X-ray binary & hmb\\
Globural cluster & glc\\
Nova & nov\\
Millisecond pulsar (MSP) & msp\\
Pulsar (PSR )& psr\\
Pulsar wind nebula (PWN) & pwn\\ 
Binary star & bin\\
Star forming-HII region (SFR) & sfr\\
{\it (for potential associations with unknown SNRs and/or PWNe and/or PSRs)} &\\
Supernova remnant (SNR) & snr\\
\hline
\noalign{\smallskip}
EXTRAGALACTIC SOURCES & \\
\noalign{\smallskip}
\hline
\noalign{\smallskip}
BL Lac object  (BZB) & bzb \\
Quasar radio loud with flat radio spectrum (BZQ) & bzq\\
Blazar of uncertain type (BZU)& bzu\\
Normal galaxy & gal\\
Radio galaxy & rdg \\
Starburst galaxy & sbg \\
Seyfert galaxy & sey \\
\hline
\noalign{\smallskip}
UNCERTAIN CLASSIFICATION & \\
\noalign{\smallskip}
\hline
\noalign{\smallskip}
Blazar candidate & bcn\\
Unclassified sources & unc\\
Unidentified gamma-ray source & ugs\\
\noalign{\smallskip}
\hline
\end{tabular}
\end{center}
\label{tab:classes}
\end{table}

\section{Correlation with existing databases}
\label{sec:crossmatches}
We searched the following major radio, infrared, optical and X-ray surveys 
and both the NASA Extragalactic Database (NED)\footnote{\underline{http://ned.ipac.caltech.edu/}} 
and the SIMBAD Astronomical Database\footnote{\underline{http://simbad.u-strasbg.fr/simbad/}}
to verify whether multifrequency information can confirm the natures of uncertain counterparts and blazar candidates.
{\bf In the following we list all the catalogs and the surveys searched for our investigation; 
the abbreviations correspond to the designators used for the multifrequency notes in Table~\ref{tab:main}.}
 
Below $\sim$1 GHz radio frequency we searched
the VLA Low-Frequency Sky Survey Discrete Source Catalog \citep[VLSS;][- V]{cohen07} 
and the recent revision VLLSr\footnote{\underline{http://heasarc.gsfc.nasa.gov/W3Browse/all/vlssr.html}}
\citep{lane14}, both the Westerbork Northern Sky Survey \citep[WENSS;][- W]{rengelink97} and the 
Westerbork in the Southern Hemisphere Source Catalog \citep[WISH;][- W]{debreuck02},
the Sydney University Molonglo Sky Survey \citep[SUMSS;][- S]{mauch03},
the Parkes-MIT-NRAO Surveys \citep[PMN;][- Pm]{wright94},
the Texas Survey of Radio Sources at 365 MHz \citep[TEXAS;][- T]{douglas96}
the Parkes Southern Radio Source catalog \citep[PKS;][- Pk]{wright90},
and the Low-frequency Radio Catalog of Flat-spectrum Sources \citep[LORCAT;][- L]{lorcat}.

For the radio counterparts (i.e., above $\sim$1 GHz), we searched the NRAO VLA Sky Survey \citep[NVSS;][- N]{condon98}, 
the VLA Faint Images of the Radio Sky at Twenty-Centimeters \citep[FIRST;][- F]{becker95,white97}, 
{\bf the 87 Green Bank catalog of radio sources \citep[87GB;][- 87]{gregory91} and the 
Green Bank 6-cm Radio Source Catalog \citep[GB6;][- GB]{gregory96},}
the Australia Telescope 20 GHz Survey \citep[AT20G;][- A]{murphy10},
the Combined Radio All-Sky Targeted Eight-GHz Survey \citep[CRATES;][- c]{healey07}
and the Australian Long Baseline Array (LBA) Calibrator Survey of southern compact extragalactic radio sources\footnote{\underline{http://astrogeo.org/lcs/}}
\citep[LCS1][- lcs]{petrov11}.

For the infrared, we queried the \wse\ all-sky survey in the  
AllWISE Source catalog\footnote{\underline{http://wise2.ipac.caltech.edu/docs/release/allwise/}} \citep[][- w]{wright10}
and the Two Micron All Sky Survey \citep[2MASS;][- M]{skrutskie06}
since each \wse\ source is automatically matched to the closest 2MASS potential counterpart \citep[see][for details]{cutri12}.

Then, we also searched for optical counterparts, with or without spectra available, 
in the Sloan Digitized Sky Survey Data Release 9 \citep[SDSS DR9; e.g.][- s]{ahn12}, 
in the Six-degree-Field Galaxy Redshift Survey \citep[6dFGS;][- 6]{jones04,jones09}.
We also queried the United States Naval Observatory (USNO)-B1 Catalog \citep[][- U]{monet03} for the optical counterparts 
of the bcn and the unc sources and reported their magnitudes 
since they could be useful to plan future follow up observations.

In X-rays, we searched the ROSAT all-sky survey 
in both the ROSAT Bright Source Catalog \citep[RBSC;][- X]{voges99} 
and the ROSAT Faint Source Catalog \citep[RFSC;][- X]{voges00}.
as well as \xmm\ Slew Survey \citep[XMMSL;][- x]{saxton08,warwick12}, 
the Deep Swift X-Ray Telescope Point Source Catalog \citep[1XSPS;][- x]{evans14}
the Chandra Source Catalog \citep[CSC;][- x]{evans10} and \swf\ X-ray survey for all the \fer\ 
UGSs \citep{stroh13}.
{\bf We used the same designator for the X-ray catalogs of \xmm, \chn\ and \swf.
These X-ray observatories perform only pointed observations thus there is the chance that the 
pointed observations present in their archives and related to the field of the \fer\ source was not requested as follow up of the $\gamma$-ray source but for a different reason.
Consequently, the discovery of the X-ray counterpart for an associated and/or identified source could be serendipitous.}

To perform the searches over all the surveys and the catalogs cited above we considered
the 1$\sigma$ positional uncertainties reported therein, with the only exceptions being the \wse\ all-sky survey
and the SDSS DR9. In these two cases we searched the closest IR and optical counterpart 
within a maximum angular separation of 3\arcsec.3 and of 1\arcsec.8
for the AllWISE survey and the SDSS DR9, respectively.
These values have been derived on the basis of the statistical analysis 
described in D'Abrusco et al. (2013) and Massaro et al. (2014b) 
developed following the approach also presented in Maselli et al. (2010) and Stephen et al. (2010).

To verify the presence of SFRs consistent 
with the positional uncertainty of the \fer\ sources, we searched in the following surveys and catalogs:
1) the Sharpless catalog of H\,{\sc ii} regions \citep{sharpless59},
2) the H\,{\sc ii} regions in the (InfraRed Astronomical Satellite (IRAS) point source catalog \citep{hughes89,codella94},
3) the radio survey of H\,{\sc ii} regions at 4.85 GHz \citep{kuchar97},
4) the new catalog of H\,{\sc ii} regions in the Milky Way \citep{giveon05},
5) the radio catalog of Galactic H\,{\sc ii} regions from decimeter to millimeter wavelength \citep{paladini03} 
6) the SCUBA imaging survey of ultra compact H\,{\sc ii} regions \citep{thompson06};
7) the recent \wse\ Catalog of Galactic H\,{\sc ii} Regions \citep{anderson14}
8) the QUaD Galactic Plane Survey \citep{culverhouse11}
9) the 5 GHz VLA survey of the Galacitc plane \citep{becker94}
and 10) the WSRTGP \citep{taylor96}.
We remark that for the cross-matching performed between the SFR catalogs and the \fer\ sources 
we took into account the size of each H\,{\sc ii} region in combination with the \fer\ uncertainty position.

To check for the presence of PSRs we used the Australia Telescope National Facility (ATNF) Pulsar Catalog \citep{manchester05},
the Catalogue of Galactic Supernova Remnants \citep{green09}, and
the census of high-energy observations of Galactic SNRs\footnote{\underline{http://www.physics.umanitoba.ca/snr/SNRcat/}} \citep{ferrand12}.

Table~\ref{tab:main} summarizes all the notes derived from the multifrequency analysis.
In particular, for the PSRs, we report the information shown in the Public List of LAT-Detected Gamma-Ray 
Pulsars\footnote{\underline{https://confluence.slac.stanford.edu/display/GLAMCOG/Public+List+of+LAT-Detected+Gamma-Ray+Pulsars}} 
and by the Pulsar Search Consortium (PSC) as well as from literature 
\citep[e.g.][]{hobbs04,espinoza13,ray12,ray13,pletsch12a,pletsch12b,pletsch13,hanabata14}. Symbols used for their 
multifrequency notes are labelled as reported in Table~\ref{tab:psr} \citep[see][for more details]{abdo13}.
\begin{table}
\caption{Multifrequency notes for pulsars.}
\begin{center}
\begin{tabular}{|cl|}
\hline
\noalign{\smallskip}
Symbol & Comment\\
\noalign{\smallskip}
\hline
\noalign{\smallskip}
b & PSR in a binary system \\
e & PSR detected in $\gamma$ rays by the Compton Gamma-ray Observatory \\
g & PSR discovered in LAT $\gamma$-ray data \\
m & MSP \\ 
p & Pulsar  discovered by the Pulsar Search Consortium (PSC) \\
r & PSR discovered in the radio \\
u & PSR discovered using a \fer-LAT seed position \\
x & PSR discovered in X-rays \\ 
\noalign{\smallskip}
\hline
\end{tabular}
\end{center}
\label{tab:psr}
\end{table}

\section{The refined associations for the \fer\ catalogs}
\label{sec:refined}
The refined associations for the 1FGL and the 2FGL catalog (hereinafter 1FGLR and 2FGLR, respectively) is presented in Table~\ref{tab:main}.
For each source we report: the 1FGL name together with the 2FGL and the 1FHL ones
as derived from the associations published in those catalogs, the counterpart name, and in 37 cases 
also an alternative association; the category of $\gamma$-ray source association; 
the class of each counterpart as found in the literature; notes derived from the multifrequency investigation; 
we also report whether the counterpart lies inside a SFR, and specifically for the PSRs in a PWN or a SNR.
In addition, in Table~\ref{tab:main} we also report if a PWN or a SNR contains a known PSR.
The coordinates of each counterpart are also included in Table~\ref{tab:main} together with  
the optical magnitudes in B and R band from the USNO-B1 Catalog \citep{monet03}.
Within the multifrequency notes in Table~\ref{tab:main} we also indicate if the spectral energy distribution (SED) of the source
is shown in Takeuchi et al. (2013), if a redshift estimate is present in literature (indicating the reference)
and if the radio counterpart has a flat radio spectrum (i.e., designated rf in the notes of Table~\ref{tab:main} 
whenever $\alpha<$0.5 in the radio band).
In the on-line version of the table we also report the infrared analysis flag \citep{wright10} for those sources, classified as blazar-like 
with a counterpart in the AllWISE catalog within 3\arcsec.3 as well as the column with the confirmed redshifts.

For the counterpart names we used the following priorities:
if the source is a known blazar we adopted the \bzcat\ nomenclature while for the known pulsars 
that of the ATNF Pulsar Catalog\footnote{\underline{http://www.atnf.csiro.au/research/pulsar/psrcat/}} 
\citep{manchester05}. Radio galaxies, Seyfert galaxies and starburst names were indicated
from the Third Cambridge Catalogue of Radio Sources and revised versions \citep[3C and 3CR][]{edge59,bennett62,spinrad85}, 
the New General catalog and the Messier catalog or using their radio names reported in one of the major surveys 
(see Section~\ref{sec:crossmatches}). Then Galactic sources as SNRs and PWNe have their names in Galactic coordinates
as reported in the Catalogue of Galactic Supernova Remnants\footnote{\underline{http://www.mrao.cam.ac.uk/surveys/snrs/}} \citep{green09},
while the most common names were used for globular clusters and binaries. 
The counterparts that were associated with SFRs were also labelled with their name in Galactic coordinates.
Finally, the remaining uncertain sources were identified by their radio names whenever possible or by 
their ROSAT name one. A handful of counterparts were indicated using the nomenclature of optical catalogs.
Adopting these choices in the names we remark that all the counterparts can be retrieved from the 
NASA/IPAC Extragalactic Database (NED) and SIMBAD archives without using their coordinates. 
\begin{table*}
\caption{Refined associations of the \fer\ catalogs (first 30 lines).}
\resizebox{\textwidth}{!}{
\begin{tabular}{|lllclllllllllllllllllll|}
\noalign{\smallskip}
\hline
\noalign{\smallskip}
1FGL  &  2FGL & 1FHL & category & counterpart1 & notes1 & z & inside & include & class (ctp1) & R.A. (ctp1) & Dec. (ctp1) & B (ctp1) & R (ctp1) & counterpart2 & notes2 & class (ctp2) & R.A. (ctp2) & Dec. (ctp2) & B (ctp2) & R (ctp2) & $\pi_{kde,bzb}$ & $\pi_{kde,bzq}$\\
name  &  name & name &               &          &                       & &             &           &             & (J2000) & (J2000) & mag & mag &                      &             & & (J2000) & (J2000) & mag & mag & & \\
\hline
Col. 1 & Col. 2 & Col. 3 & Col. 4 & Col. 5 & Col. 6 & Col. 7 & Col. 8 & Col. 9 & Col. 10 & Col. 11 & Col. 12 & Col. 13 & Col. 14 & Col. 15 & Col. 16 & Col. 17 & Col. 18 & Col. 19 & Col. 20 & Col. 21 & Col. 22 & Col. 23 \\ 
\hline
\noalign{\smallskip}
  1FGL J0000.8+6600c &  &  & U &  &  &  &  &  & ugs &  &  &  &  &  &  &  &  &  &  &  &  & \\
  1FGL J0000.9$-$0745 & 2FGL J0000.9$-$0748 &  & A & BZBJ0001$-$0746 &  & 0.0 &  &  & bcn & 00:01:18.01 & $-$07:46:26.90 & 18.14 & 17.19 &  &  &  &  &  &  &  & 84.2 & 22.5\\
  1FGL J0001.9$-$4158 & 2FGL J0001.7$-$4159 &  & C & 1RXS J000135.5$-$415519 & w,M,U,X &  &  &  & unc & 00:01:35.50 & $-$41:55:18.98 & 20.53 & 19.08 & SUMSS J000133$-$415524 & S,M,x $-$ SED in Takeuchi+13 & unc & 00:01:33.05 & $-$41:55:24.31 & 18.95 & 18.01 & 0.3 & 0.0\\
  1FGL J0003.1+6227 & 2FGL J0002.7+6220 &  & C & NVSS J000225+622937 & N,M,U &  &  &  & unc & 00:02:25.24 & +62:29:37.10 & 17.63 & 20.17 &  &  &  &  &  &  &  &  & \\
  1FGL J0004.3+2207 & 2FGL J0004.2+2208 &  & U &  &  &  &  &  & ugs &  &  &  &  &  &  &  &  &  &  &  &  & \\
  1FGL J0004.7$-$4737 & 2FGL J0004.7$-$4736 &  & A & BZQJ0004$-$4736 &  & 0.88 &  &  & bzq & 00:04:35.68 & $-$47:36:18.61 &  &  &  &  &  &  &  &  &  &  & \\
  1FGL J0005.1+6829 & 2FGL J0007.7+6825c &  & C & WISE J000345.96+681640.9 & w &  &  &  & unc & 00:03:45.96 & +68:16:40.91 & 17.63 & 15.63 &  &  &  &  &  &  &  & 75.1 & 0.3\\
  1FGL J0005.7+3815 & 2FGL J0006.1+3821 &  & A & BZQJ0005+3820 &  & 0.229 &  &  & bzq & 00:05:57.18 & +38:20:15.11 &  &  &  &  &  &  &  &  &  &  & \\
  1FGL J0006.9+4652 &  &  & U &  &  &  &  &  & ugs &  &  &  &  &  &  &  &  &  &  &  &  & \\
  1FGL J0007.0+7303 & 2FGL J0007.0+7303 & 1FHL J0007.3+7303 & I & PSR J0007+7303 & g &  & GBH41.83992+0.02242 and G110.21+2.63 & SNR G119.5+10.2 & psr in snr & 00:06:34.20 & +73:11:06.60 &  &  &  &  &  &  &  &  &  &  & \\
  1FGL J0008.3+1452 &  &  & U &  &  &  &  &  & ugs &  &  &  &  &  &  &  &  &  &  &  &  & \\
  1FGL J0008.9+0635 & 2FGL J0009.0+0632 &  & A & BZBJ0009+0628 & V,T,N,87,w,M,s,U,g,c,rf $-$$-$> (z=? $-$ BL Lac $-$ Ahn+12,Massaro+14) & 0.0 &  &  & bcn & 00:09:03.93 & +06:28:21.24 & 19.06 & 18.42 &  &  &  &  &  &  &  & 64.9 & 21.3\\
  1FGL J0009.1+5031 & 2FGL J0009.1+5030 & 1FHL J0009.2+5032 & A & NVSS J000922+503028 & L,N,U,x $-$ SED in Takeuchi+13 $-$$-$> (z=? $-$ BL Lac $-$ Shaw+13) & 0.0 &  &  & bcn & 00:09:22.53 & +50:30:28.91 & 19.74 & 19.35 &  &  &  &  &  &  &  & 45.5 & 0.0\\
  1FGL J0011.1+0050 & 2FGL J0011.3+0054 &  & A & BZQJ0011+0057 &  & 1.492 &  &  & bzq & 00:11:30.40 & +00:57:51.80 &  &  &  &  &  &  &  &  &  &  & \\
  1FGL J0013.1$-$3952 & 2FGL J0012.9$-$3954 &  & A & BZBJ0012$-$3954 &  & 0.0 &  &  & bcn & 00:12:59.91 & $-$39:54:26.10 & 18.14 & 18.05 &  &  &  &  &  &  &  & 63.1 & 74.3\\
  1FGL J0013.7$-$5022 &  &  & A & BZBJ0014$-$5022 &  & 0.0 &  &  & bcn & 00:14:11.22 & $-$50:22:32.59 & 19.36 & 18.82 &  &  &  &  &  &  &  & 13.1 & 0.2\\
  1FGL J0016.6+1706 &  &  & U &  &  &  &  &  & ugs &  &  &  &  &  &  &  &  &  &  &  &  & \\
  1FGL J0017.4$-$0510 & 2FGL J0017.6$-$0510 &  & A & BZQJ0017$-$0512 &  & 0.227 &  &  & bzq & 00:17:35.82 & $-$05:12:41.69 &  &  &  &  &  &  &  &  &  &  & \\
  1FGL J0017.7$-$0019 & 2FGL J0017.4$-$0018 &  & A & BZQJ0016$-$0015 &  & 1.577 &  &  & bzq & 00:16:11.09 & $-$00:15:12.38 &  &  &  &  &  &  &  &  &  &  & \\
  1FGL J0018.6+2945 & 2FGL J0018.5+2945 & 1FHL J0018.6+2946 & A & BZBJ0018+2947 &  & 0.1 &  &  & bzb & 00:18:27.75 & +29:47:30.41 &  &  &  &  &  &  &  &  &  &  & \\
  1FGL J0019.3+2017 &  &  & A & BZBJ0019+2021 &  & 0.0 &  &  & bzb & 00:19:37.85 & +20:21:45.61 &  &  &  &  &  &  &  &  &  &  & \\
  1FGL J0021.7$-$2556 & 2FGL J0021.6$-$2551 &  & A & BZBJ0021$-$2550 &  & 0.0 &  &  & bcn & 00:21:32.50 & $-$25:50:49.20 & 16.86 & 17.51 &  &  &  &  &  &  &  & 93.3 & 2.4\\
  1FGL J0022.2$-$1850 & 2FGL J0022.2$-$1853 & 1FHL J0022.2$-$1853 & A & NVSS J002209$-$185332 & N,6,U,g,X $-$$-$> (z=? $-$ BL Lac $-$ Shaw+13) & 0.0 &  &  & bcn & 00:22:09.16 & $-$18:53:32.78 & 18.07 & 16.95 &  &  &  &  &  &  &  & 99.4 & 3.6\\
  1FGL J0022.5+0607 & 2FGL J0022.5+0607 & 1FHL J0022.5+0607 & A & BZBJ0022+0608 &  & 0.0 &  &  & bzb & 00:22:32.44 & +06:08:04.31 &  &  &  &  &  &  &  &  &  &  & \\
  1FGL J0023.0+4453 & 2FGL J0023.2+4454 &  & A & BZQJ0023+4456 &  & 1.062 &  &  & bzq & 00:23:35.44 & +44:56:35.81 &  &  &  &  &  &  &  &  &  &  & \\
  1FGL J0023.5+0930 & 2FGL J0023.5+0924 &  & A & PSR J0023+0923 & mbrup &  &  &  & msp & 00:23:16.89 & +09:23:24.18 &  &  &  &  &  &  &  &  &  &  & \\
  1FGL J0023.9$-$7204 & 2FGL J0023.9$-$7204 &  & A & GCl 1 (47 Tuc) &  &  &  &  & glc & 00:24:05.36 & $-$72:04:53.20 &  &  &  &  &  &  &  &  &  &  & \\
  1FGL J0024.6+0346 & 2FGL J0024.5+0346 &  & A & BZQJ0024+0349 &  & 0.545 &  &  & bzq & 00:24:45.21 & +03:49:03.50 &  &  &  &  &  &  &  &  &  &  & \\
  1FGL J0028.9$-$7028 &  &  & U &  &  &  &  &  & ugs &  &  &  &  &  &  &  &  &  &  &  &  & \\
  1FGL J0029.9$-$4221 & 2FGL J0030.2$-$4223 &  & A & BZQJ0030$-$4224 &  & 0.495 &  &  & bzq & 00:30:17.58 & $-$42:24:46.01 &  &  &  &  &  &  &  &  &  &  & \\
\hline
\noalign{\smallskip}
\end{tabular}}
Column description. 
(1): 1FGL name;
(2): 2FGL name;
(3): 1FHL name;
(4): $\gamma$-ray source category (I=identified; A=associated; C=candidate; U=unidentified; see Section~\ref{sec:cat} for details);
(5): name of the first low-energy counterpart;
(6): multifrequency notes on the first low-energy counterpart; 
(7): redshift estimate if the source is extragalactic (in the on line version of the catalog there is also a note if the $z$ measurement is uncertain);
(8): notes if the first low-energy counterpart lies inside a SFR/SNR/PWN;
(9): notes if the first low-energy counterpart includes a PSR/MSP;
(10): source class for the first low-energy counterpart (see Section~\ref{sec:assoc} for details);
(11): R.A. (J2000) for the first low-energy counterpart;
(12): Dec. (J2000) for the first low-energy counterpart;
(13): B magnitude for the first low-energy counterpart taken from the USNO-B1 Catalog \citep{monet03}, only for the bcn and the unc sources;
(14): R magnitude for the first low-energy counterpart taken from the USNO-B1 Catalog \citep{monet03}, only for the bcn and the unc sources;
(15): name of the second low-energy counterpart;
(16): multifrequency notes on the second low-energy counterpart; 
(17): source class for the second low-energy counterpart (see Section~\ref{sec:assoc} for details);
(18): R.A. (J2000) for the second low-energy counterpart;
(19): Dec. (J2000) for the second low-energy counterpart;
(20): B magnitude for the second low-energy counterpart taken from the USNO-B1 Catalog \citep{monet03}, only for the bcn and the unc sources;
(21): R magnitude for the second low-energy counterpart taken from the USNO-B1 Catalog \citep{monet03}, only for the bcn and the unc sources;
(22): probability that the first low-energy counterpart has infrared colors similar to the known 1FGL BZBs estimated with the KDE method (see Section~\ref{sec:kde} for details);
(23): probability that the first low-energy counterpart has infrared colors similar to the known 1FGL BZQs estimated with the KDE method (see Section~\ref{sec:kde} for details).

Symbols used for the multifrequency notes are all reported in Section~\ref{sec:crossmatches} 
together with the references of the catalogs/surveys.

References for the optical spectra reported in the table: 
Ackermann et al. (2011), 
Ahn et al. (2012), 
Baker et al. (1995),
Bauer et al. (2000), 
Bade et al. (1995), 
Beuermann et al. (1999), 
Bikmaev et al. (2008), 
Britzen et al. (2007), 
Drinkwater et al. (1997), 
Healey et al. (2008),
Hewett \& Wild (2010), 
Hewitt \& Burbidge (1989), 
Johnston et al. (1995), 
Jones et al. (2004), 
Jones et al. (2009), 
Landoni et al. (2014) 
Landt et al. (2001), 
Lister et al. (2011), 
Mao (2011), 
Maza et al. (1995), 
Marti et al. (2004), 
Masetti et al. (2013), 
Massaro et al. (2014c) 
Mitton et al. (1977), 
Paggi et al. (2013), 
Quintana \& Ramirez (1995), 
Schwope et al. (2000), 
Shaw et al. (2013a), 
Shaw et al. (2013b), 
Stern \& Assef (2013), 
Stickel \& Kuehr (1996a,b) 
Titov et al. (2011), 
Vandenbroucke et al. (2010), 
Vettolani et al. (1989), 
Takeuchi et al. (2013), 
Thompson et al. (1990), 
Tsarevsky ert al. (2005), 
White et al. (1988). 
\label{tab:main}
\end{table*}

\subsection{The refined associations of the 1FGL}
\label{sec:1FGLR}
\begin{figure*}[] 
          \includegraphics[height=6.8cm,width=9.5cm,angle=0]{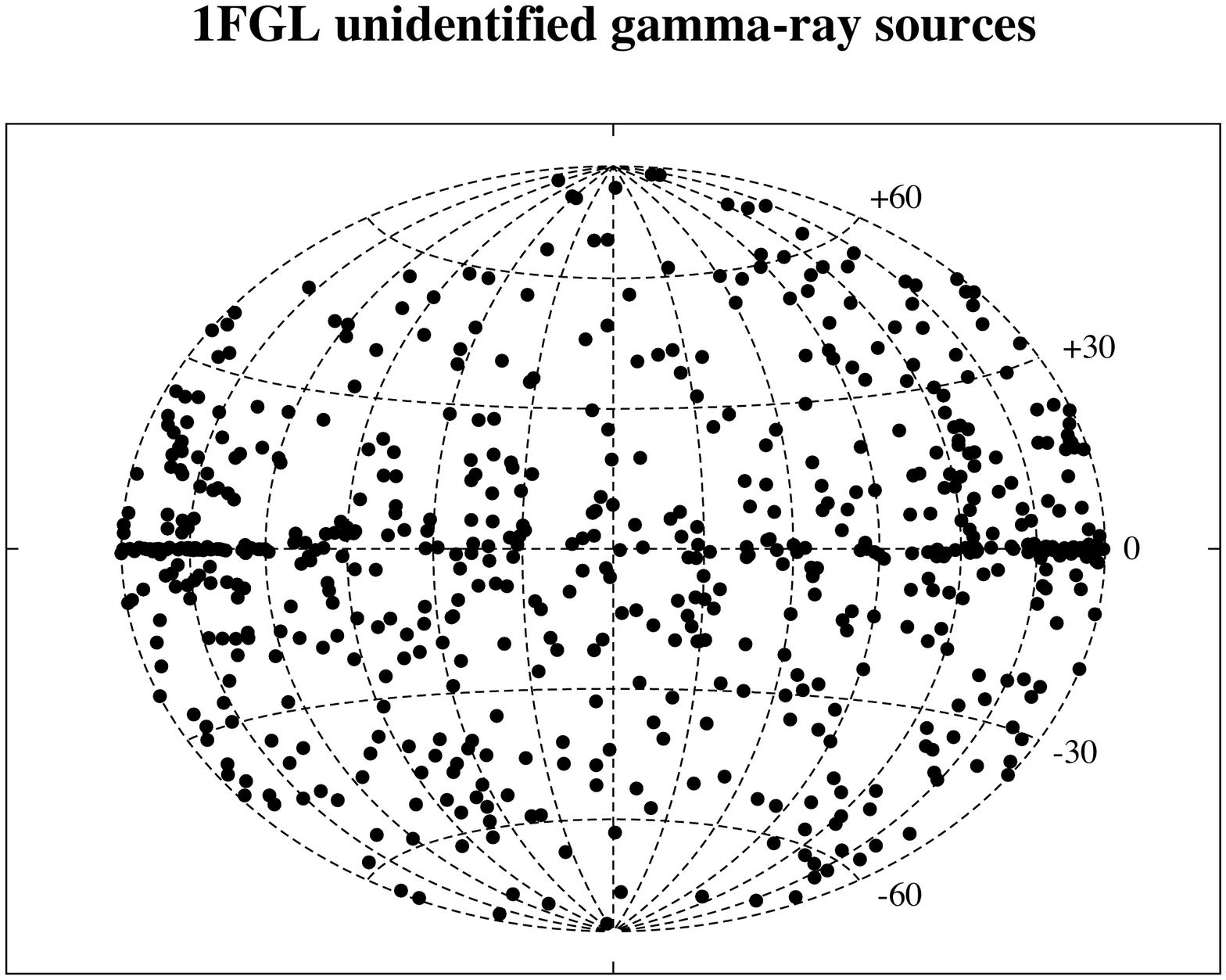}
          \includegraphics[height=6.8cm,width=9.5cm,angle=0]{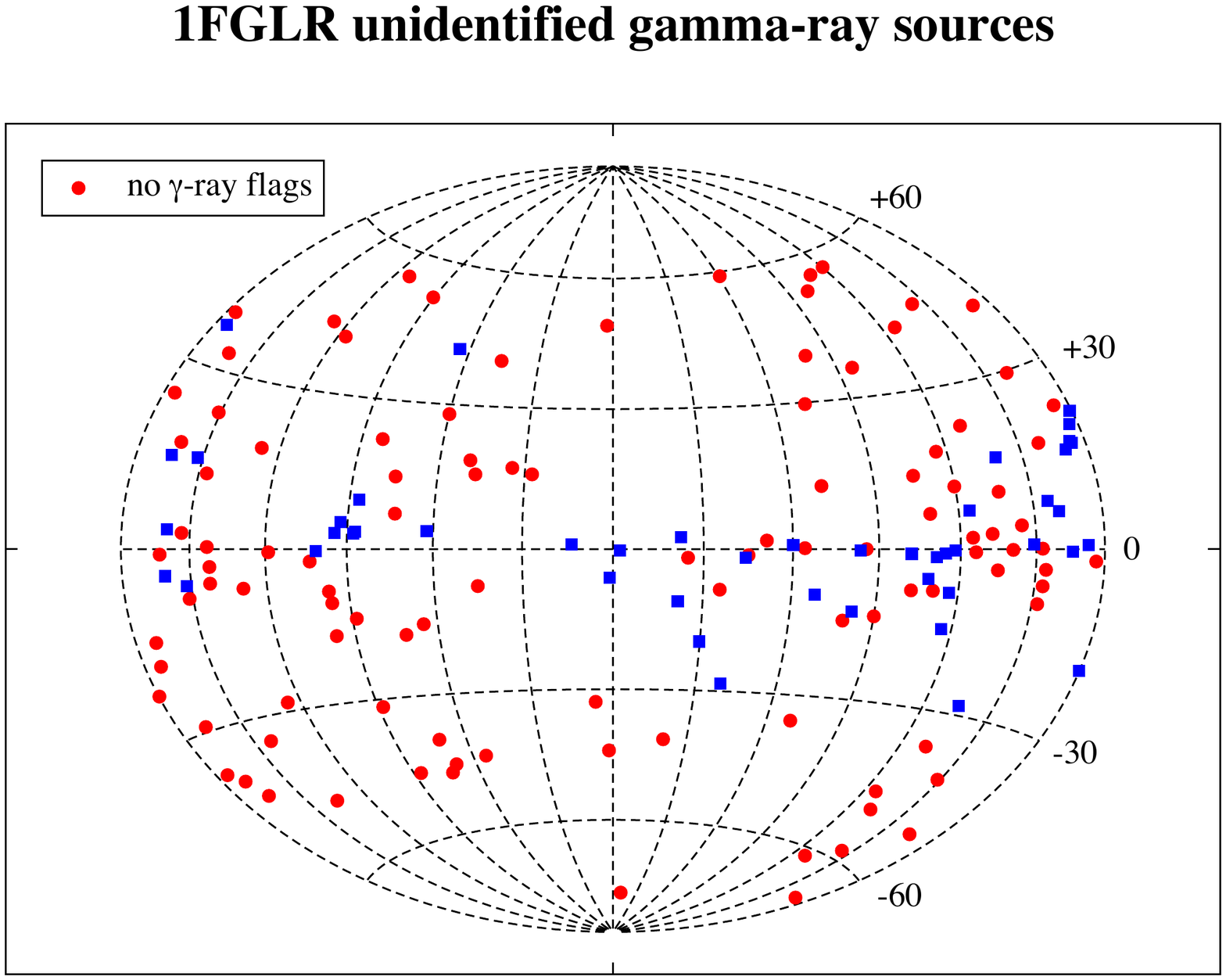}
           \caption{The all-sky distribution of the UGSs in the 1FGL (left) 
                        in comparison with those listed in the 1FGLR (right panel). 
                        {\bf Red circles mark those UGSs with no $\gamma$-ray analysis flags and blue squares those with a flag.}
                        The projection is Hammer-Aitoff projection in Galactic coordinates }
           \label{fig:skydist1}
          \end{figure*}

The 1FGLR lists 111 identified sources, 880 associated, 306 candidate associations, and 154 UGSs.
The sky distribution of the UGSs is shown in Figure~\ref{fig:skydist1} in comparison with the 1FGL associations in which
there were 66 identified objects, 755 associated, and 630 unidentified. 
The results of our multifrequency analysis for the 1FGLR are summarized in Table~\ref{tab:results} 
where we list all the sources identified, associated, and candidate associations for each class separately.
We remark that the globular cluster GCl 94 (alias M28) associated with 1FGL J1824.5$-$2449
contains a known MSP PSR J1824$-$2452A while 2MASS$-$GC01 lies in an SFR.

Among the PSRs and the MSPs are 21 sources that lie within a known PWN 
and 4 in SNRs as reported in the notes of Table~\ref{tab:main}.
Twenty PSRs and MSPs out of 39 indicated as candidate associations in both \fer\ catalogs 
have been identified after the release of the 2FGL catalog, 
as presented in the Public List of LAT-Detected Gamma-Ray Pulsars.
This strongly supports our introduction of this new category of $\gamma$-ray sources.
In five of the 10 associations of PWNe a PSR 
also has been found within the nebula
and the same situation occurs for 8 SNRs.

We find that 113 1FGL sources have an SFR within their positional uncertainty regions 
obtained by combining in quadrature the $\gamma$-ray localization at 95\% level of confidence with the size of the SFR.
Three of these sources are the HMBs and one is the globular cluster 2MASS$-$GC01.
Fifty-six of these 113 have been tentatively associated (i.e., indicated as candidate associations)
with previously UGSs while the remaining \fer\ objects all have PSRs, PWNe and/or a SNRs 
embedded in gas clouds and/or interacting with the interstellar 
medium\footnote{see also \underline{http://astronomy.nju.edu.cn/~ygchen/others/bjiang/interSNR6.htm}}. 
Among the 54 candidate associations with SFRs, two contain SNRs, one a PWN, and one a known PSR.
However, the potential associations with SFRs could be used to refine the $\gamma$-ray Galactic diffuse emission models 
and the associations in future releases of the \fer\ catalogs,
as well as lead to the discovery of new PWNe and/or unknown SNRs.
None of the MSPs has been found lying within a known SFR 
while this occurs for 21 PSRs, 7 PWNe, and 18 SNRs.

It is worth noting that 1FGL J0503.2+4526, associated according to the 2FGL and the 2LAC analyses
with an unclassified source, has as an alternative potential counterpart a SNR that includes a PSR.
A similar situation occurs for 1FGL J1837.5$-$0659c and 1FGL J1846.8$-$0233c
whose $\gamma$-ray emission could be ascribed to a PSR in a PWN. 
In addition, 1FGL J0622.2+3751 has as an alternative association with a PSR. 

Within the unclassified candidate associations (i.e., unc), 
we noticed that 8 1FGLR sources are positionally consistent with SFRs.
In addition, we also listed in the 1FGLR 
the tentative association of 1FGL J1653.6$-$0158 (also known as 2FGL J1653.6$-$0159) 
with the binary system that includes PSR J1653$-$0158 \citep{romani14}. 

{\bf Finally, we remark that in the extragalactic sky 
the Small Magellanic Cloud (alias NGC 292) is identified with the \fer\ source
1FGL J0101.3$-$7257 and classified as a normal galaxy while the Large Magellanic Cloud is associated 
with all the remaining five \fer\ sources listed as normal galaxies.}

\begin{table}
\tiny
\caption{Summary of categories for the $\gamma$-ray source associations in the refined association list of both \fer\ catalogs.}
\begin{center}
\begin{tabular}{|l|rrrr|rrrr|}
\hline
  &  & 1FGLR & & & & 2FGLR & & \\
\hline
\hline
Class  &  ID. & ASS. & CAN. & Tot. &  ID. & ASS. & CAN. & Tot.\\
\hline
\noalign{\smallskip}
GALACTIC & & & & & & & & \\
\noalign{\smallskip}
\hline
\noalign{\smallskip}
bin  &  0  &  1  &  0  &  1  &  0  &  1  &  0  &  1 \\  
hmb  &  3  &  0  &  1  &  4  &  4  &  0  &  0  &  4 \\  
glc  &  0  &  9  &  0  &  9  &  0  & 11  &  0  & 11 \\  
nov  &  0  &  0  &  0  &  0  &  1  &  0  &  0  &  1 \\  
msp  & 26  & 16  &  8  & 50  & 26  & 17  &  8  & 51 \\  
psr  & 53  &  8  & 31  & 92  & 56  & 12  & 30  & 98 \\  
pwn  &  0  &  8  &  2  & 10  &  2  &  6  &  1  &  9 \\  
snr  &  3  & 38  &  5  & 46  &  6  & 55  &  5  & 66 \\  
sfr  &  0  &  0  & 54  & 54  &  0  &  0  & 63  & 63 \\  
\hline
\noalign{\smallskip}
EXTRAGALACTIC & & & & & & & & \\
\noalign{\smallskip}
\hline
\noalign{\smallskip}
bzb  &  7  & 231  &  1  & 239  &  7  & 270  &  0  & 277 \\  
bzq  & 15  & 271  &  1  & 287  & 18  & 320  &  0  & 338 \\  
bzu  &  2  & 46  &  0  & 48  &  2  & 52  &  0  & 54 \\  
gal  &  1  &  5  &  0  &  6  &  2  &  4  &  0  &  6 \\  
rdg  &  0  &  5  &  0  &  5  &  1  &  7  &  0  &  8 \\  
sbg  &  0  &  3  &  0  &  3  &  0  &  3  &  0  &  3 \\  
sey  &  0  &  2  &  0  &  2  &  0  &  3  &  0  &  3 \\  
\hline
\noalign{\smallskip}
UNCERTAIN & & & & & & & &\\
\noalign{\smallskip}
\hline
\noalign{\smallskip}
bcn  &  0  &188  & 25  & 213  &  0  & 242  & 29  & 271 \\ 
unc  &  1  & 49  &178  & 228  &  1  & 173  &146  & 320 \\ 
ugs  &  0  &  0  &  0  & 154  &  0  &  0  &  0  & 289 \\  
\noalign{\smallskip}
\hline
\end{tabular}
\end{center}
Col. (1) Source class.
Col. (2) ID.= identified. 
Col. (3) ASS.= associated. 
Col. (4) CAN.= candidate associations.
\label{tab:results}
\end{table}

\subsection{The refined associations of the 2FGL}
\label{sec:2FGLR}
In the analysis of the 1FGLR we have already considered the associations for 1099 \fer\ sources 
that are in common between the 1FGL and the 2FGL, however, 
we report the results for the whole refined 2FGL associations (hereinafter 2FGLR) in Table~\ref{tab:results}.
As for the 1FGLR, the 2FGLR catalog
is dominated by blazar-like sources in the extragalactic sky and by pulsars around the Galactic plane.
          \begin{figure*}[] 
          \includegraphics[height=6.8cm,width=9.5cm,angle=0]{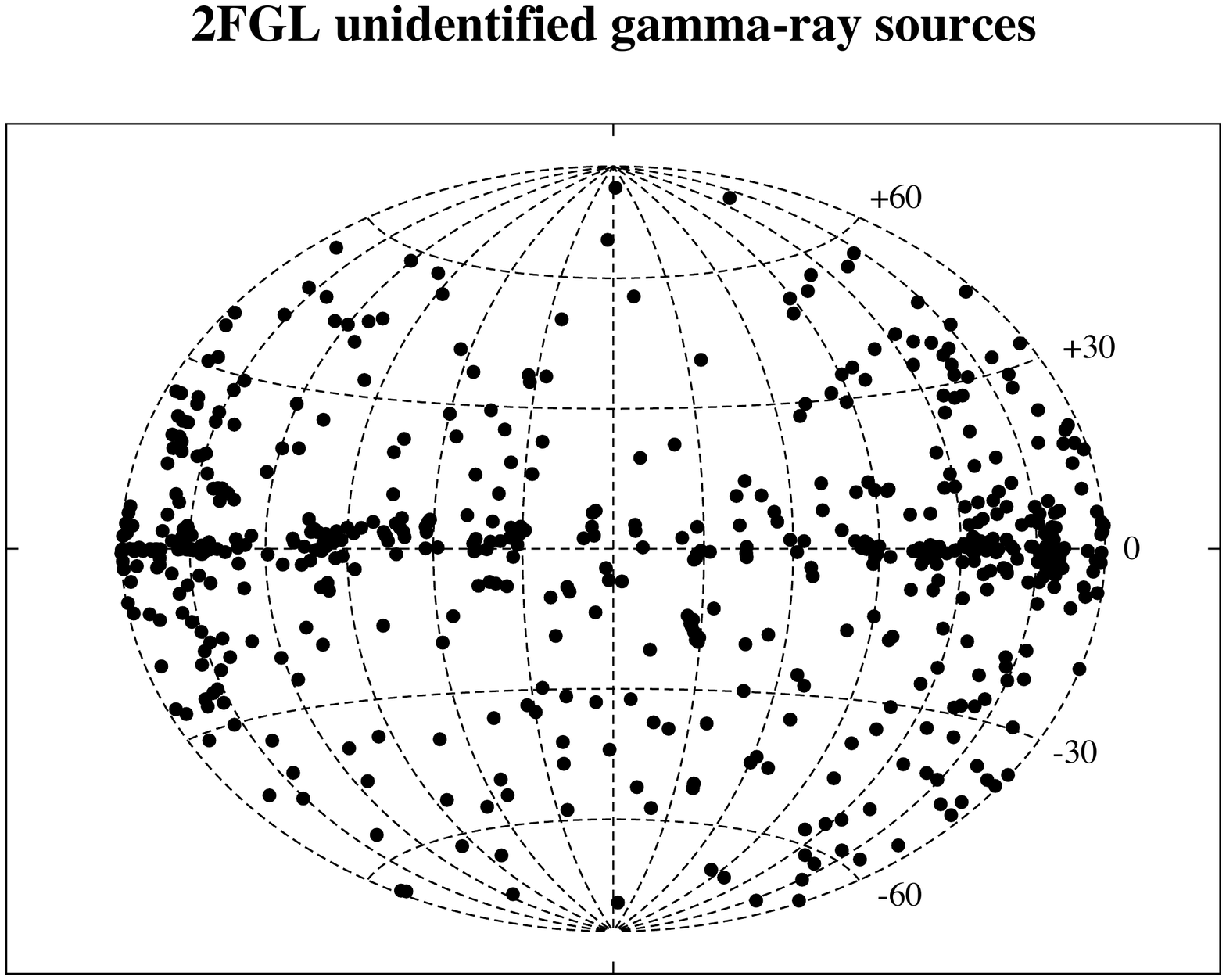}
          \includegraphics[height=6.8cm,width=9.5cm,angle=0]{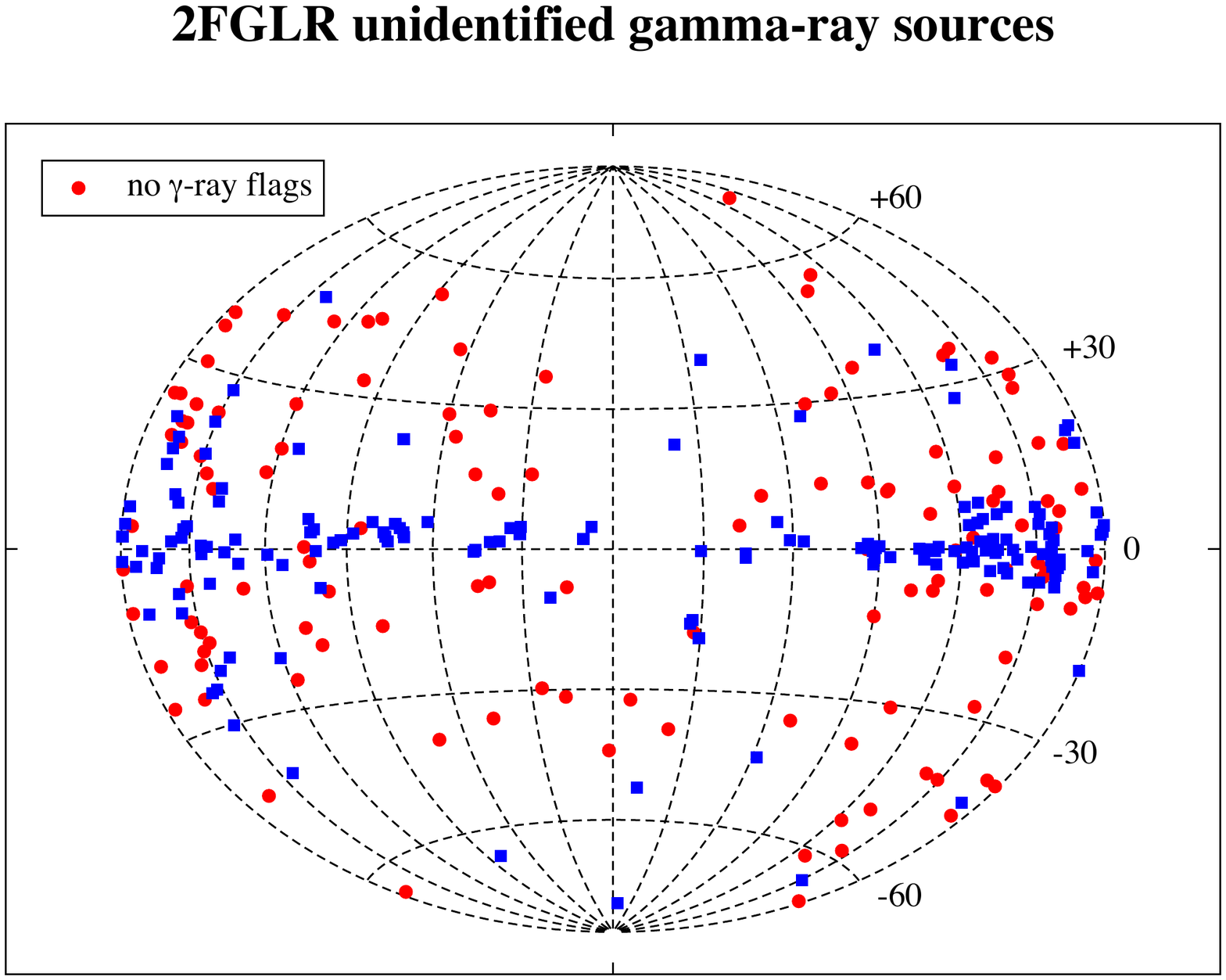}
           \caption{All-sky distribution of the UGSs in the 2FGL (left) 
                        in comparison with those listed in the 2FGLR (right panel).
                        Red circles mark those UGSs with no $\gamma$-ray analysis flags and blue squares those with a flag.}
           \label{fig:skydist2}
          \end{figure*}
The 2FGLR lists 126 identified sources as in the original catalog, 1176 associated, 282 candidate associations and 289 UGSs.
{\bf The 2FGLR sources are classified as}:
4 high mass X-ray binaries,11 globular clusters, 51 MSPs, and 98 PSRs,
21 that lie in a PWN and 6 in an SNR, and 17 associated with a SFR indicated in our Table~\ref{tab:main}. 
In the case of 2FGLR, among the PSR candidate associations that are not in the 1FGLR only one source
has been recently identified according to the Public List of LAT-Detected Gamma-Ray 
Pulsars \citep[see also][]{abdo13}.
There are also 9 PWNe and 66 sources associated with SNRs,
11 of them with a known PSR lying inside the remnant. 
Then we list 63 \fer\ sources as candidate associations with SFRs, one also including a SNR.
With respect to the 1FGL there is also a new association with a nova \citep[e.g.,][]{cheung14}.

The extragalactic sky includes 277 BZBs, 338 BZQs, 54 BZUs, 3 Seyfert galaxies, and 3 starburst galaxies. 
Four \fer\ sources are associated with the $\gamma$-ray emission
arising from the Large Magellanic cloud, as occurred for the objects classified as galaxy in the 1FGLR, 
and one with M31 (i.e., the Andromeda Galaxy) \citep{abdo10d}.
In addition we find 271 blazar candidates (bcn), 320 \fer\ unclassified sources and 289 UGSs. 
In Figure~\ref{fig:skydist2} we report the comparison between the sky distributions of all the UGSs previously listed in the 2FGL
and the remaining ones determined in this refined association analysis.

\section{Infrared colors of $\gamma$-ray blazar candidates and unclassified sources}
\label{sec:kde}          
As already done for the AGUs \citep{paper3} 
and the UGSs \citep[e.g.,][]{paggi13} listed in the 2FGL
we performed a non-parametric analysis 
of the infrared colors for the \fer\ sources classified as bcn and unc
 in the merged \fer\ catalog 1FGLR+2FGLR.
We used  kernel density estimation \citep[KDE; see e.g.,][and reference therein]{dabrusco09,laurino11}
to verify the consistency of their 
infrared colors with the so-called \wse\ Gamma-ray Strip \citep[e.g.,][]{paper1,paper3,ugs5} 
The KDE technique allows us to estimate the probability function of a multivariate
distribution and does not require any assumption about the shape of the ``parent" distributions. 

Consequently, for all the bcn and unc sources with a \wse\ counterpart 
we can provide an estimate of the probability 
$\pi_{kde}$ that such source is consistent with the \wse\ Gamma-ray Strip.
In particular we differentiated between $\pi_{kde,bzb}$ 
and $\pi_{kde,bzq}$ considering the comparison with the infrared colors of the 
BZB and BZQ subclasses, respectively. 
We considered as training samples to build the \wse\ 
Gamma-ray Strip\citep{paper1,ugs2} and to compare the infrared colors, 
the \bzcat\ blazars listed in both 1FGLR and 2FGLR.
We remark that \wse\ counterparts of the bcn and of the unc 
sources with the infrared analysis flags\footnote{As for example contamination and confusion from nearby bright sources, see e.g., \underline{http://wise2.ipac.caltech.edu/docs/release/allsky/expsup/sec6\_3a.html} for additional details} \citep{wright10}
have been considered in our analysis and their flags are reported in Table~\ref{tab:main} 
together with the probabilities derived from the KDE analysis.

\begin{figure}[]
\includegraphics[height=7.cm,width=9.6cm,angle=0]{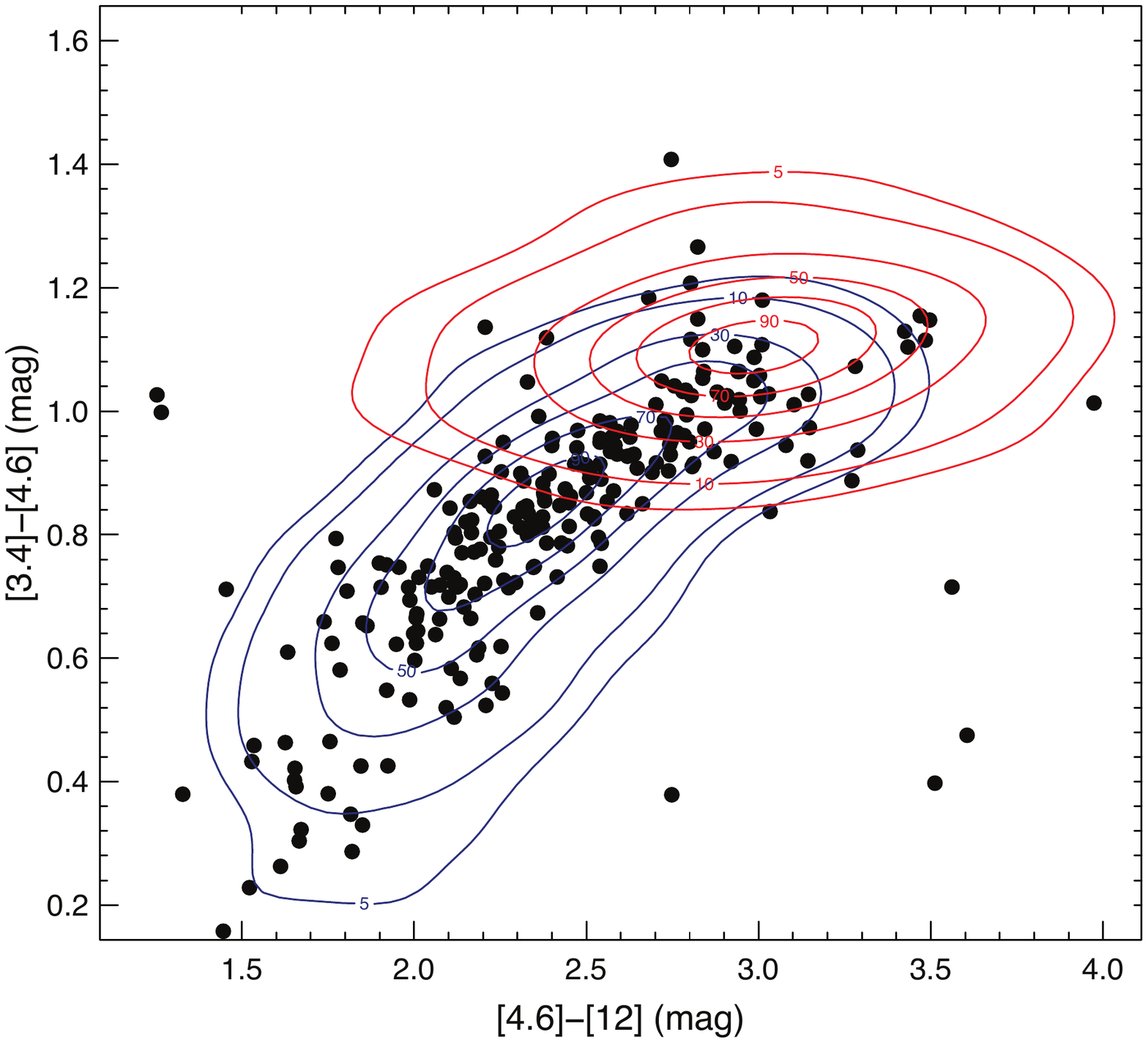}
\includegraphics[height=7.cm,width=9.6cm,angle=0]{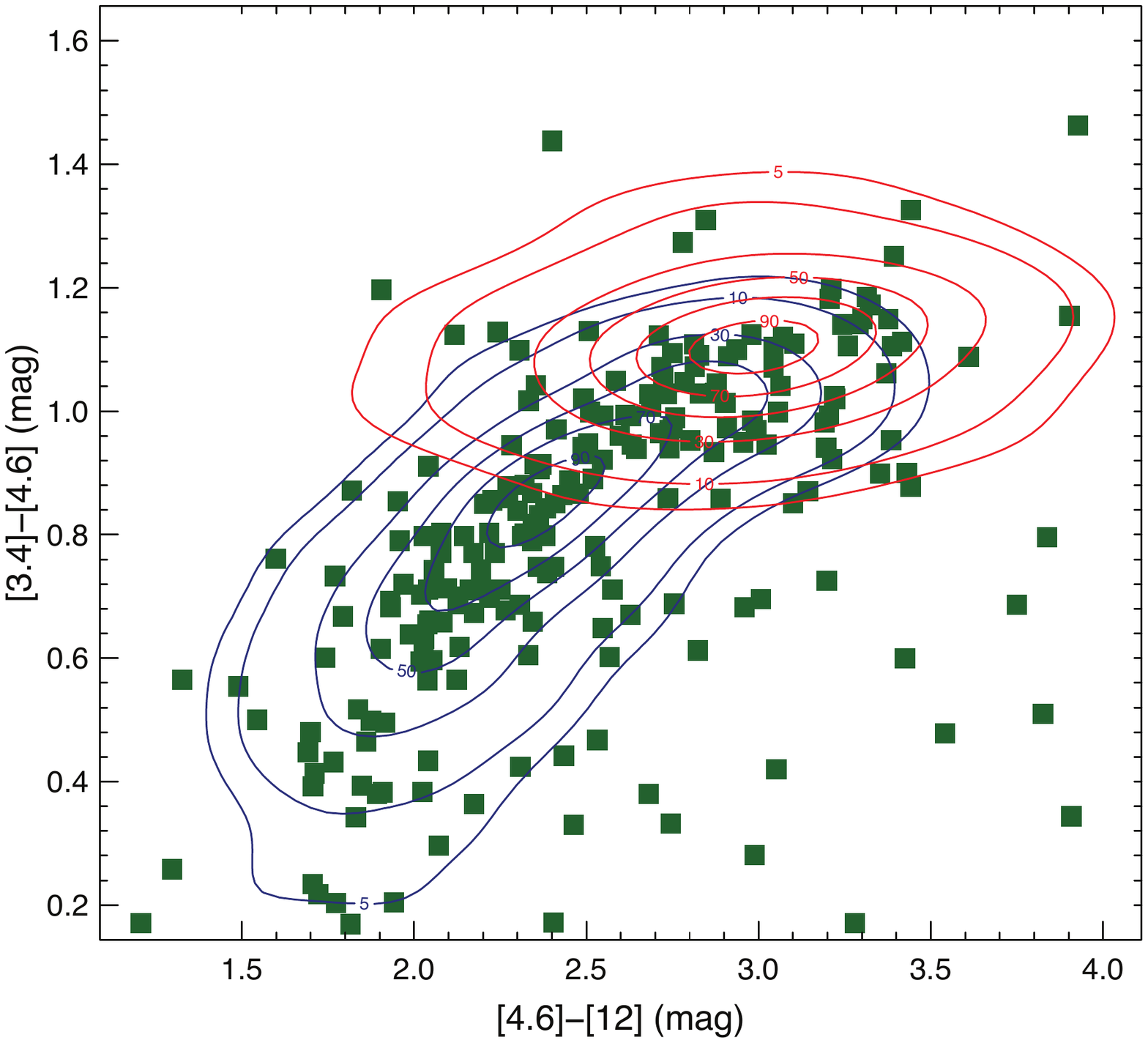}
\caption{The isodensity contours drawn from the KDE technique 
for the \fer-\bzcat\ blazars (i.e., training sample) that constitute the \wse\ Gamma-ray Strip \citep{paper1}
in the [3.4]-[4.6]-[12] $\mu$m color-color plot. 
Blue contours were computed from the infrared colors of the BZBs while the red ones were derived from the BZQs.
The infrared colors of the sources classified as bcn (top - black circles) and unc (bottom - green squares) 
are also reported to show their consistency with the \wse\ Gamma-ray Strip.
The numbers appearing close to each contour correspond to the values of $\pi_{kde}$ in both panels.
The last contour correspond to 95\% confidence level (see Section~\ref{sec:kde} for details).
A similar analysis for the active galaxies of uncertain type listed in the 2FGL was already performed in Massaro et al. (2012a).
Here this study has been repeated for both the
1FGLR and 2FGLR lists.
 refined association list of the \fer\ catalogs.
 }
\label{fig:kde}
\end{figure}

In Figure~\ref{fig:kde}, the isodensity contours drawn from the KDE density probabilities 
are plotted for the \fer-\bzcat\ blazars in the [3.4]-[4.6]-[12] $\mu$m color-color diagram. 
The IR colors of the sources listed in the refined association list of the \fer\ catalogs classified as  bcn and unc are also shown.
Values of  $\pi_{kde,bzb}$ and $\pi_{kde,bzq}$ greater than 5\% 
indicate that the source has infrared colors consistent with the portion \wse\ Gamma-ray Strip 
constituted by the BZBs and by the BZQs, respectively, at 95\% confidence level.

Our analysis confirms that 419 out of 499 total bcn (262) and unc (237) sources in the 
refined and merged associations list of the \fer\ catalogs
are consistent with the \wse\ Gamma-ray Strip at 95\% confidence level.
For the unc sources this situation occurs for 172 sources out of 237 objects while for the bcn
in 242 out of 262 cases.
We remark that 374 out of the 499 sources analyzed with the KDE are associated, 1 is identified, and 
the remaining 124 are candidate associations.
This analysis supports the introduction of the category for the candidate associations, 
since a large fraction of the bcn and the unc 
with a \wse\ counterpart are likely to be blazar-like sources being 
consistent with the \wse\ Gamma-ray Strip. 

In our analysis, we took into account the correction for Galactic extinction
for all the \wse\ magnitudes according to the Draine (2003) relation.
As shown in D'Abrusco et al. (2013), this correction affects only marginally the [3.4]-[4.6] $\mu$m color,
mostly at low Galactic latitudes (i.e., $|b|<$15 \degr).

\section{Optical spectroscopic observations of $\gamma$-ray blazar candidates}
\label{sec:optical}          
The total number of \fer\ sources listed uniquely in the 1FGLR and in the 2FGLR are 2219. 
We found spectroscopic information for 177 of them not reported in 
the previous 1FGL and 2FGL catalogs, including 8 observed during our spectroscopic campaign (see below).

We have been able to confirm spectroscopically 117 BL Lacs, 21 with firm redshift estimates and 
27 quasars (QSOs); an additional eight sources appear to have a BL Lac nature 
but the lack of good optical spectra in the literature or, if a spectrum is available, a low signal to noise ratio
did not allow us to verify their natures. Thus 24 out of these 177 sources still have an uncertain classification
and an uncertain redshift estimate. In particular among them we found 11 sources listed as blazars in Ackermann et al. (2011)
plus one listed in Healey et al. (2007) with no optical spectra published or described in literature. 
These 177 sources are all listed in Table~\ref{tab:main} as bcn' 
with the only exception being one radio source that appears to be a radio galaxy; in the table notes we report their possible classifications
and their redshift estimates with a question mark (?) indicating that further investigation is required.

Optical spectroscopic observations of four candidate associations
and of three associated with unclassified sources (unc) were performed with the 2.1-m telescope of the 
Observatorio Astron\'omico Nacional (OAN) in San Pedro M\'artir (M\'exico) 
on the nights between 28 June and 2 July 2014 (UT) as reported in Table~\ref{tab:optical}.  
The telescope carries a Boller \& Chivens spectrograph and a 1024$\times$1024 pixel E2V-4240 CCD. 
The spectrograph was tuned in the 4000$\div$8000 \AA~range (grating 300 l/mm), 
with a resolution of 4.5 \AA~per pixel, which corresponds to 8 \AA~(Full-width-half-maximum), and a 2$\farcs$5 slit. 
Data were wavelength calibrated using Copper-Helium-Neon-Argon lamps, 
while for flux calibration spectrophotometric standard stars were observed twice during every night of the observing run.

In addition we also observed the $\gamma$-ray blazar candidate BZB J2340+8015
associated with 1FGL J2341.6+8015 at the Observatorio Astrofisico Guillermo Haro (OAGH) located at 
Cananea, Sonora in M\'exico. The telescope was equipped with a spectrograph Boller \& Chivens  
with a CCD SITe 1k$\times$1k, tuned in the 4000$\div$7000 \AA~range (grating 150 l/mm). We used a slit width of 
2$\farcs$5 with a resolution $\sim$15\AA~(FWHM).

The data reduction for both telescopes and instruments 
was carried out using the Image Reduction and Analysis Facility (\textsc{IRAF}) package developed by the
National Optical Astronomy Observatory, including bias subtraction, 
spectroscopic flat fielding, optimal extraction of the spectra and interpolation of the wavelength 
solution. All spectra were reduced and calibrated employing standard techniques in \textsc{IRAF} 
and our own IDL routines \citep[see also][for additional details]{matheson08}.

\begin{figure}[]
\begin{center}
\includegraphics[height=6.8cm,width=9.4cm,angle=0]{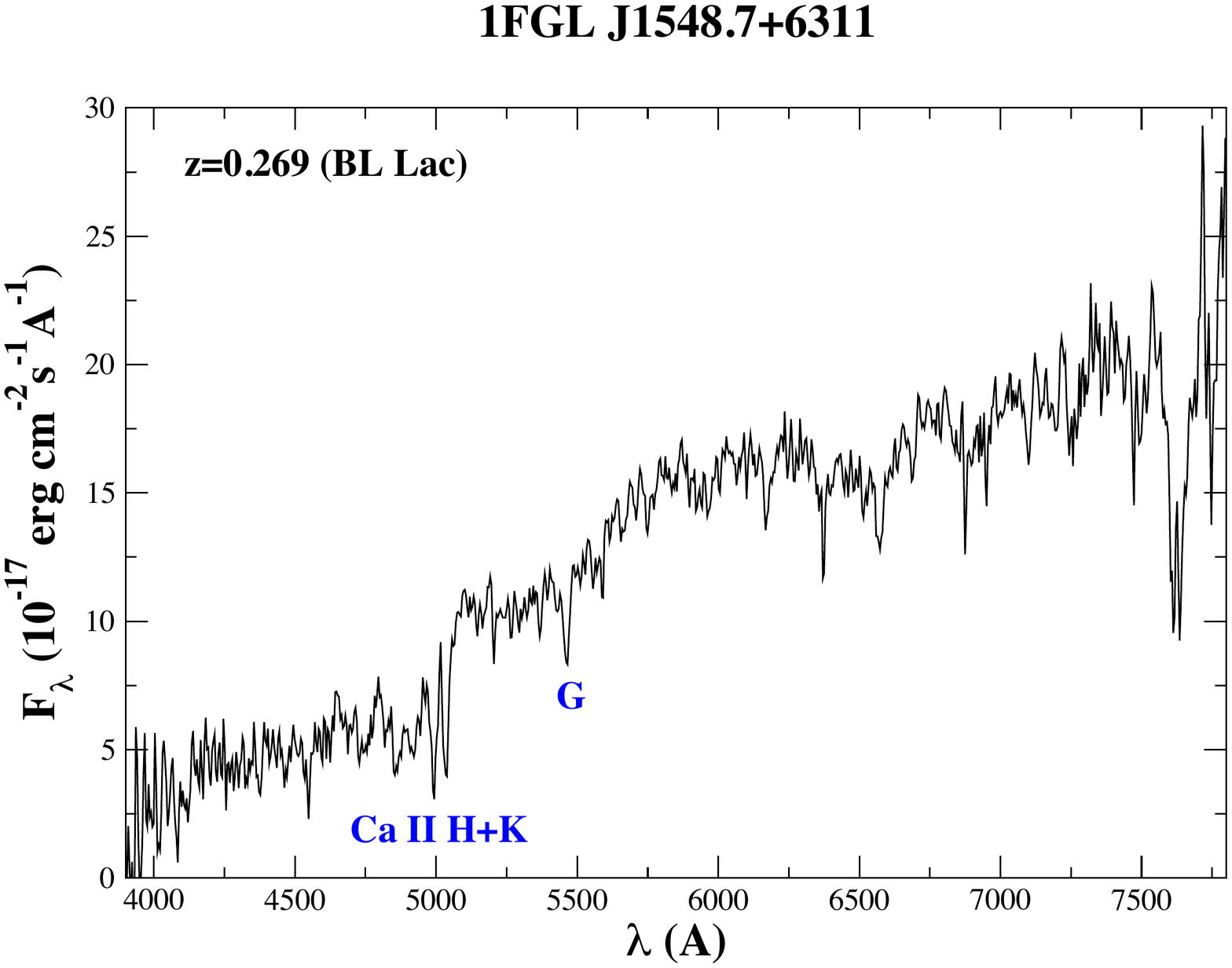}
\includegraphics[height=4.6cm,width=6.6cm,angle=0]{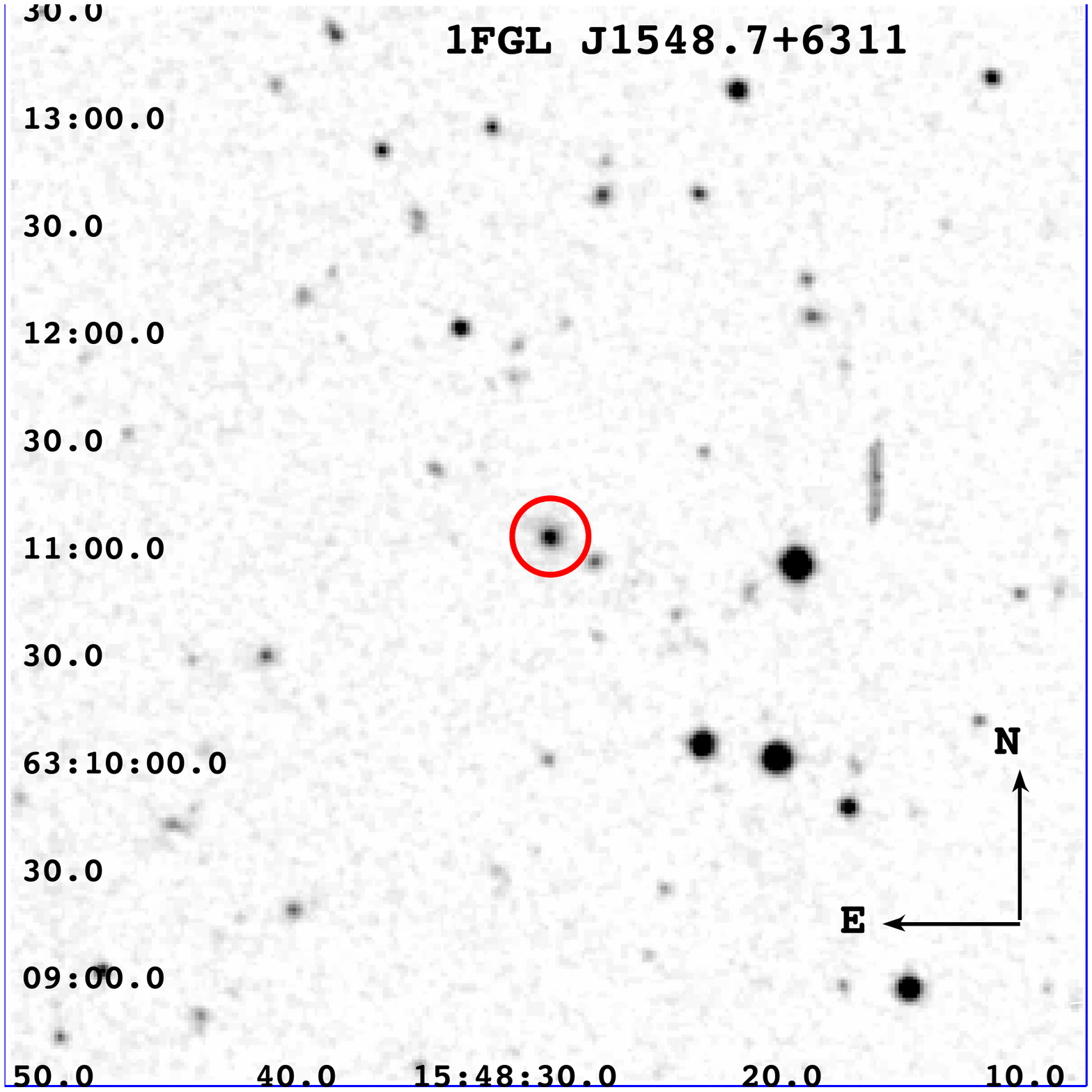}
\end{center}
\caption{Upper panel) The optical spectra of the counterpart associated with 
1FGL J1548.7+6311 observed at OAN in San Pedro M\'artir (M\'exico) on 28 June 2014.
The absorption features identified as Ca\,{\sc ii} H\&K and the G band used to determine its redshift are marked.
The source has been classified as a BL Lac.
(Lower panel) The 5\arcmin\,$\times$\,5\arcmin\ finding chart from the Digitized Sky Survey (red filter). 
The potential counterpart of 1FGL J1548.7+6311, the target of our observation,
is indicated by the red circle.}
\label{fig:J1548}
\end{figure}

\begin{figure}[]
\begin{center}
\includegraphics[height=6.8cm,width=9.4cm,angle=0]{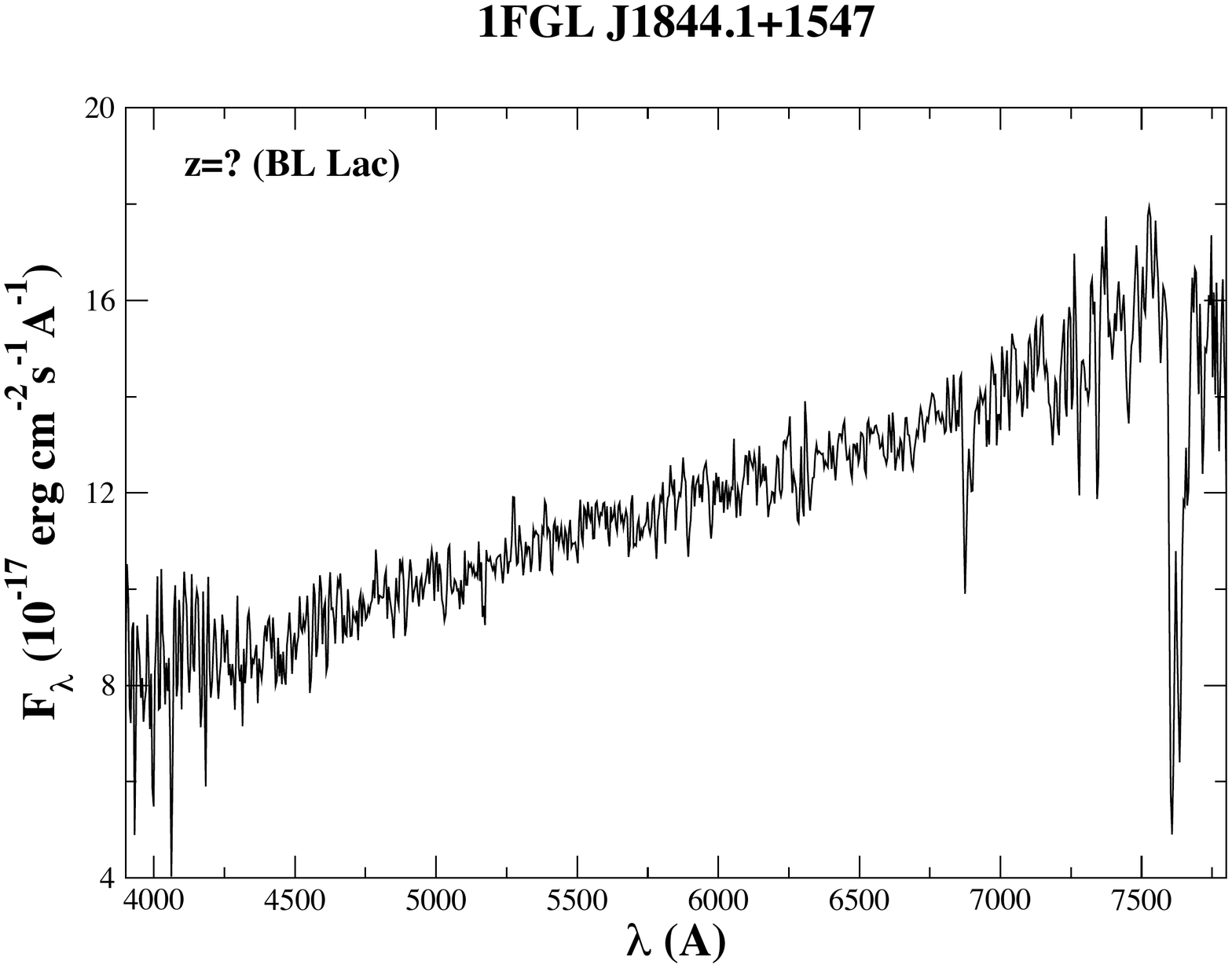}
\includegraphics[height=4.6cm,width=6.6cm,angle=0]{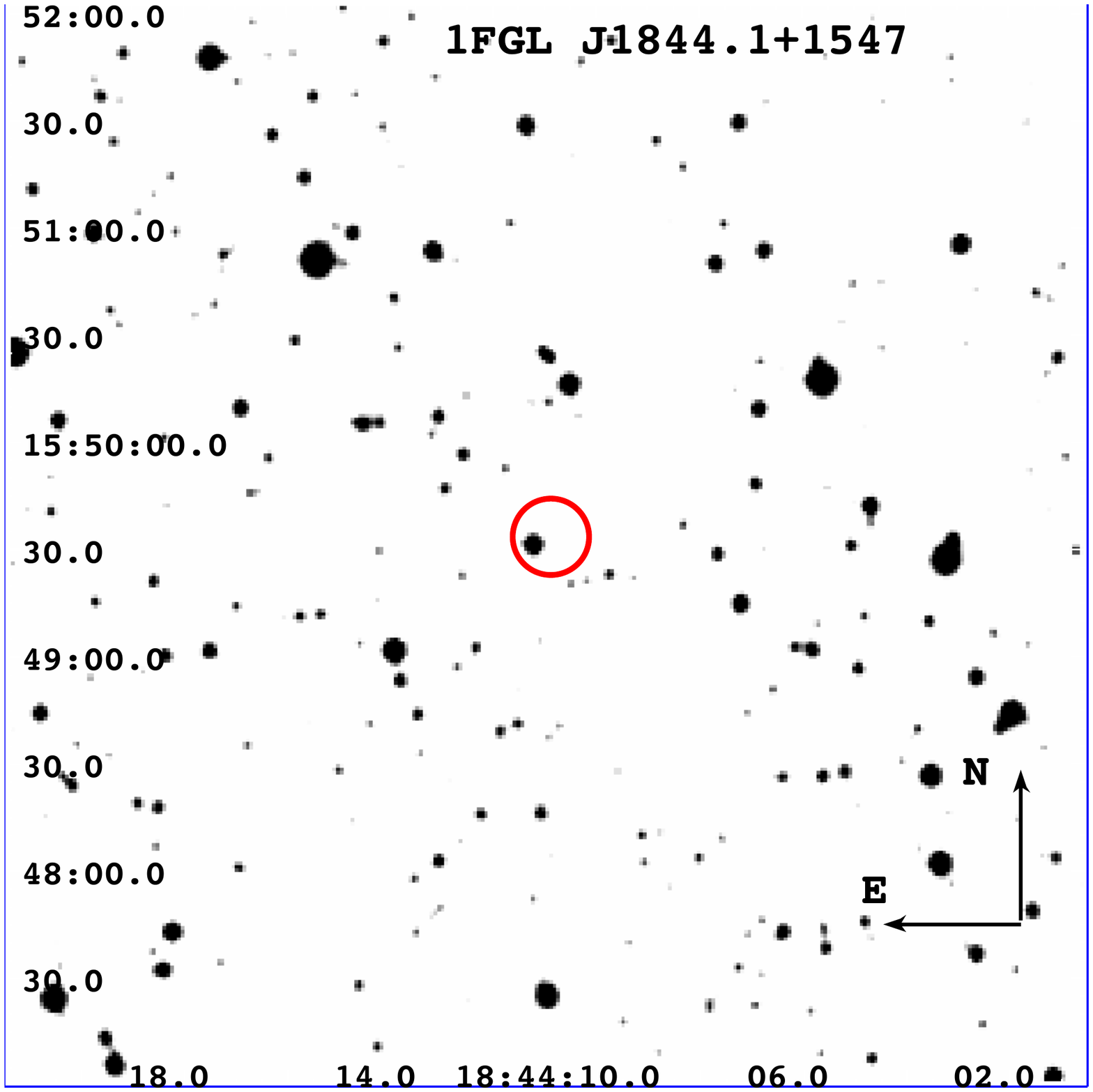}
\end{center}
\caption{Upper panel) The optical spectra of the counterpart associated with 
1FGL J1844.1+1547 observed at OAN in San Pedro M\'artir (M\'exico) on 29 June 2014.
The source has been classified as a BL Lac on the basis of its featureless continuum.
(Lower panel) The 5\arcmin\, $\times$ \,5\arcmin\ finding chart from the Digitized Sky Survey (red filter). 
The potential counterpart of 1FGL J1844.1+1547, the target of our observation, is indicated by the red circle.}
\label{fig:J1844}
\end{figure}

\begin{figure}[]
\begin{center}
\includegraphics[height=6.8cm,width=9.4cm,angle=0]{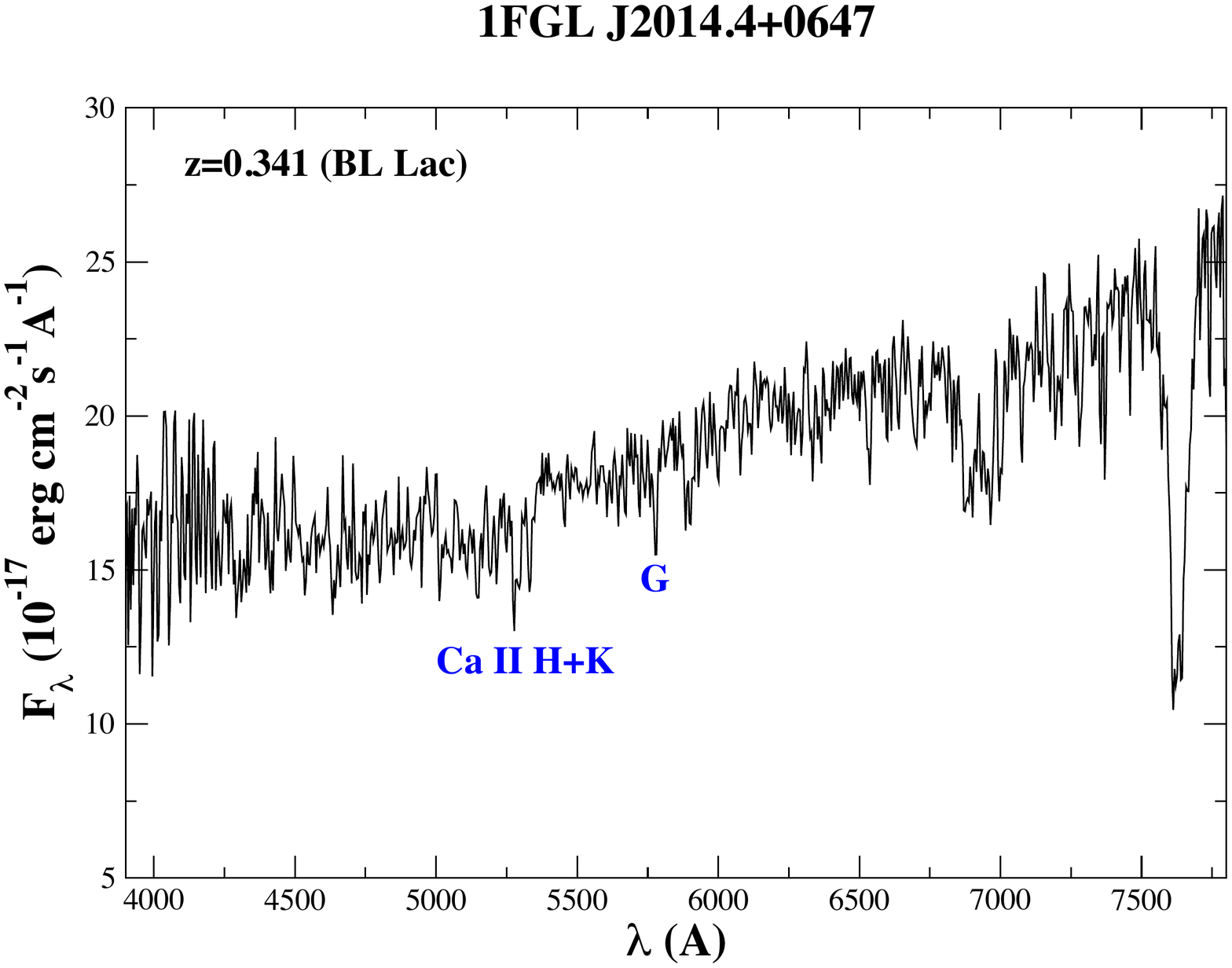}
\includegraphics[height=4.6cm,width=6.6cm,angle=0]{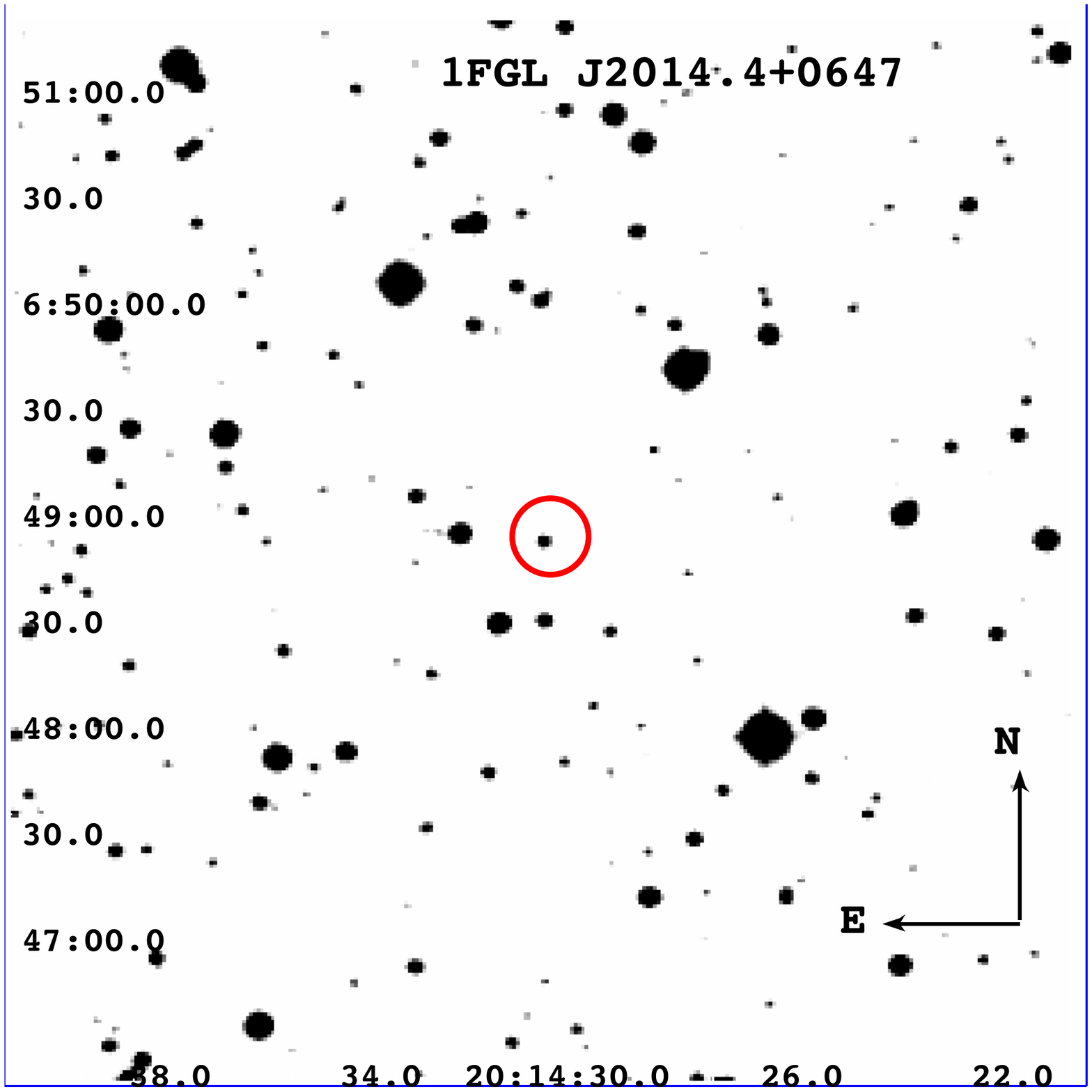}
\end{center}
\caption{Upper panel) The optical spectra of the counterpart associated with 
1FGL J2014.4+0647 observed at OAN in San Pedro M\'artir (M\'exico).
The source was observed twice on 29 June 2014 and on 30 June 2014
and both spectra are shown.
The absorption features identified as Ca\,{\sc ii} H\&K and the G band used to determine its redshift are marked.
The source has been classified as a BL Lac.
Lower panel) The 5\arcmin\, $\times$ \,5\arcmin\ finding chart from the Digitized Sky Survey (red filter). 
The potential counterpart of 1FGL J2014.4+0647
is indicated by the red circle.}
\label{fig:J2014}
\end{figure}

\begin{figure}[]
\begin{center}
\includegraphics[height=6.8cm,width=9.4cm,angle=0]{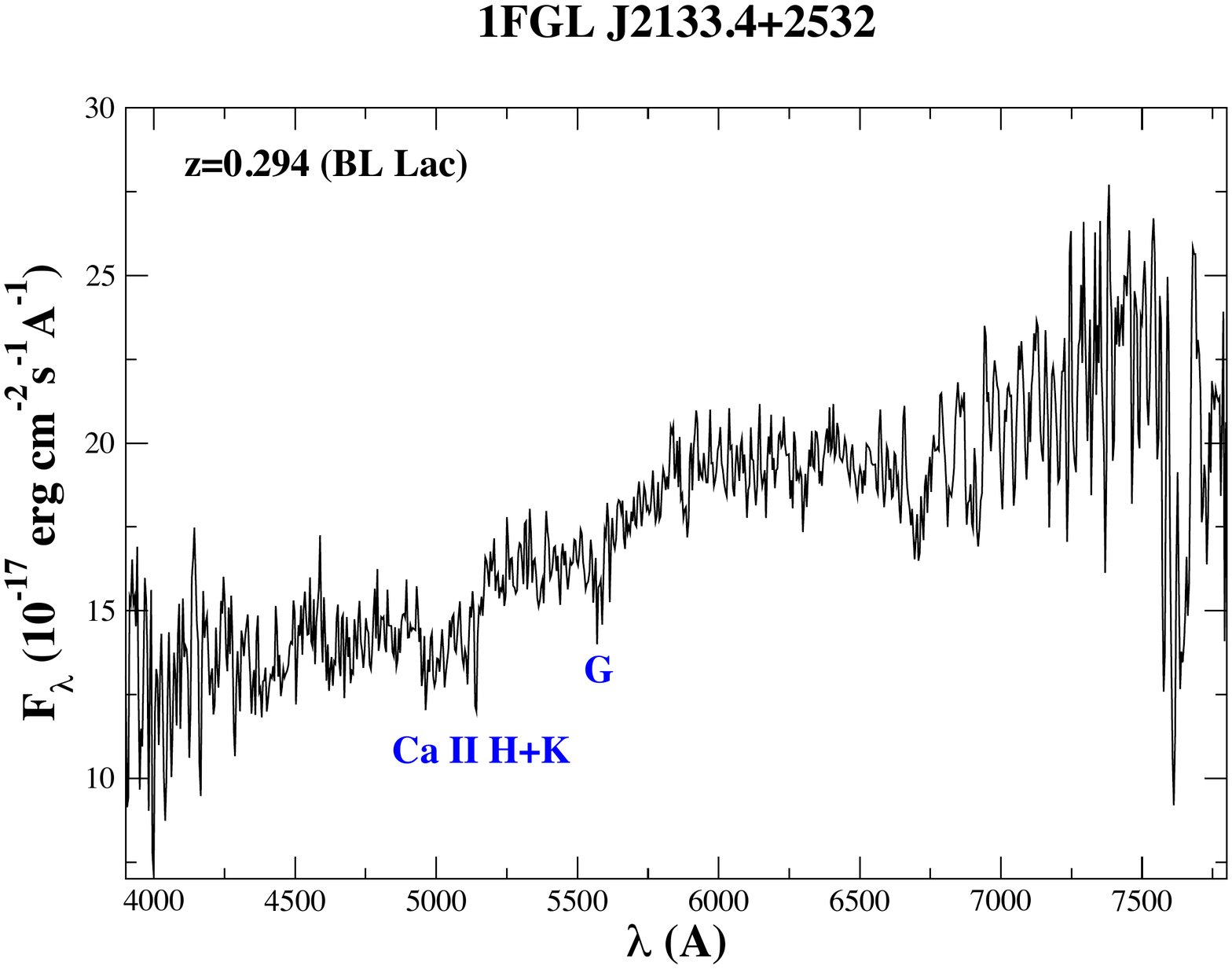}
\includegraphics[height=4.6cm,width=6.6cm,angle=0]{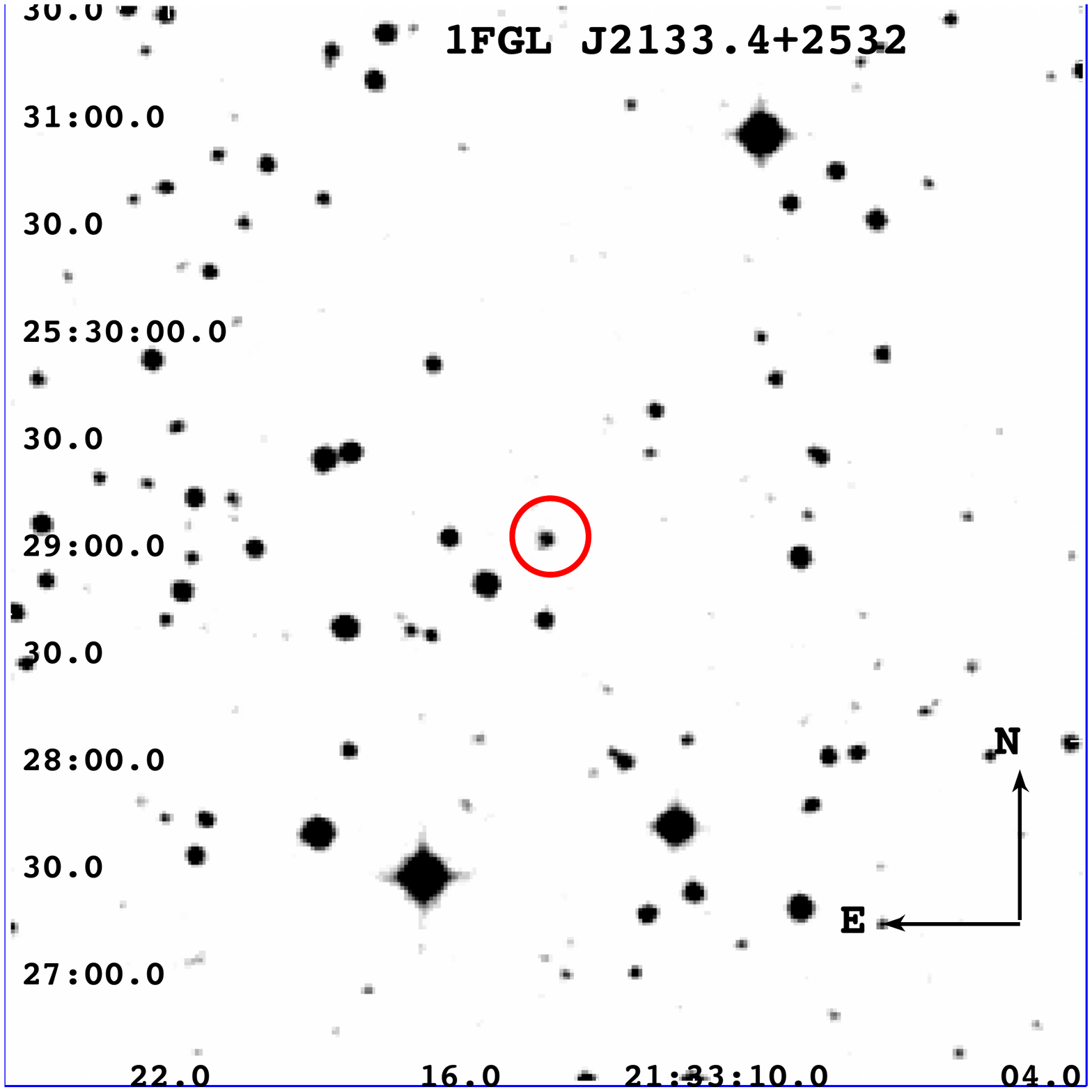}
\end{center}
\caption{Upper panel) The optical spectra of the counterpart associated with 
1FGL J2133.4+2532 observed at OAN in San Pedro M\'artir (M\'exico) on 29 June 2014.
The absorption features identified as Ca\,{\sc ii} H\&K and the G band used to determine its redshift are marked.
The source has been classified as a BL Lac.
Lower panel) The 5\arcmin\, $\times$ \,5\arcmin\ finding chart from the Digitized Sky Survey (red filter).  
The potential counterpart of 1FGL J2133.4+2532 
is indicated by the red circle.}
\label{fig:J2133}
\end{figure}

\begin{figure}[]
\begin{center}
\includegraphics[height=6.8cm,width=9.4cm,angle=0]{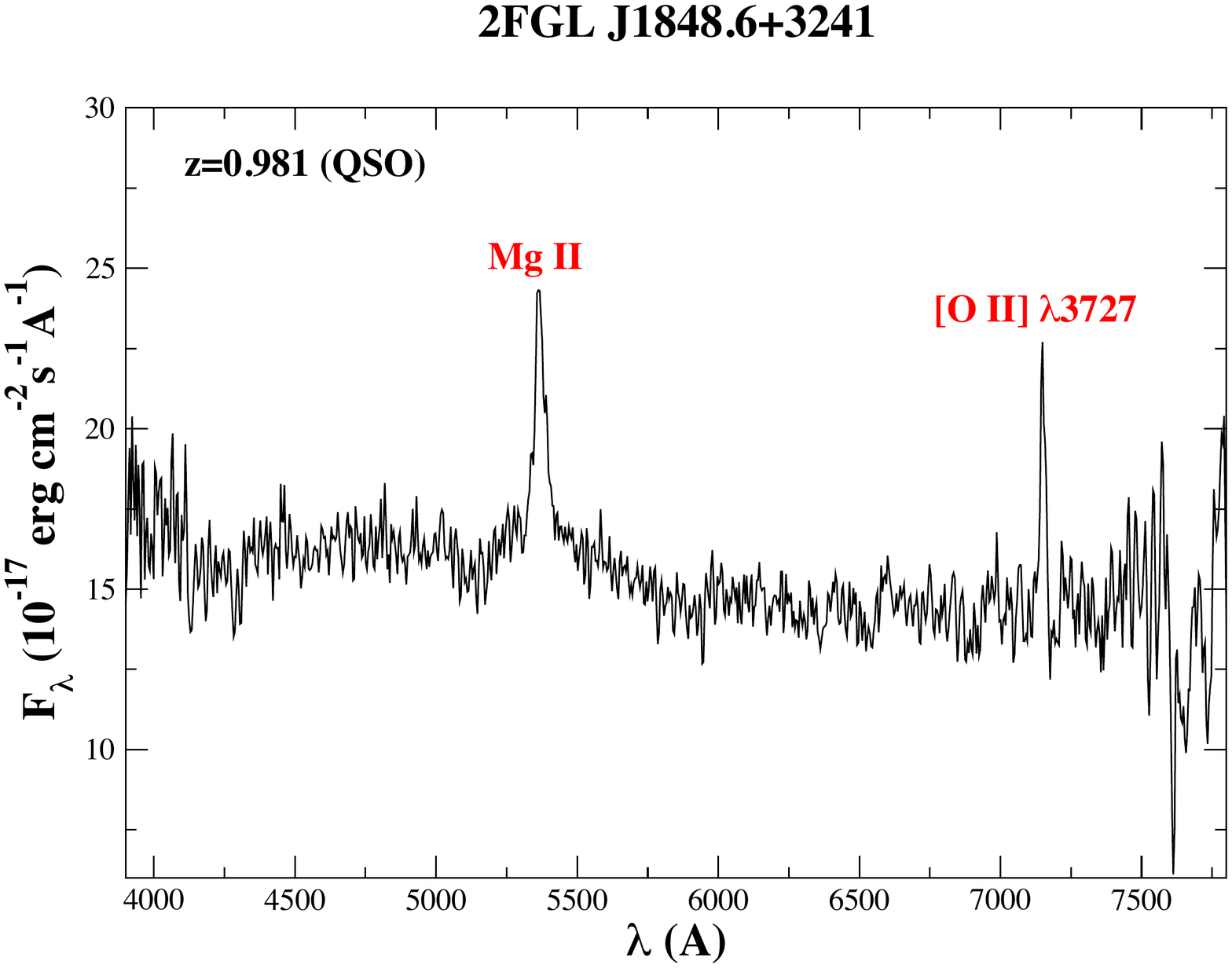}
\includegraphics[height=4.6cm,width=6.6cm,angle=0]{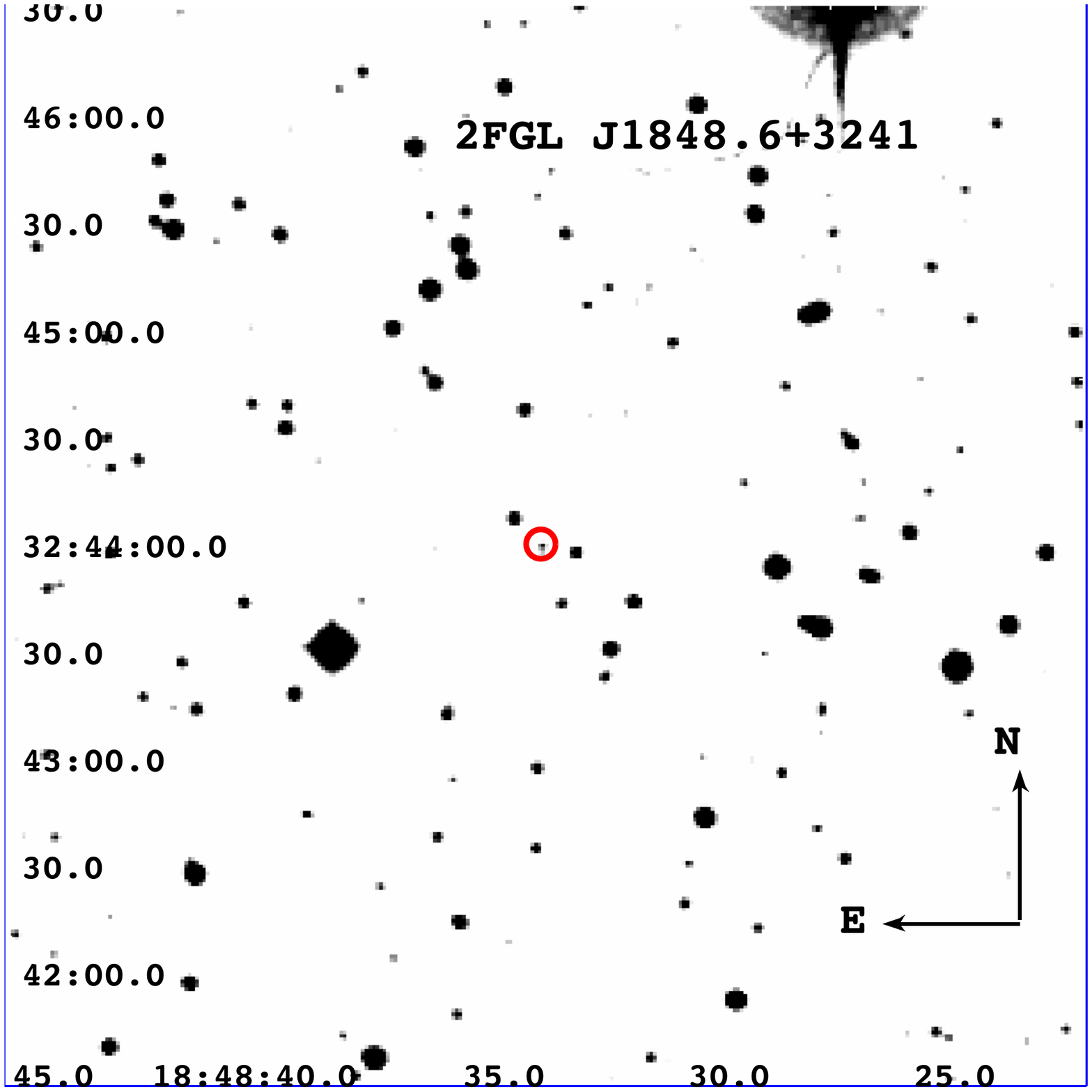}
\end{center}
\caption{Upper panel) The optical spectra of the counterparts associated with 
2FGL J1848.6+3241 observed at OAN in San Pedro M\'artir (M\'exico) on 2 July 2014.
The emission lines marked allowed us to estimate their redshift. The source has been classified as a QSO.
Lower panel) The 5\arcmin\, $\times$ \,5\arcmin\ finding chart from the Digitized Sky Survey (red filter). 
The potential counterpart of 2FGL J1848.6+3241
is indicated by the red circle.}
\label{fig:J1848}
\end{figure}

\begin{figure}[]
\begin{center}
\includegraphics[height=6.8cm,width=9.4cm,angle=0]{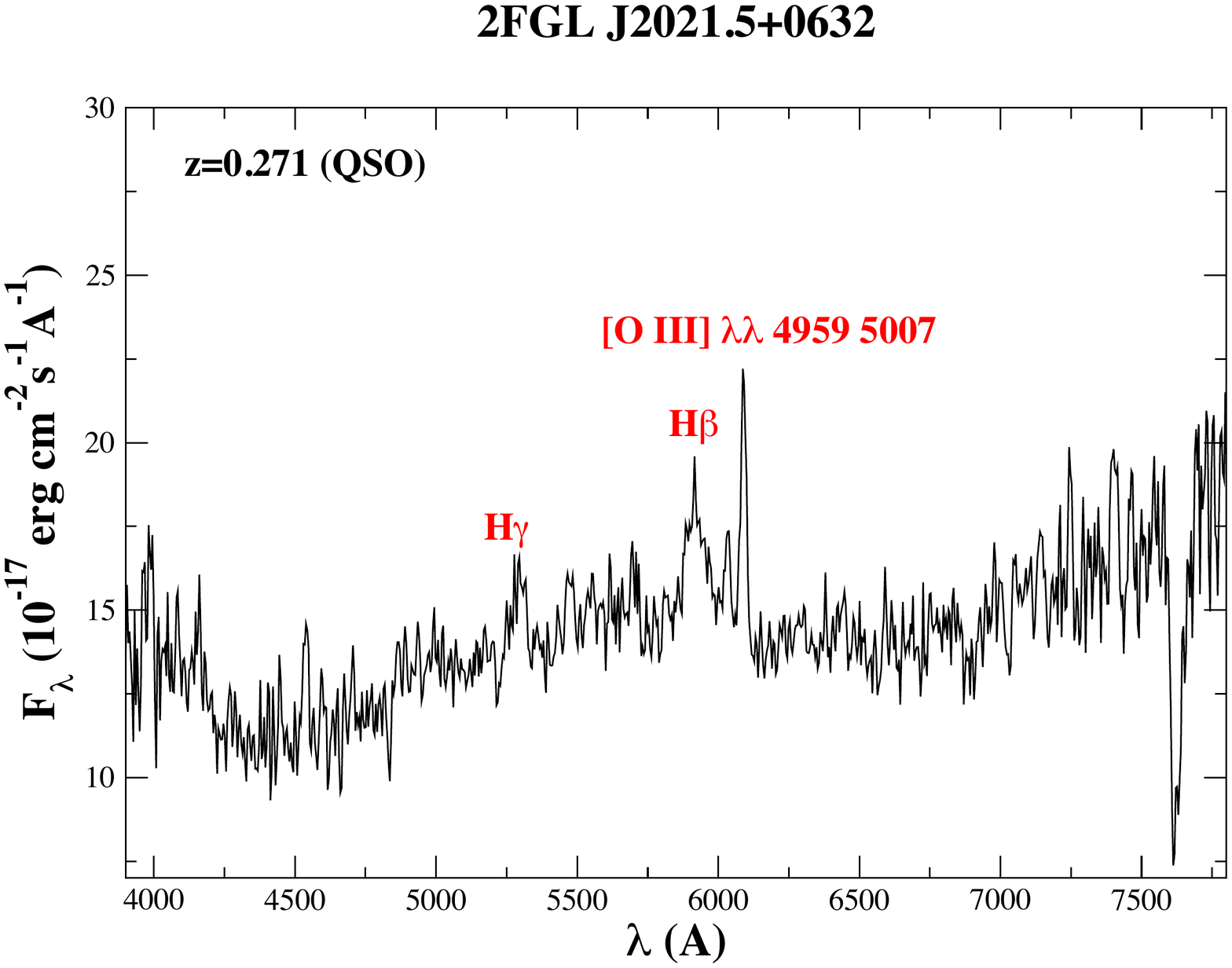}
\includegraphics[height=4.6cm,width=6.6cm,angle=0]{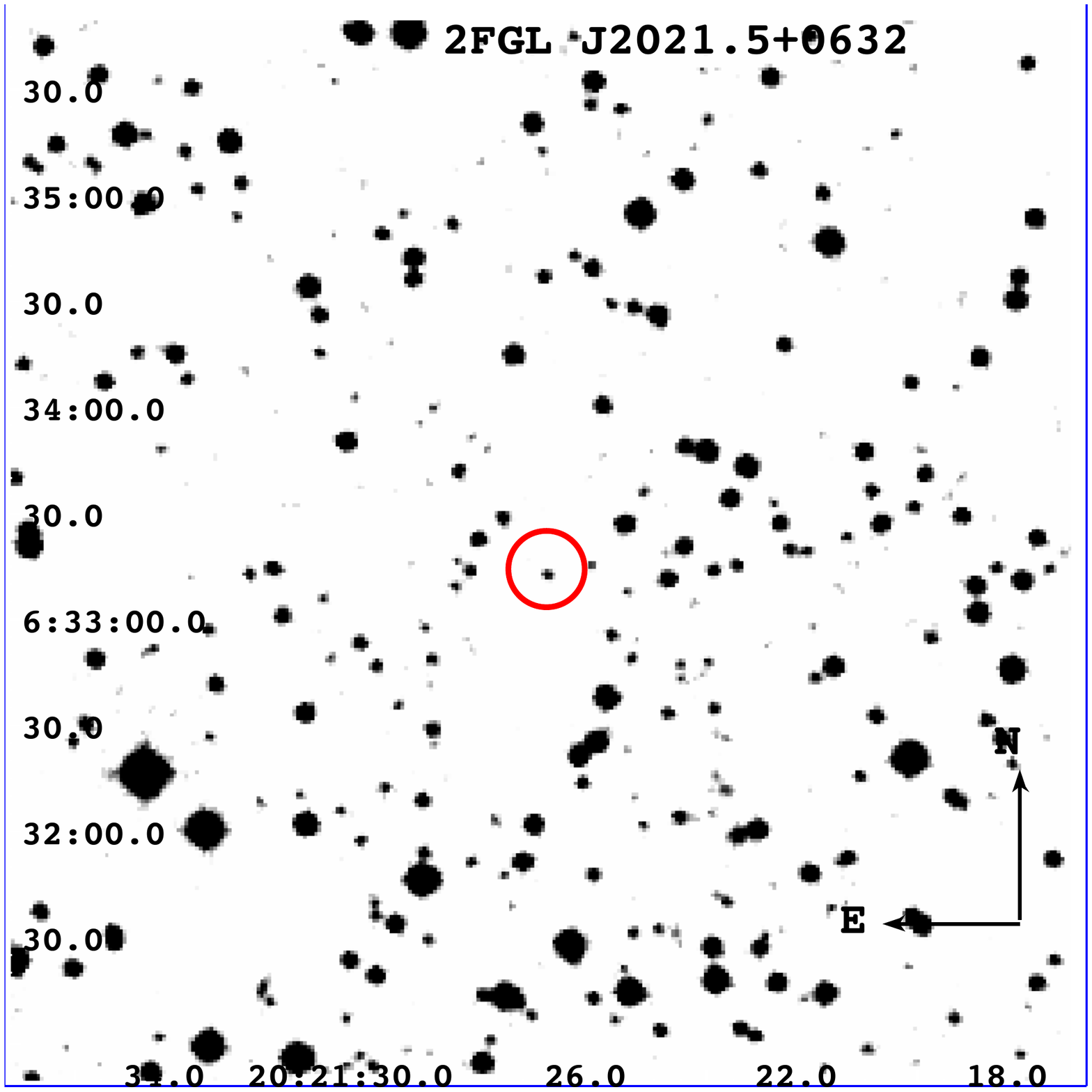}
\end{center}
\caption{Upper panel) The optical spectra of the counterparts associated with 
2FGL J2021.5+0632 observed at OAN in San Pedro M\'artir (M\'exico) on 2 July 2014.
The emission lines marked allowed us to estimate their redshift. The source has been classified as a QSO.
Lower panel) The 5\arcmin\, $\times$ \,5\arcmin\ finding chart from the Digitized Sky Survey (red filter). 
The potential counterpart of 2FGL J2021.5+0632
is indicated by the red circle.}
\label{fig:J2021}
\end{figure}

\begin{figure}[]
\begin{center}
\includegraphics[height=6.8cm,width=9.4cm,angle=0]{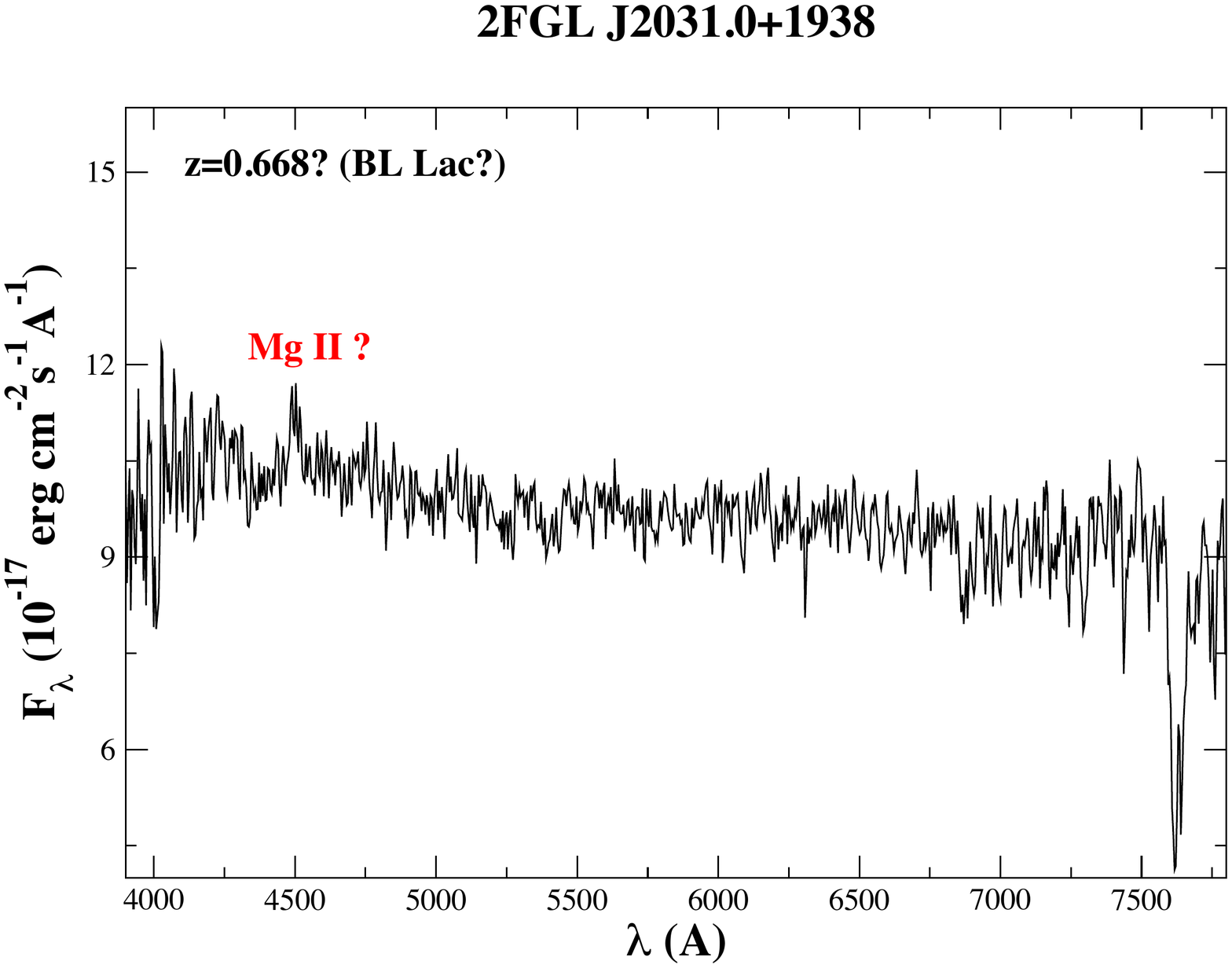}
\includegraphics[height=4.6cm,width=6.6cm,angle=0]{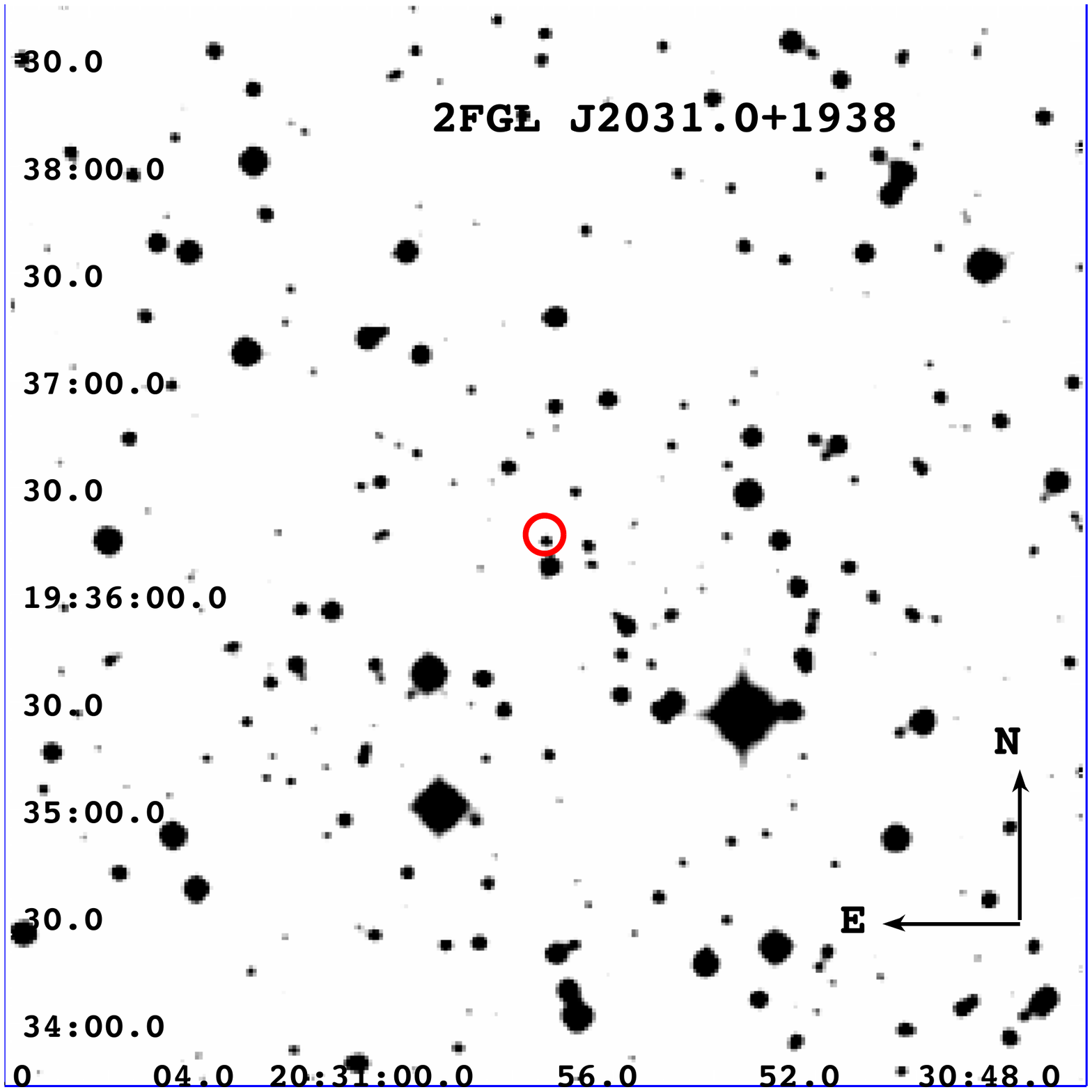}
\end{center}
\caption{Upper panel) The optical spectra of the counterpart associated with 
2FGL J2031.0+1938 observed at OAN in San Pedro M\'artir (M\'exico) on 1 July 2014.
The emission line potentially identified as Mg\,{\sc i} is marked. The source has been classified as a BL Lac.
Lower panel) The 5\arcmin\, $\times$ \,5\arcmin\ finding chart from the Digitized Sky Survey (red filter).  
The potential counterpart of 2FGL J2031.0+1938
is indicated by the red circle.}
\label{fig:J2031}
\end{figure}

These eight sources for which new spectroscopic data are provided in our analysis
are listed in Table~\ref{tab:main} as $\gamma$-ray blazar candidates (i.e., bcn).
Five of them belong to the 1FGLR and are all classified as BL Lac objects according to our analysis; 
for three of them the presence of several spectral features allowed us to determine their redshifts.
The log of the spectroscopic observations and the results of our analysis are reported in Table~\ref{tab:optical},
while in Figure~\ref{fig:J1548}, Figure~\ref{fig:J1844}, Figure~\ref{fig:J2014} and Figure~\ref{fig:J2133} 
we show their optical spectra together with their finding charts.
The remaining three sources are two quasars corresponding to 2FGL J1848.6+3241 and 2FGL J2021.5+0632 
as shown in Figures~\ref{fig:J1848} and \ref{fig:J2021} at redshifts 0.981 and 0.217, respectively,
and 2FGL J2031.0+1938, a BL Lac object with an uncertain redshift 
since there is a unique emission line visible, potentially identified as Mg\,{\sc ii} (see Figure~\ref{fig:J2031}).

\begin{figure}[]
\begin{center}
\includegraphics[height=6.8cm,width=9.4cm,angle=0]{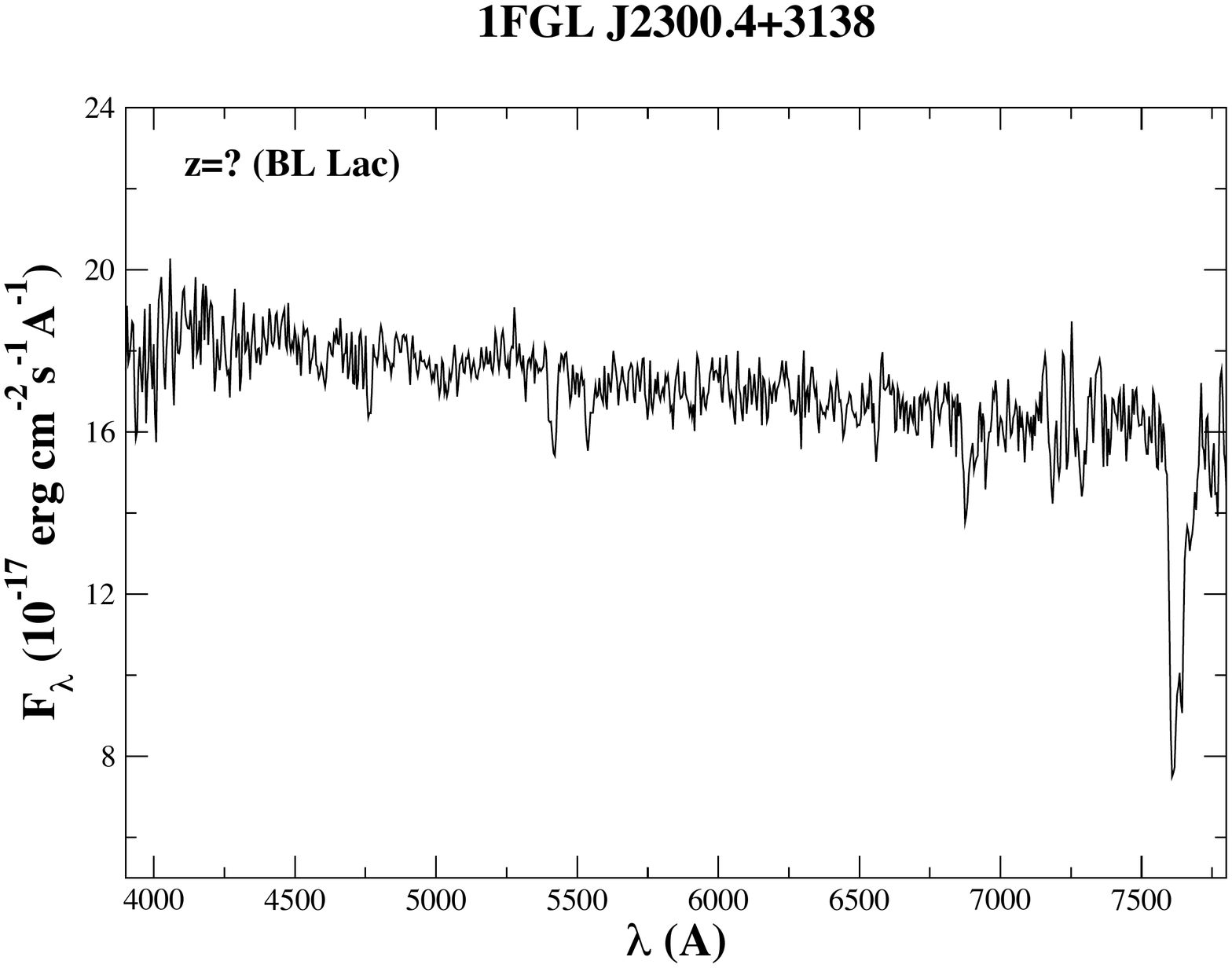}
\includegraphics[height=4.6cm,width=6.6cm,angle=0]{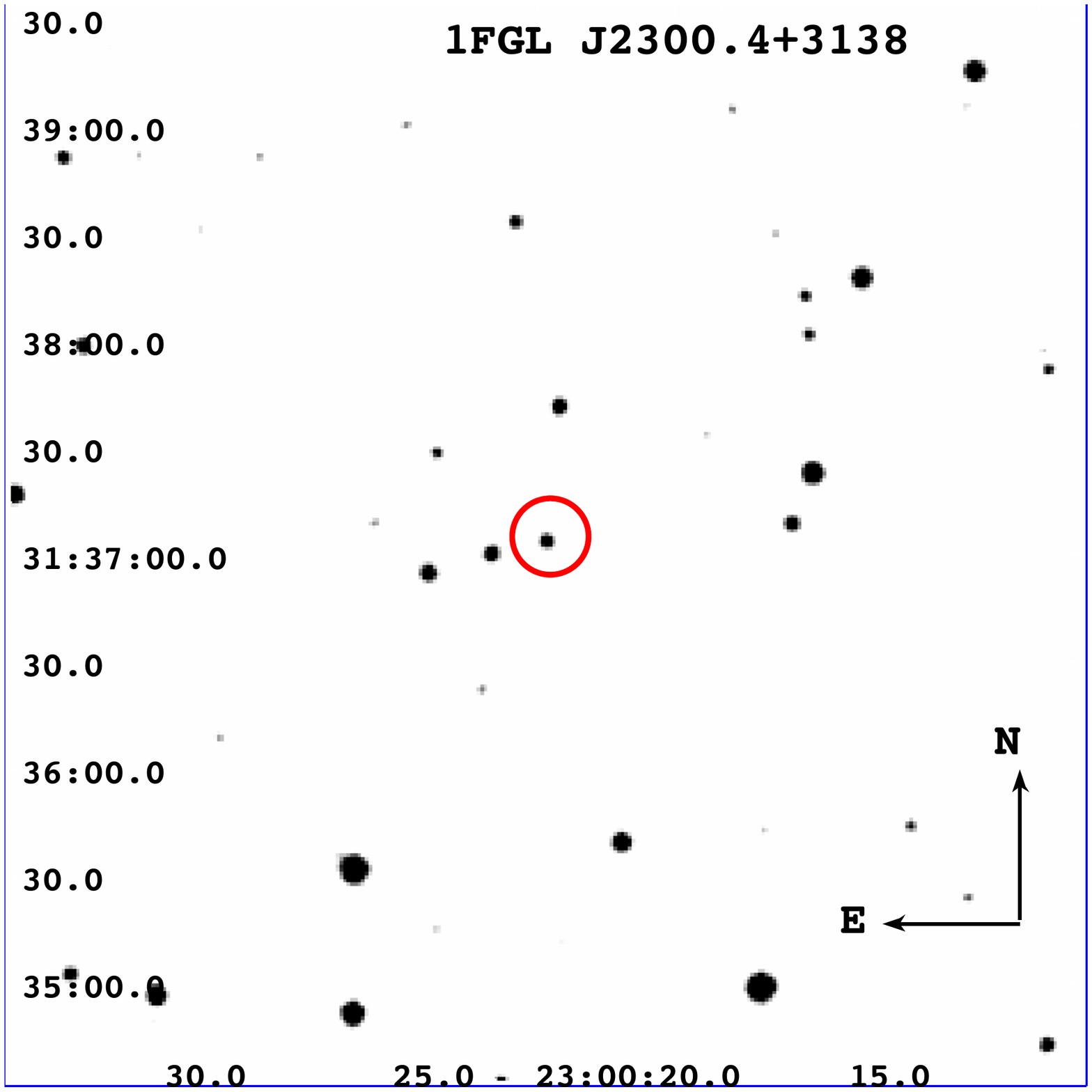}
\end{center}
\caption{Upper panel) The optical spectra of the counterparts associated with 
2FGL J2300.4+3138 observed at OAN in San Pedro M\'artir (M\'exico) on 1 July 2014.
Several unidentified absorption features superimposed to the optical continuum are visible; however,
we were not able to identify the C\,{\sc iv}feature previously reported by Shaw et al. (2013a).
(Lower panel) The 5\arcmin\, $\times$ \,5\arcmin\ finding chart from the Digitized Sky Survey (red filter). 
The potential counterpart of 1FGL J2300.4+3138
pointed during our observations is indicated by the red circle.}
\label{fig:J2300}
\end{figure}

\begin{figure}[]
\begin{center}
\includegraphics[height=6.8cm,width=9.4cm,angle=0]{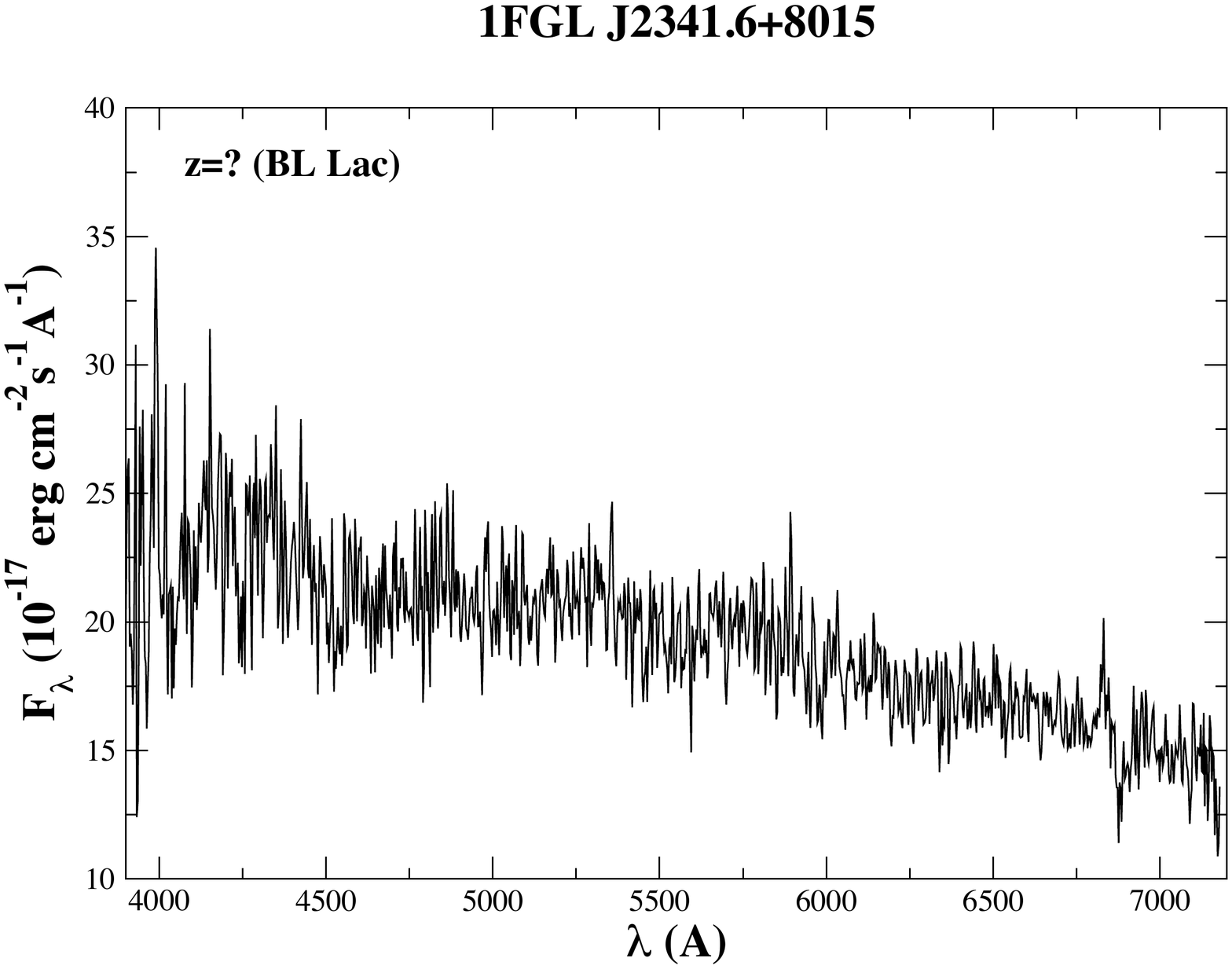}
\includegraphics[height=4.6cm,width=6.6cm,angle=0]{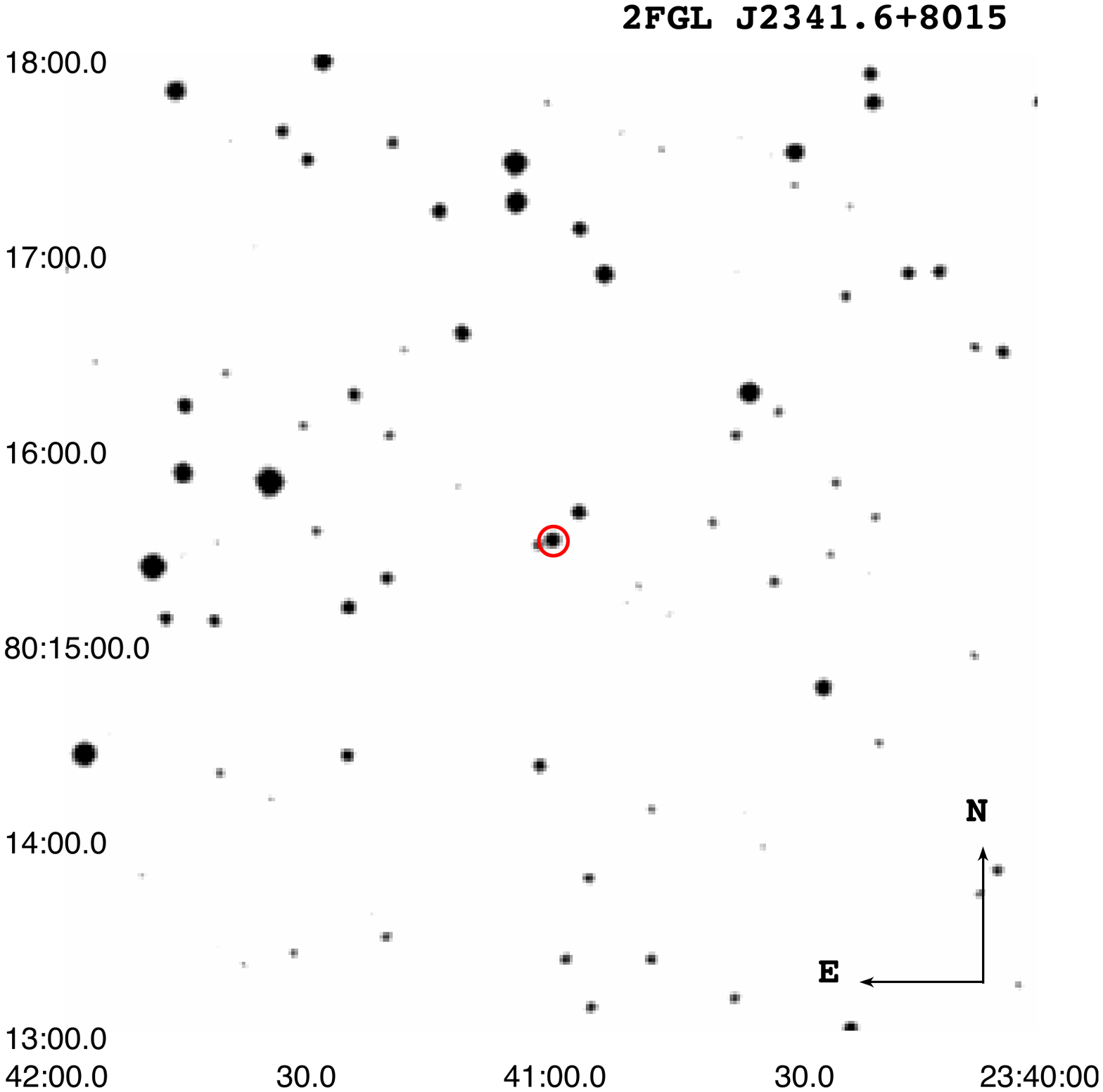}
\end{center}
\caption{Upper panel) The optical spectra of the counterparts associated with 
1FGL J2341.6+8015 observed at OAGH on 28 August 2014.
The source has been classified as a BL Lac on the basis of its featureless continuum.
Lower panel) The 5\arcmin\, $\times$ \,5\arcmin\ finding chart from the Digitized Sky Survey (red filter). 
The potential counterpart of 1FGL J2341.6+8015
is indicated by the red circle.}
\label{fig:J2341}
\end{figure}

During our observing nights we also obtained the spectra for two 2FGL sources, 2FGL J1719.3+1744 and 2FGL J1801.7+4405,
each with uncertain redshift as reported in the \bzcat\ (see the figures in the Appendix).
We were able to confirm the redshift for 2FGL J1801.7+4405 while 2FGL J1719.3+1744 (alias BZB J1719+1745) is completely featureless.
Results for these two additional spectra are also reported in Table~\ref{tab:optical}.
We also observed 1FGL J1942.7+1033, which is a BL Lac object already classified in literature 
\citep[][and Appendix for the figure]{tsarevsky05,masetti13},
and 1FGL J2300.4+3138 for which a series of unidentified absorption features are clearly visible in its optical spectrum. 
Assuming that these unidentified absorption features are due to Mg\,{\sc ii} intervening systems along the line of sight,
the source should lie at redshift $\geq$0.96. This source was in the sample observed by Shaw et al. (2013a)
but we cannot confirm their results since we did not find any spectral feature identifiable as C\,{\sc iv} (see Figure~\ref{fig:J2300}). 
Finally we also report the spectrum of BZB J2340+8015 associated with 1FGL J2341.6+8015
observed at OAGH that was originally classified as a BL Lac candidate in the \bzcat\ at redshift 0.274.
We have been able to confirm the BL Lac nature of this source but not its redshift due to its
featureless spectrum (see Figure~\ref{fig:J2341}).
All our finding charts are from the Digitized Sky Survey\footnote{\underline{http://archive.eso.org/dss/dss}}.
\begin{table}
\tiny
\caption{Log and results of optical spectroscopic observations.}
\begin{center}
\begin{tabular}{|llclll|}
\hline
\fer\  & Obs. Date  & Exposure & Class & $z$ & Spectral \\ 
Name   & yyyy-mm-dd &  (sec)   &       &          & Features   \\
\hline
\noalign{\smallskip}
1FGL J1548.7+6311   & 2014-06-28 & 2x1800 & BL Lac & 0.269 & Ca\,{\sc ii} H\&K, G, H$\beta$, MgI\\
1FGL J1844.1+1547   & 2014-06-29 & 2x1800 & BL Lac & ?    & none                       \\
1FGL J2014.4+0647   & 2014-06-29 & 2x1800 & BL Lac & 0.341 & Ca\,{\sc ii} H\&K, G, H$\beta$, MgI\\
1FGL J2133.4+2532   & 2014-06-29 & 2x1800 & BL Lac & 0.294 & Ca\,{\sc ii} H\&K, G, H$\beta$, MgI\\
2FGL J1848.6+3241   & 2014-07-02 & 2x1800 & QSO & 0.981 & Mg\,{\sc ii}, [OII] $\lambda$3727 \\
2FGL J2021.5+0632   & 2014-07-02 & 1800 & QSO & 0.217 & H$\gamma$, H$\beta$, [OIII] $\lambda$4959 \& $\lambda$5007 \\
2FGL J2031.0+1938   & 2014-07-01 & 2x1800 & BL Lac & 0.668? & Mg\,{\sc ii}? \\
\hline
\noalign{\smallskip}
2FGL J1719.3+1744   & 2014-06-30 & 2x1800 & BL Lac & ?    & none                       \\
2FGL J1801.7+4405   & 2014-06-30 & 2x1800 & QSO    & 0.663 & Mg\,{\sc ii},  [OII] $\lambda$3727  \\
1FGL J1942.7+1033   & 2014-06-29 & 2x1800 & BL Lac & ?    & none                       \\
1FGL J2300.4+3138   & 2014-07-01 & 2x1800 & BL Lac & ? & unid. abs. systems \\
1FGL J2341.6+8015   & 2014-08-28 & 3x1800 & BL Lac & ?    & none                       \\
\hline
\noalign{\smallskip}
\end{tabular}
\end{center}
\label{tab:optical}
\end{table}

\section{Comparison with statistical analyses}
\label{sec:statistical}          
We compared our multifrequency analysis on the candidate associations
with the results obtained via statistical analyses based on the Classification Tree (CT) and on the Logistic Regression (LR)
procedures performed on the 1FGL sources \citep{ackermann12} and using the Random Forest algorithm,
named {\it sybil}, presented by Mirabal et al. (2012) \footnote{\underline{http://www.gae.ucm.es/~mirabal/sibyl.html}} for the 2FGL sources.

For the 1FGL sources the analysis executed with both the CT and the LR statistical methods assesses the probability 
of correct classification based on fitting a model form to the \fer\ data. The result is a 
PSR-like or an AGN-like classification, occurring when the $\gamma$-ray source properties are likely more consistent with 
those of a pulsar or an active galaxy, respectively; a similar prediction is also made for the 2FGL sources analyzed 
by the {\it sybil} algorithm.
Thus, all these methods provide a classification of the potential counterpart on the basis 
of the $\gamma$-ray properties (e.g., spectral shape, variability, etc.)
but cannot permit to locate it. Consequently we can only verify if a candidate association that appear to be a blazar or a pulsar
have the corresponding $\gamma$-ray source classified as AGN-like or PSR-like by the statistical analyses cited above.

There are 286 candidate associations of type bzb in the 1FGLR that were analyzed in Ackermann et al. (2012).
Within this sample we found that there is only one bzb and one hmb, 
the first statistically classified as AGN-like while the second as a PSR-like.  
{\bf Then there are 24 sources classified as bcn, with 23 of them indicated as AGN-like while the last is classified as PSR-like.}

Among the Galactic sources, there are:
\begin{itemize}
\item 7 sources, classified by us as potential MSPs, four appear to be PSR-like
while 3 show $\gamma$-ray properties more similar to AGNs (i.e., AGN-like).
\item 25 out of 29 Galactic sources classified are in our analysis as PSRs are confirmed PSR-like by statistical methods while 4 are indicated as AGN-like;
\item 2 PWN candidate associations, both resembling PSR-like sources in $\gamma$-rays.  
\item 5 sources reported in our Table~\ref{tab:main} as SNRs, four are classified as PSR-like objects and one has an 
uncertain statistical classification (i.e., indicated as conflict in Ackermann et al. 2012).
\item 33 out of 47 candidate associations classified as SFRs are PSR-like 
while only 9 appear to be classifiable as AGN-like by the statistical methods; the remaining five are unclassified.
\end{itemize}

The comparison between our candidate associations in the Galactic plane and the results of the statistical analyses
supports the relevance of searching for the presence of SFRs within the \fer\ positional uncertainty region
where PSRs, SNRs, PWNe could be embedded therein or shocked regions could be potential sources of $\gamma$-rays not yet confirmed. 
Finally, in the sample of 170 unc sources, classified as candidate associations by our analysis, 125 of them appearing to be AGN-like
while 40 are indeed PSR-like; the remaining five are conflicts between the LR and the CT methods. 
In particular, four out of eight unc sources having a SFR consistent with the \fer\ position are classified as PSR-like
on the basis of their $\gamma$-ray behavior.

In the case of the 2FGLR, we investigated 133 candidate associations with the {\it sybil} procedure, 
and classified them accordingly.
Of these, in the sample of 107 \fer\ sources that are classified as AGN-like by {\it sybil}: 24 are bcn, 78 are unc type; then
the remaining objects are two SFRs, three PSRs, seven MSPs (three expected to be PSR-like $\gamma$-ray sources)
and five SFRs all classified as AGN-like.

Finally, we highlight that the results of the statistical analyses are in agreement with the classification proposed 
in the refined association lists of both \fer\ catalogs.

\section{Gamma-ray connections}
\label{sec:connections}

\subsection{The radio-$\gamma$-ray connection}
\label{sec:radio}          
Many attempts were made in the past to
test for correlations between the radio and $\gamma$-ray emissions 
of AGNs and in particular for blazars \citep[e.g.,][]{stecker93,padovani93,salamon94,taylor07}.
This connection was also used before the \fer\ era to search for counterparts of the $\gamma$-ray sources \citep{mattox97},
so motivating all the past and present radio follow up campaigns for the \fer\ sources \citep[e.g.,][]{petrov13,schinzel14}.

However, biases and selection effects have to be taken into account to prove this correlation properly,
since it is important to address intrinsic source variability, biases due to redshift dependence \citep{elvis78}, 
the ``common distance'' bias \citep{pavlidou11}, problems related to source misidentifications 
and incorrect associations, to name a few. All these issues can mimic a correlation \citep{mucke97}.

Since the launch of \fer\ several investigations have also been performed to search for a definitive answer regarding the
existence of the radio-$\gamma$-ray connection \citep{ghirlanda10,ghirlanda11,mahony10} 
until it was proved and described accurately in Ackermann et al. (2011b).
Recently we also suggested that a link between the radio and the $\gamma$-ray emissions 
in blazars can be extended well below $\sim$1 GHz \citep{ugs3,ugs6,74mhz}.

\begin{figure}[] 
          \includegraphics[height=6.8cm,width=9.5cm,angle=0]{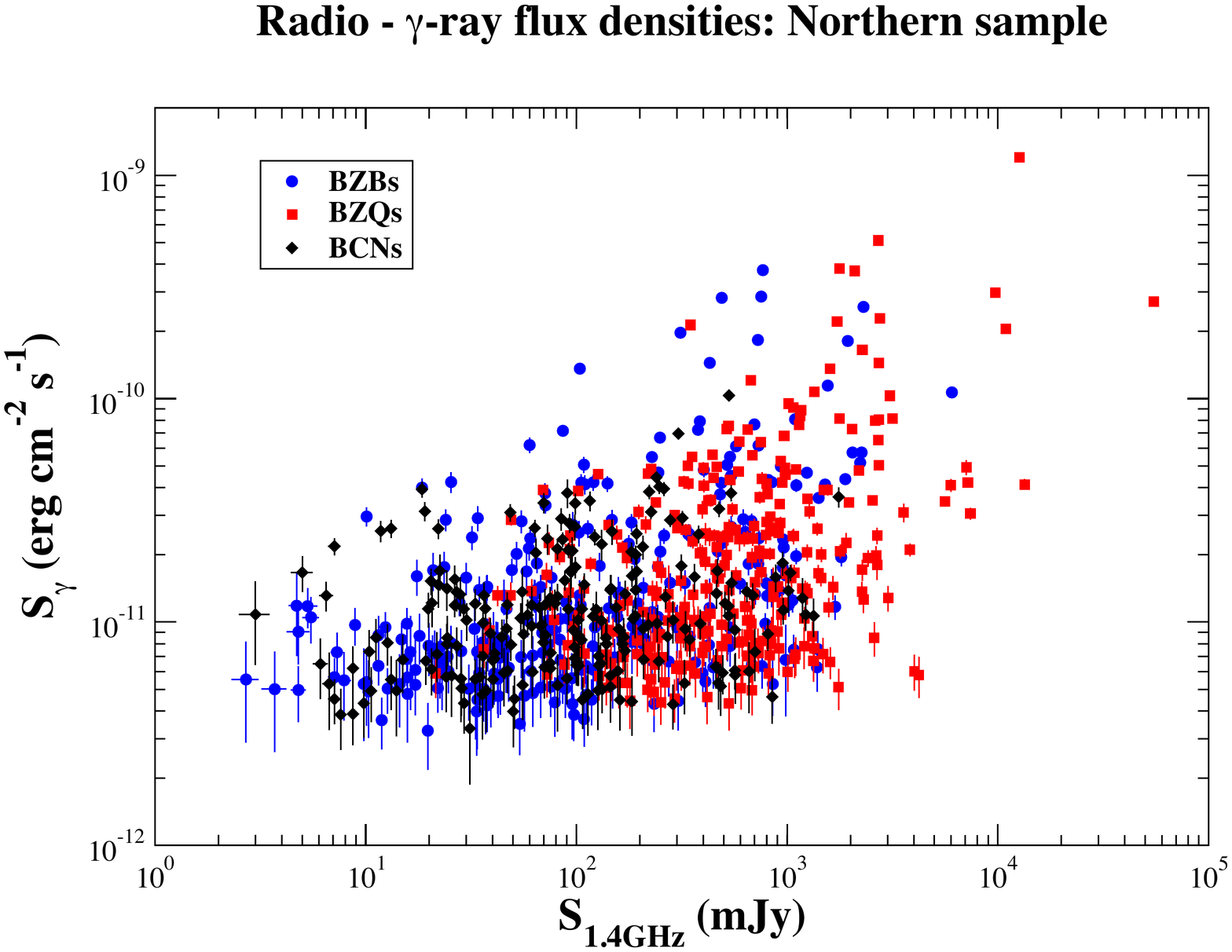}
          \includegraphics[height=6.8cm,width=9.5cm,angle=0]{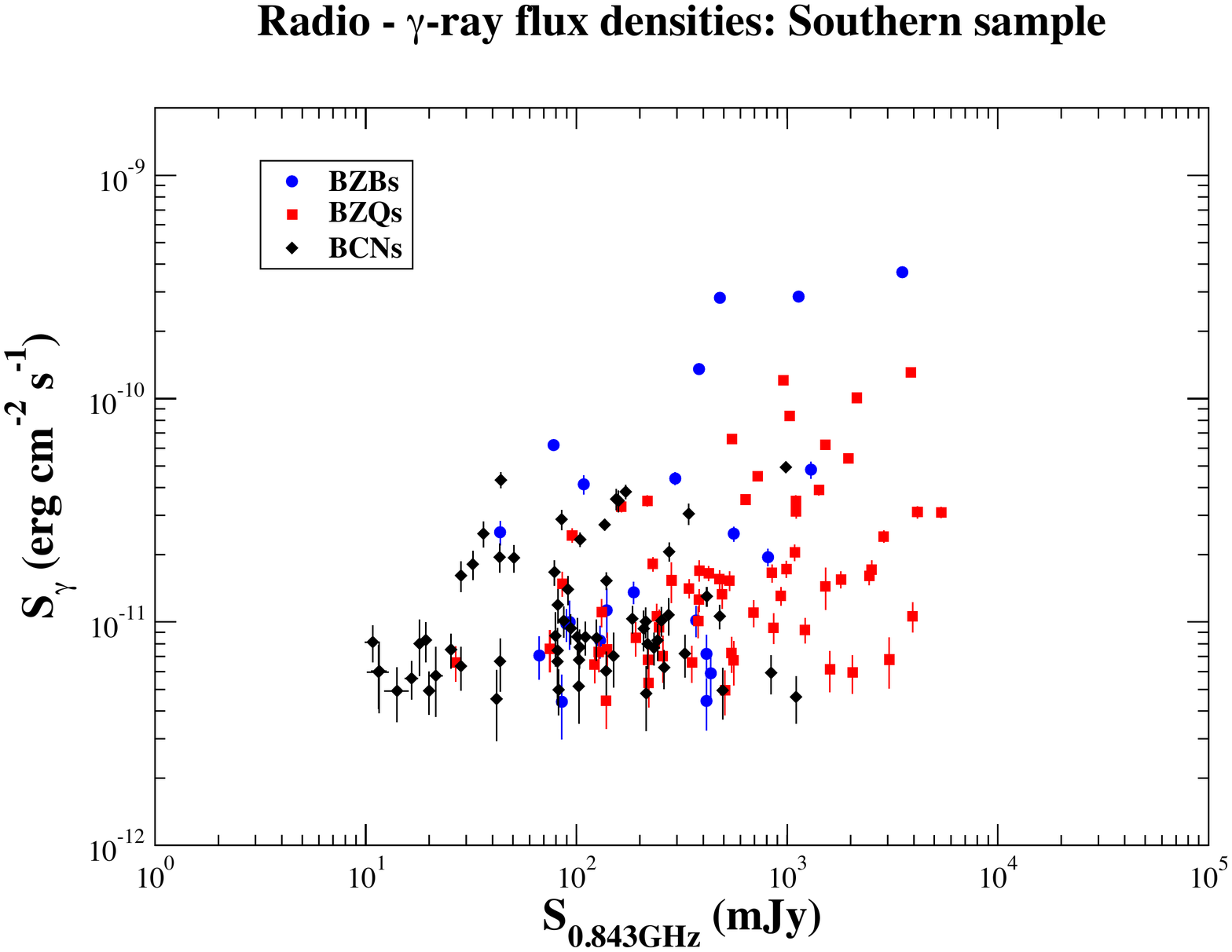}
           \caption{Scatter plot of the radio flux densities vs. the $\gamma$-ray flux for blazar-like sources.
                         Spectroscopically confirmed BZBs and BZQs are shown in blue circles and red squares, respectively, while bcn sources
                         are black diamonds. Sources with a radio counterpart in the NVSS are in the upper panel
                         while the sources in SUMSS are  in the lower panel.}
           \label{fig:radioFluxes}
          \end{figure}

\begin{figure}[] 
          \includegraphics[height=6.8cm,width=9.5cm,angle=0]{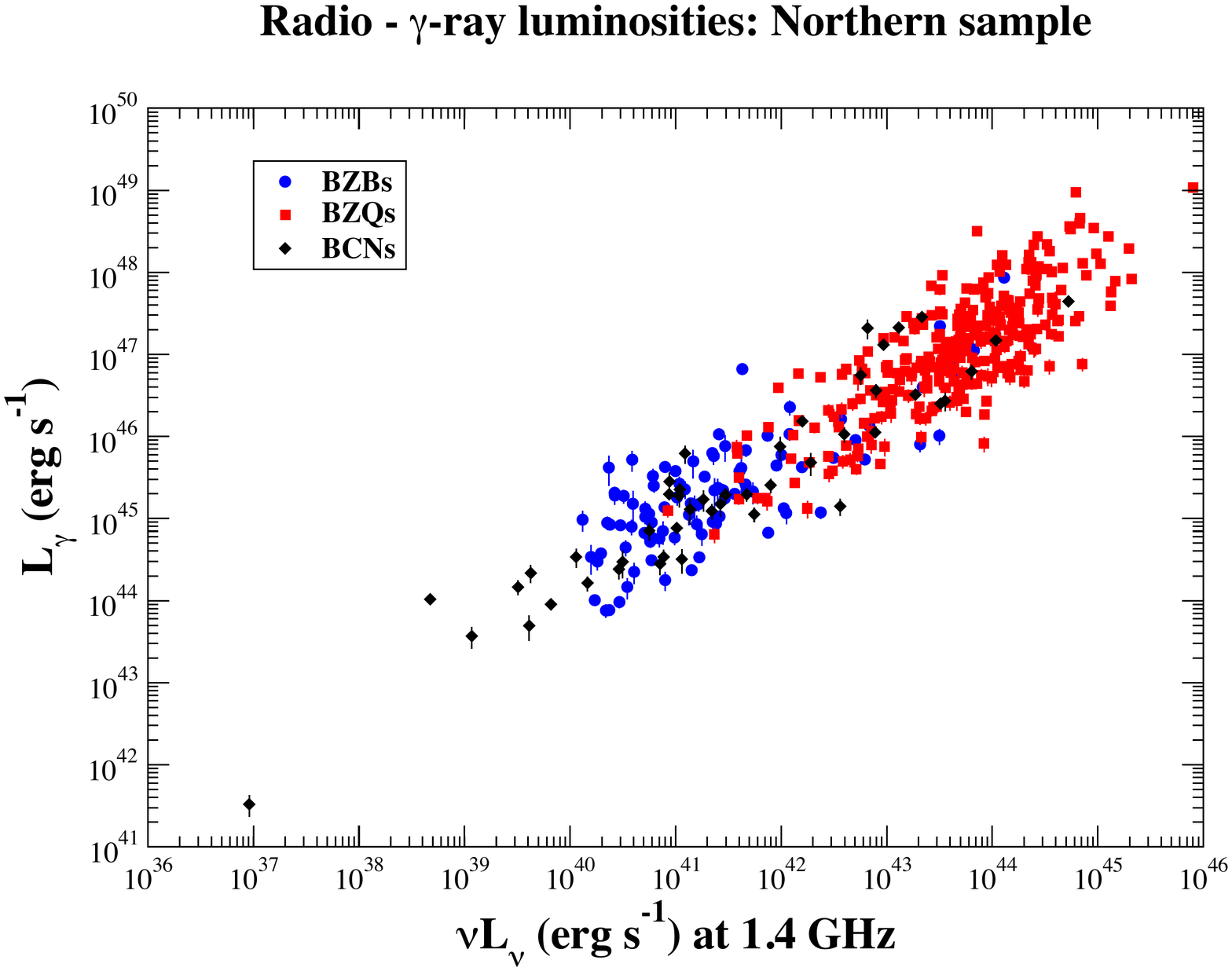}
          \includegraphics[height=6.8cm,width=9.5cm,angle=0]{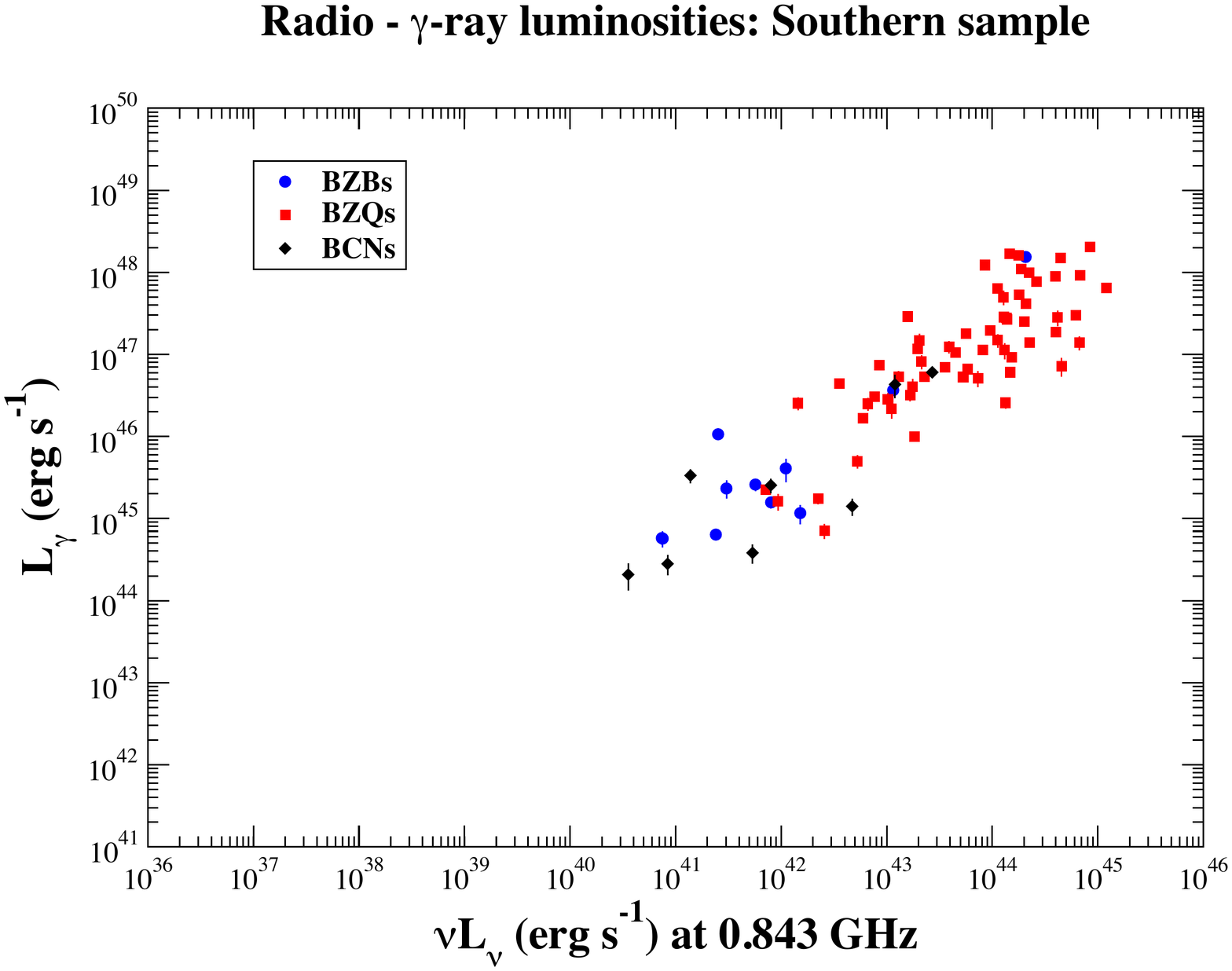}
           \caption{Scatter plot of the radio - $\gamma$-ray luminosities for blazar-like sources.
                         Spectroscopically confirmed BZBs and BZQs are shown in blue circles and red squares, respectively, while bcns sources
                         are black diamonds. Sources with a radio counterpart in the NVSS are in the upper panel
                         while the sources in SUMSS are in the lower panel.}
           \label{fig:radioLuminosities}
          \end{figure}

Given the new multifrequency analysis carried out here and the candidate associations,
in particular for the bcn class we illustrate the current status of the radio-$\gamma$-ray connection.
To this end we show both the flux-flux and the luminosity-luminosity scatter plots for the blazars in our merged 
refined list of \fer\ associations (see Figure~\ref{fig:radioFluxes} and Figure~\ref{fig:radioLuminosities}).

A study of the correlations is out of the scope of this paper. Our main goal is verify that
sources classified as bcn are in agreement with the  behavior of blazars in these parameter spaces.
It is worth noting that we divided our sample of blazar-like sources in two subsamples: the northern one and the southern one.
the former includes all the blazars (i.e., bzb, bzq, and bcn classes) that have a radio counterpart within the NVSS footprint
while in the latter the subsample includes those in the area covered by the SUMSS. 
We distinguish these two samples since the radio surveys were carried out at different frequencies.

Finally, we remark that only blazars with a firm redshift estimate as reported in our merged list of refined associations are shown in
Figure~\ref{fig:radioLuminosities}.
According to this figure we could expect most of the bcn sources to be BL Lac objects, although this has to be confirmed with optical 
spectroscopic observations.

\subsection{The infrared-$\gamma$-ray connection}
\label{sec:infrared}          
D'Abrusco et al. (2012) studied the \fer\ blazar sample listed in the 2FGL and 
lying in the footprint of the \wse\ Preliminary survey and reported the discovery 
of a correlation between their infrared and their $\gamma$-ray spectral indices.
This is directly related to their peculiar infrared colors and the \wse\ Gamma-ray Strip \citep{paper1}.

\begin{figure}[] 
          \includegraphics[height=6.8cm,width=9.5cm,angle=0]{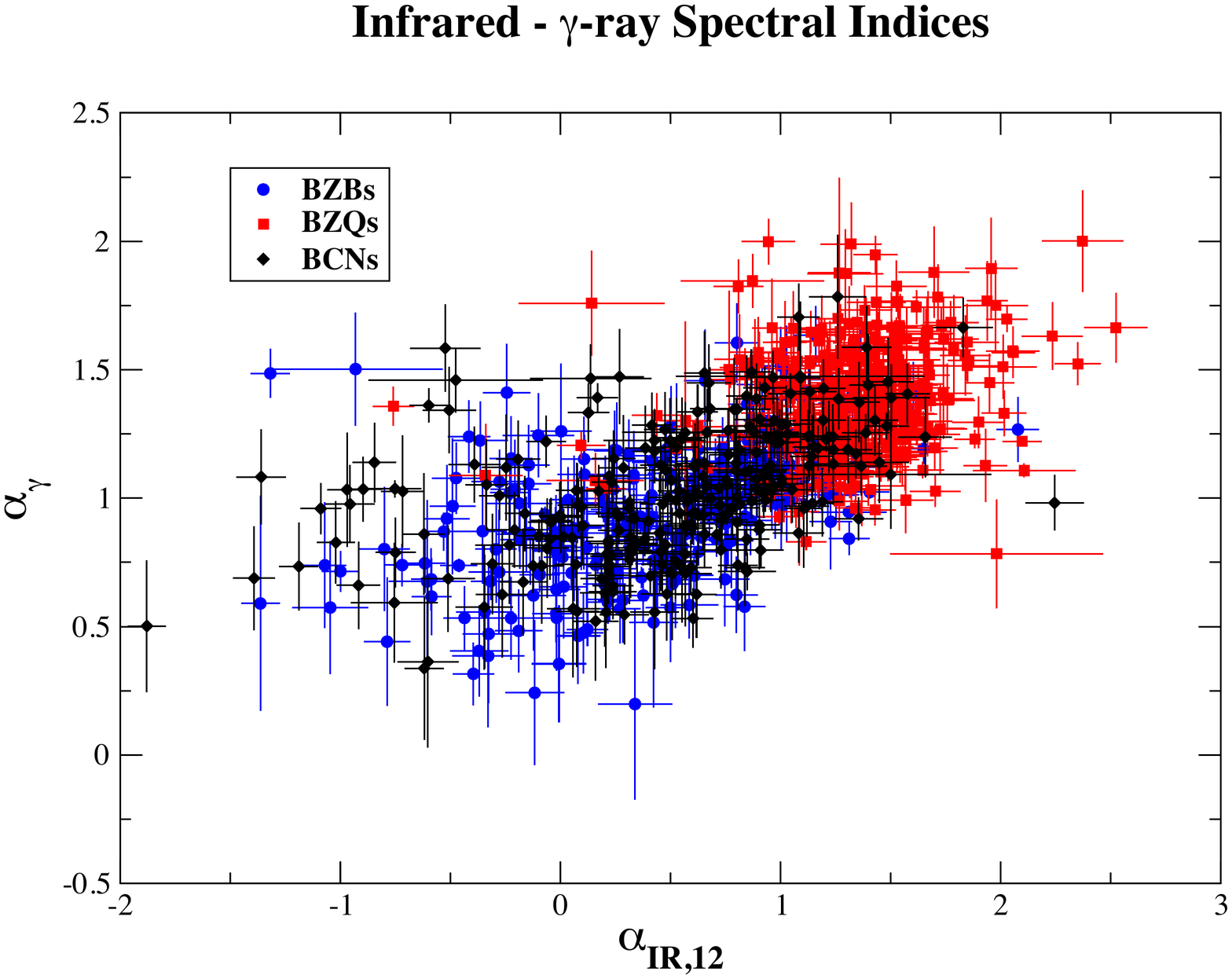}
           \caption{Scatter plot of the infrared spectral index evaluated using the first two \wse\ bands vs. the $\gamma$-ray index
                        for the blazar-like sources in our merged list of refined \fer\ associations.
                        Spectroscopically confirmed BZBs and BZQs are shown in blue circles and red squares, respectively, while bcn sources
                        are black diamonds.}
           \label{fig:indices}
          \end{figure}

\begin{figure}[] 
          \includegraphics[height=6.8cm,width=9.5cm,angle=0]{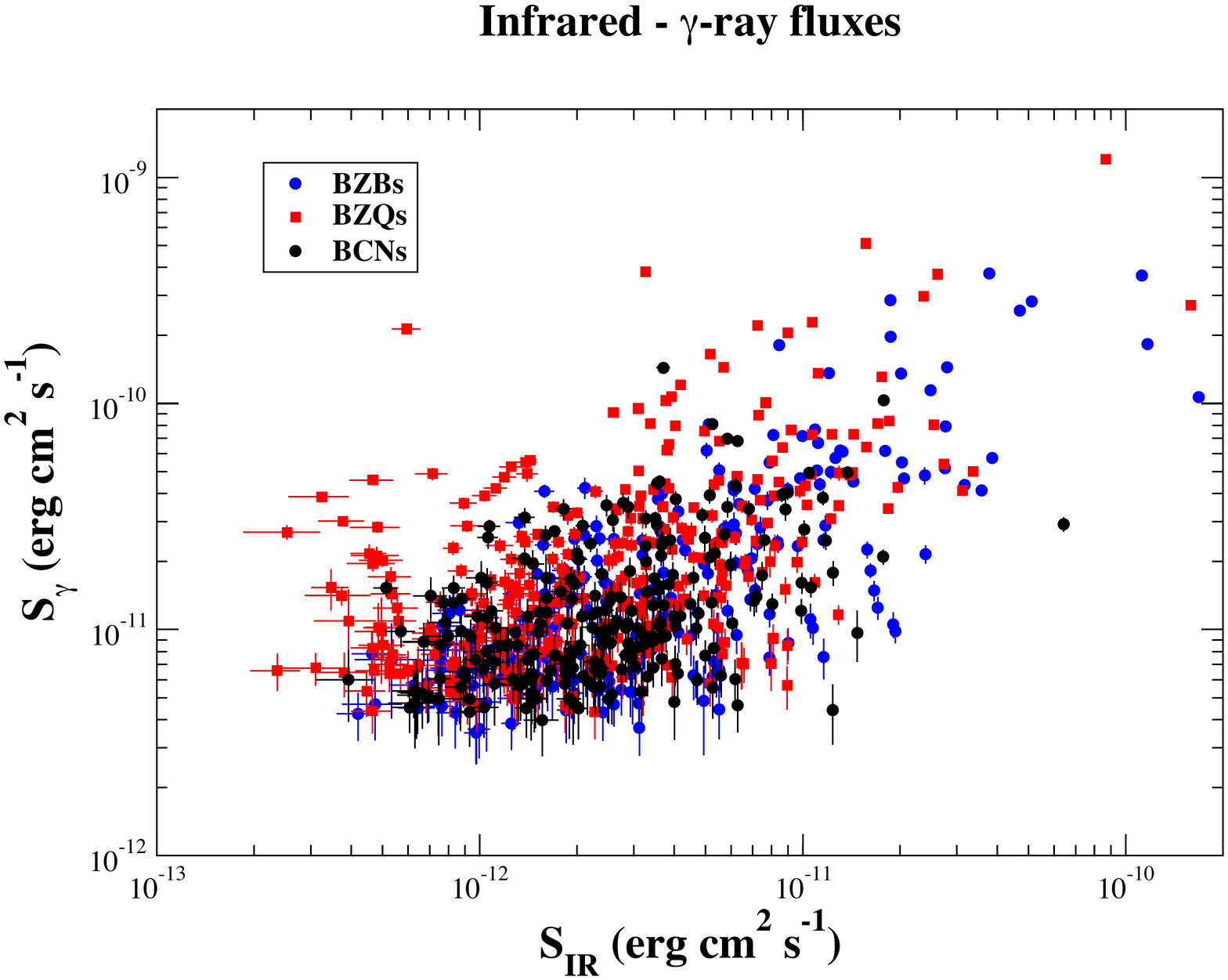}
          \includegraphics[height=6.8cm,width=9.5cm,angle=0]{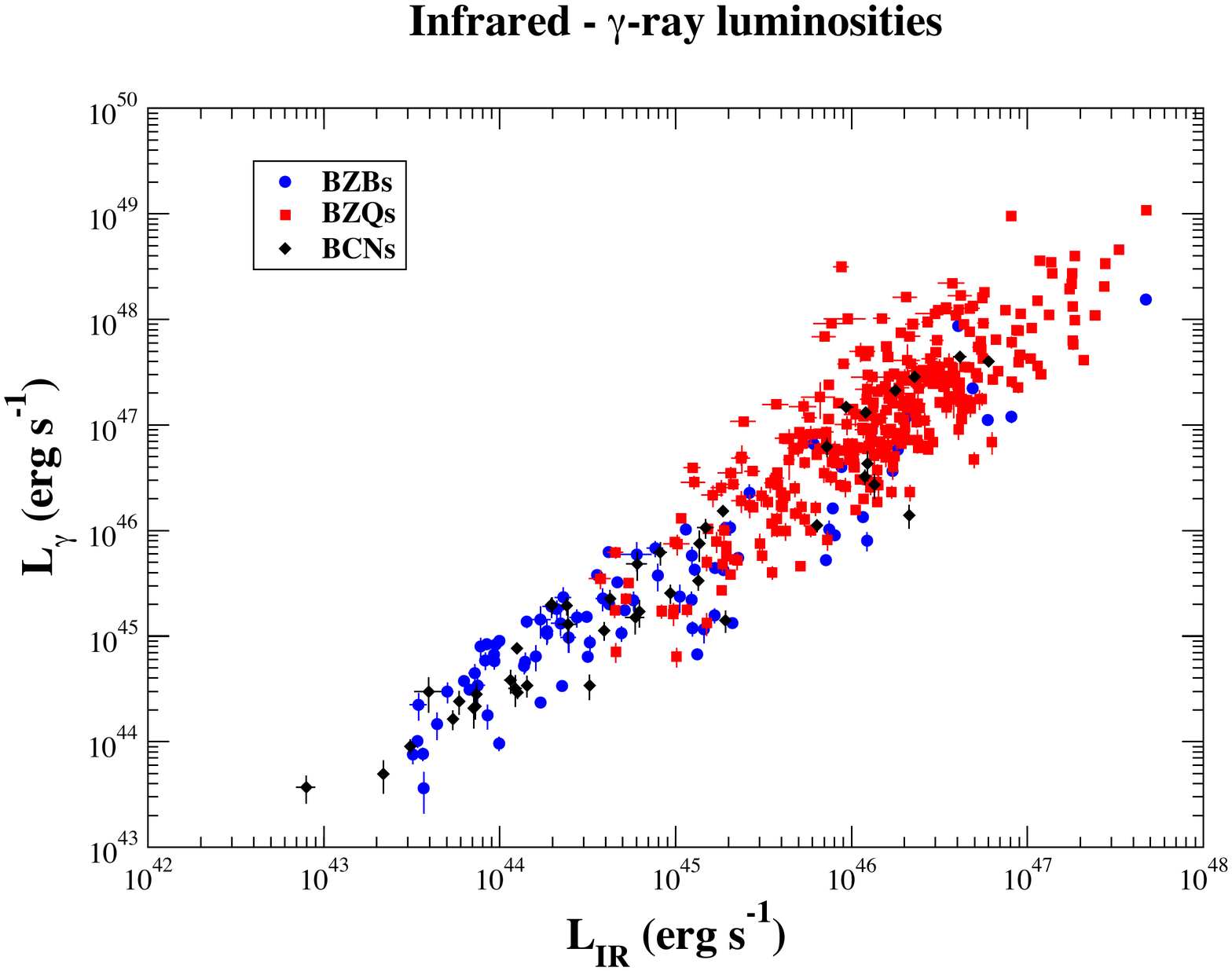}
           \caption{The scatterplot of the infrared - $\gamma$-ray fluxes (upper panel) and luminosities (lower panel) for the blazar-like sources.
                         Spectroscopically confirmed BZBs and BZQs are shown in blue circles and red squares, respectively, while bcn sources
                         are black diamonds.}
           \label{fig:infrared}
          \end{figure}

Here we report an updated scatterplot for the spectral indices (see Figure~\ref{fig:indices})
as originally presented in D'Abrusco et al. (2012). The correlation the spectral shapes
in the infrared and in $\gamma$-rays is still present and as occurred in the case of the radio - $\gamma$-ray scatter plot 
(see Section~\ref{sec:radio}) the locations of the bcn sources appear to be more consistent with the BL Lac population.
The linear correlation coefficient for the whole data set is 0.64.
Finally, we also present the connection between infrared and $\gamma$-ray fluxes and luminosities in Figure~\ref{fig:infrared}
that is again in good agreement with our previous results \citep{paper2,ugs1}.

\section{Summary and conclusions}
\label{sec:conclusions}
Two years after the release of the second \fer\ source catalog we present 
a comprehensive multifrequency investigation of all the $\gamma$-ray associations 
listed in both the 1FGL and the 2FGL catalogs.

First, we introduced a new category of $\gamma$-ray source associations in addition to {\it identified} and {\it associated} sources.
We label as {\it candidate associations}
those $\gamma$-ray sources having a potential low-energy counterpart of few specific classes of well-known $\gamma$-ray emitters 
lying within the \fer\ positional uncertainty region and/or with angular separations between the \fer\ and the counterpart position
smaller than the maximum one for all the associated sources of the same class. 
Then we provided a new classification scheme for the low-energy counterparts of the \fer\ sources
based on the multifrequency observations and on the optical spectroscopic information now available.
We also presented here a cross-matching between the \fer\ catalogs and several surveys of SFRs 
(see Section~\ref{sec:crossmatches} for details)
to highlight the possibility that an unknown SNR, PWN and/or PSR is embedded therein, 
and to provide information that could be used to refine models of diffuse Galactic $\gamma$-ray for future releases of the \fer\ catalogs.

The total number of $\gamma$-ray sources considered in our investigation 
for both the 1FGL and the 2FGL comprises 2219 unique \fer\ objects, all listed in Table~\ref{tab:main}
with each assigned counterpart and their main multifrequency properties.
Overall, in the refined association list of the \fer\ catalogs, we found 174 \fer\ sources with a SFR consistent with their $\gamma$-ray positions.
In particular, 60 \fer\ objects out of these 174 do not have the ``c'' flag in at least one of the \fer\ names and include: 
13 identified sources, 17 associated, and 30 candidate associations.
Their counterparts are also classified as four HMBs; 17 PSRs, 
among which three lie in a PWN and two in a SNR; three PWNe, one with a PSR; one radio galaxy;
thirteen SNRs, two with a PSR included; one binary star; three unclassified sources; and 19 SFRs.

We found spectroscopic information for 177 extragalactic $\gamma$-ray sources not reported in the previous version of the \fer\ catalogs.
We included analyses of eight new optical spectroscopic observations performed with the 
2.1-meter telescope of the OAN in San Pedro M\'artir and with the
OAGH to confirm the natures of these blazar-like sources found in our multifrequency investigation.
Blazar candidates and counterparts with unknown origin (i.e., unc) were also 
analyzed with the KDE technique to determine the fraction having \wse\ IR colors 
similar to known $\gamma$-ray blazars. 
We compared the refined associations listed in both the \fer\ catalogs with the classification proposed 
on the basis of statistical investigations \citep[][]{ackermann12,mirabal12} 
and we found a good agreement with our results.  

\begin{figure}[] 
          \includegraphics[height=6.8cm,width=9.5cm,angle=0]{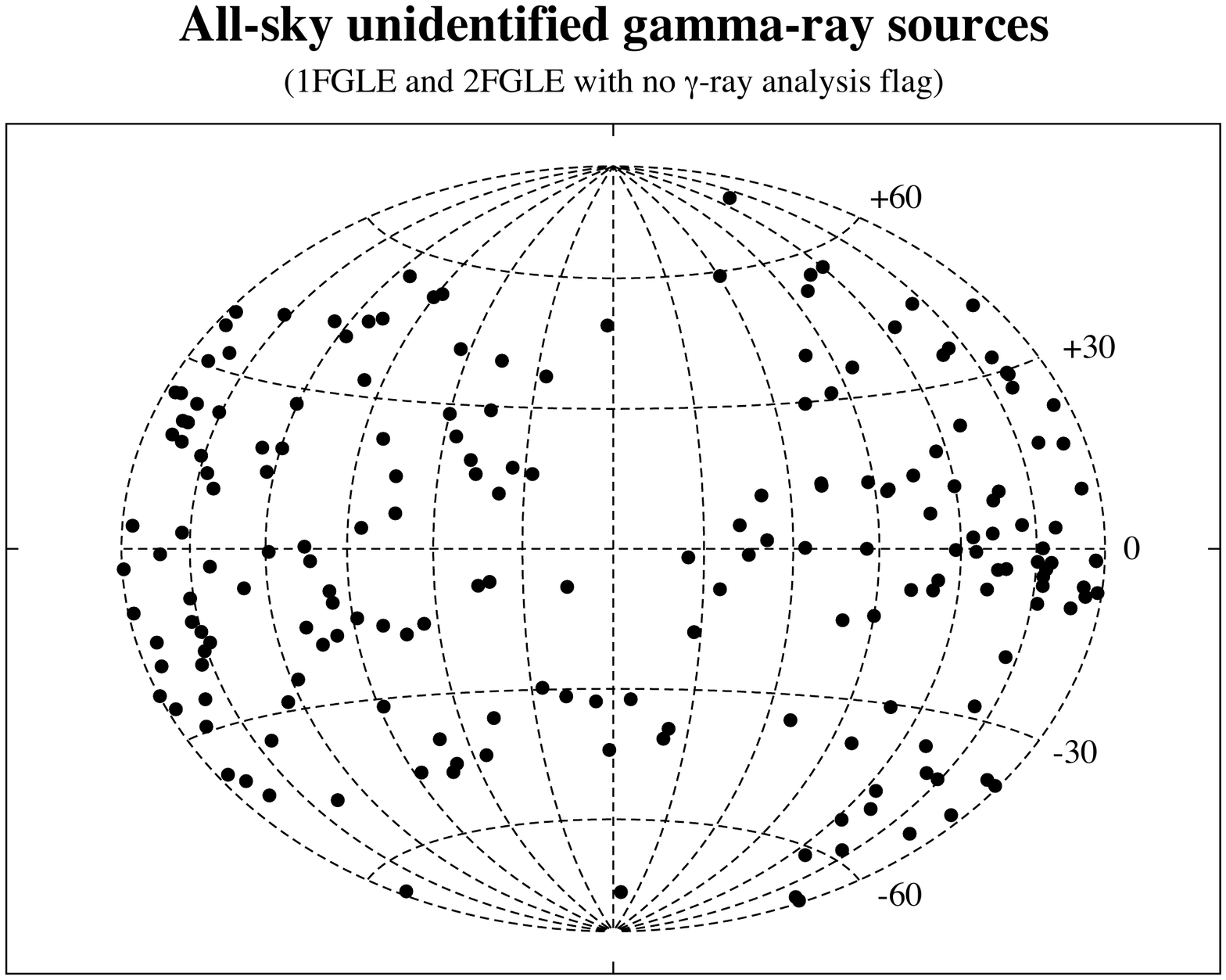}
           \caption{All-sky distribution of the remaining UGSs
                        with no $\gamma$-ray analysis flags in the refined association list of the \fer\ catalogs.}
           \label{fig:skydistugs}
          \end{figure}

We note that in the combined list of refined associations for 1FGL and 2FGL, among 2219 sources, 
394 are still UGSs (i.e., $\sim$18\% of the entire sample) 
with 191 of them having no $\gamma$-ray analysis flags. This clean sample of 191 UGSs
appears to have a uniform distribution in the sky with a small excess 
toward the Galactic plane, as shown in Figure~\ref{fig:skydistugs}.
Moreover, we conclude that the fraction of \fer\ sources with plausible counterparts, combining identifications, associations and candidate associations,
within their $\gamma$-ray positional uncertainty regions, 
is $\sim$80\% and up to $\sim$90\% when considering sources with no $\gamma$-ray analysis flags. 

\begin{figure}[] 
          \includegraphics[height=6.8cm,width=9.5cm,angle=0]{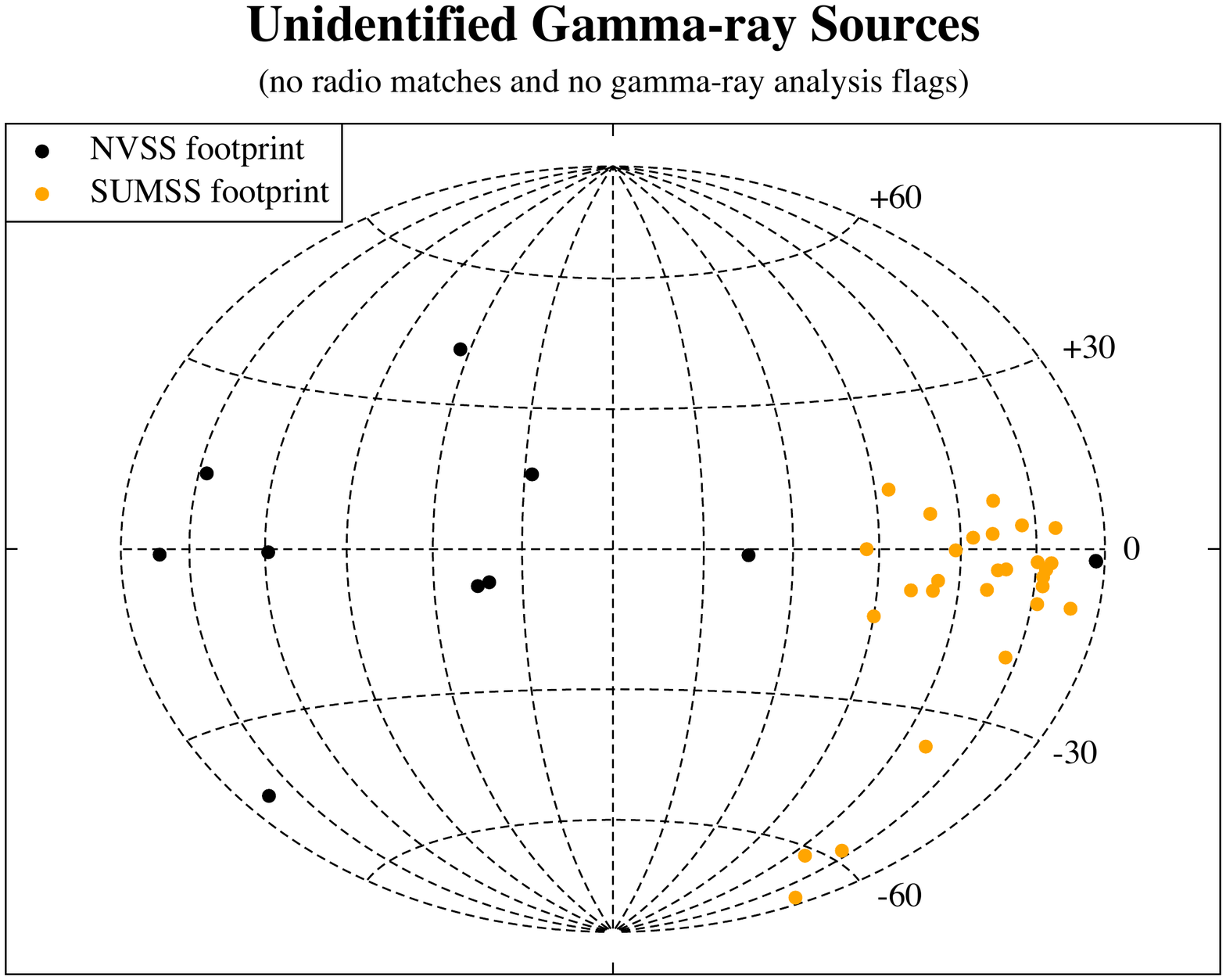}
           \caption{All-sky distribution of the 40 UGSs
                        that do not have any NVSS and/or SUMSS radio source within their positional uncertainty regions at 95\% level of confidence
                        and do not present any $\gamma$-ray analysis flag.
                        We show those lying in the NVSS footprint (i.e., declination greater than $-$40 \degr) as black circles while the others in the Southern hemisphere
                        are indicated as orange circles.}
           \label{fig:skynomatches}
          \end{figure}

\begin{figure}[] 
          \includegraphics[height=6.4cm,width=9.4cm,angle=0]{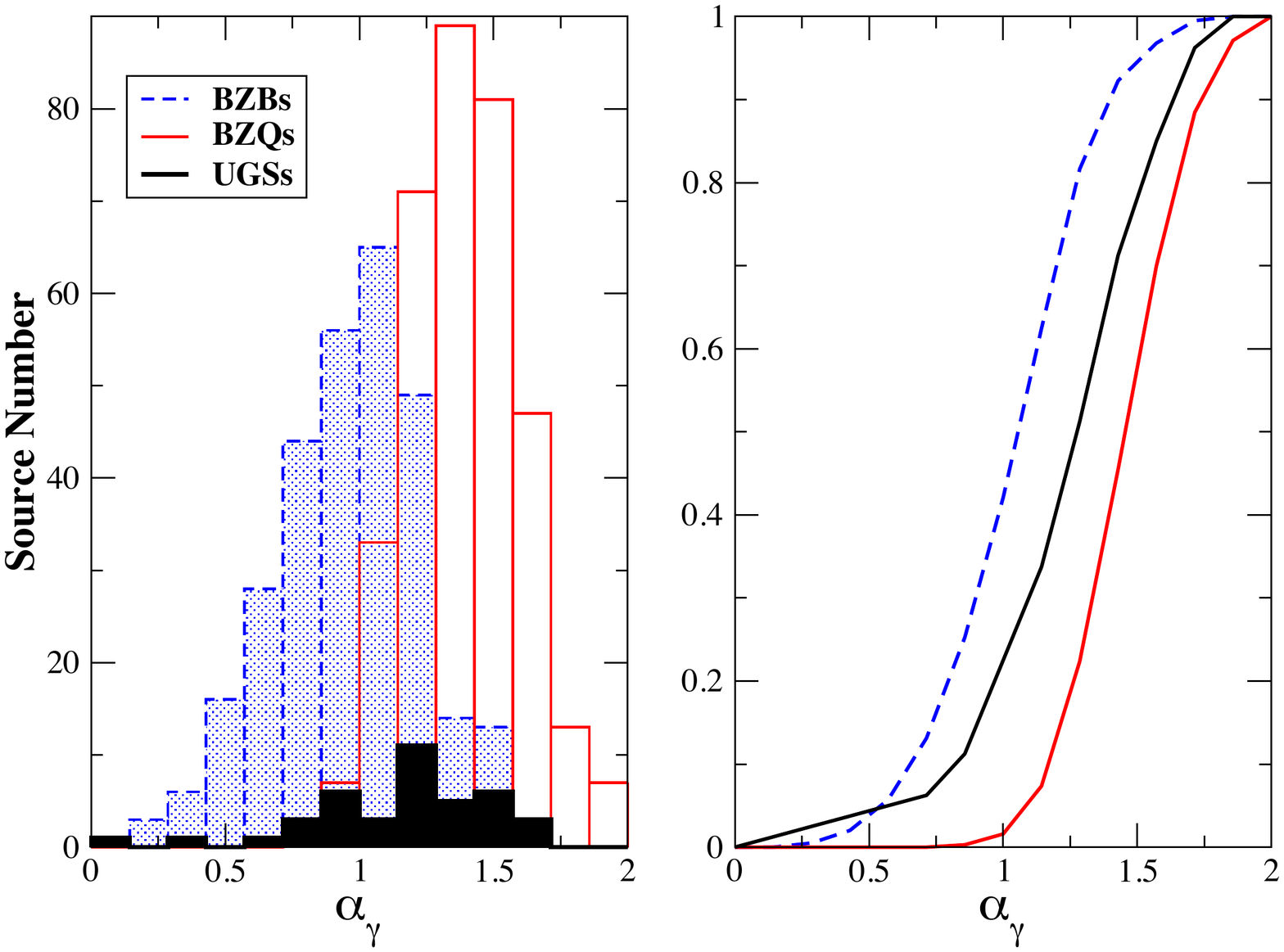}
          \includegraphics[height=6.4cm,width=9.4cm,angle=0]{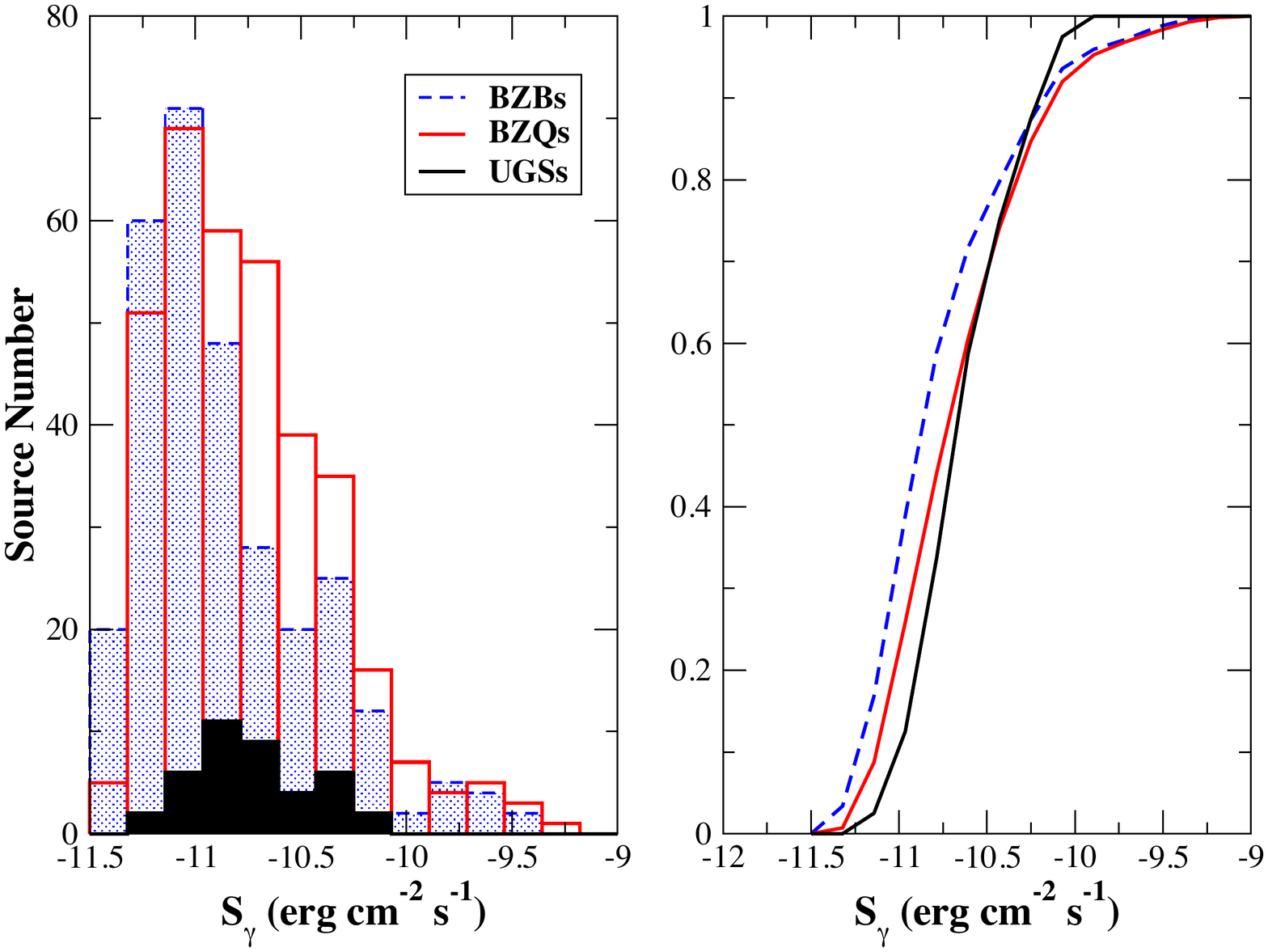}
           \caption{The distribution of the $\gamma$-ray spectral index (top) and that of the energy flux (bottom) for the 40 UGSs (straight black line)
                        that do not have any NVSS and/or SUMSS radio source within their positional uncertainty regions 
                        and without $\gamma$-ray analysis flags in comparison with those of the BZBs (dashed blue line) and the BZQs (straight red line).
                        Cumulative distributions are also reported on the right panels.
                        The energy flux is the one reported in both the 1FGL and the 2FGL catalogs.}
           \label{fig:comparing}
          \end{figure}

Finally, we remark that within the sample of 191 UGSs that do not have any $\gamma$-ray analysis flags 
 are 40 sources (listed in Table~\ref{tab:special}) that do not have any
NVSS and/or SUMSS radio source within their positional uncertainty regions at 95\% level of confidence.
Their all-sky distribution, separated into those lying in the NVSS footprint (i.e., declination greater than $-$40 \degr)
and the others observable from the Southern hemisphere is shown in Figure~\ref{fig:skynomatches}.
The excess in the subsample of those with declinations less than $-$40\degr, clearly visible in Figure~\ref{fig:skynomatches},
could be  due to shallower coverage by SUMSS relative to the NVSS catalog.
It is worth mentioning that for these 40 UGSs the distributions of the $\gamma$-ray spectral index and of
the $\gamma$-ray fluxes do not show any significant differences from those of BZQs and the BZBs
(see Figure~\ref{fig:comparing}). However it is intriguing that these UGSs tend to be brighter than the BZB population in the gamma-rays.
It is unlikely that this small UGS subsample is constituted by blazars since given their $\gamma$-ray fluxes $S_\gamma$ and assuming 
the typical radio-$\gamma$-ray fluxes of the blazar population, 
the expected radio flux densities should be greater than $\sim$ 50 mJy for a large fraction, well above the NVSS and the SUMSS flux thresholds.
On the other hand we also conclude that their steep values of $\gamma$-ray spectral indices are not compatible 
with sources emitting via dark matter annihilation \citep[e.g.,][]{belikov12,drlica14}.

\begin{table}
\tiny
\begin{center}
\caption{Unidentified Gamma-ray Sources with no radio matches and no $\gamma$-ray analysis flags.}
\begin{tabular}{|ll|}
\hline
1FGL & 2FGL \\
name & name \\
\hline
\noalign{\smallskip}
  1FGL J0032.7-5519 & 2FGL J0032.7-5521\\
  1FGL J0212.3+5319 & 2FGL J0212.1+5318\\
  1FGL J0545.6+6022 & 2FGL J0545.6+6018\\
  1FGL J0709.0-1116 & \\
  1FGL J0802.4-5622 & 2FGL J0802.7-5615\\
  1FGL J0854.6-4504 & 2FGL J0854.7-4501\\
  1FGL J0933.9-6228 & 2FGL J0934.0-6231\\
  1FGL J1036.2-6719 & 2FGL J1036.1-6722\\
  1FGL J1117.0-5339 & 2FGL J1117.2-5341\\
  1FGL J1306.4-6038 & 2FGL J1306.2-6044\\
  1FGL J1405.5-5846 & \\
  1FGL J1441.8-6404 & \\
  1FGL J1518.0-5233 & 2FGL J1518.4-5233\\
  1FGL J1639.5-5152 & \\
  1FGL J1650.3-5410 & \\
  1FGL J1702.4-5653 & 2FGL J1702.5-5654\\
  1FGL J1743.5-3314 & \\
  1FGL J1806.2+0609 & 2FGL J1805.8+0612\\
  1FGL J1825.7-1410c & \\
  1FGL J1948.6+2437 & \\
  1FGL J2112.5-3044 & 2FGL J2112.5-3042\\
  1FGL J2227.4-7804 & \\
  1FGL J2325.8-4043 & \\
  1FGL J2333.0-5535 & 2FGL J2332.5-5535\\
   & 2FGL J0237.9+5238\\
   & 2FGL J1016.4-4244\\
   & 2FGL J1058.7-6621\\
   & 2FGL J1208.5-6240\\
   & 2FGL J1404.0-5244\\
   & 2FGL J1410.4+7411\\
   & 2FGL J1422.3-6841\\
   & 2FGL J1507.0-6223\\
   & 2FGL J1617.3-5336\\
   & 2FGL J1626.4-4408\\
   & 2FGL J1639.8-5145\\
   & 2FGL J1641.8-5319\\
   & 2FGL J1643.3-4928\\
   & 2FGL J1741.1-6750\\
   & 2FGL J1742.5-3323\\
   & 2FGL J1753.8-4446\\
\noalign{\smallskip}
\hline
\end{tabular}
\end{center}
\label{tab:special}
\end{table}

\acknowledgements
We thank the anonymous referee for useful comments that led to improvements in the paper.
F. Massaro and G. Tosti also thank L. Costamante.
This investigation is supported by the NASA grants NNX12AO97G and NNX13AP20G.
The work by G. Tosti is supported by the ASI/INAF contract I/005/12/0.
H. A. Smith acknowledges partial support from NASA/JPL grant RSA 1369566.
HOF was funded by a postdoctoral UNAM grant and is currently granted by a C\'atedra CONACyT para J\'ovenes Investigadores.
V. Chavushyan acknowledges funding by CONACyT research grant 151494 (M\'exico).
We thank the staff at the Observatorio Astron\'omico Nacional in San Pedro M\'artir (M\'exico) for all their help during the observation runs.
Part of this work is based on archival data, software or on-line services provided by the ASI Science Data Center.
This research has made use of data obtained from the high-energy Astrophysics Science Archive
Research Center (HEASARC) provided by NASA's Goddard Space Flight Center; 
the SIMBAD database operated at CDS,
Strasbourg, France; the NASA/IPAC Extragalactic Database
(NED) operated by the Jet Propulsion Laboratory, California
Institute of Technology, under contract with the National Aeronautics and Space Administration.
Part of this work is based on the NVSS (NRAO VLA Sky Survey):
The National Radio Astronomy Observatory is operated by Associated Universities,
Inc., under contract with the National Science Foundation and on the VLA Low-frequency Sky Survey (VLSS).
The Molonglo Observatory site manager, Duncan Campbell-Wilson, and the staff, Jeff Webb, Michael White and John Barry, 
are responsible for the smooth operation of Molonglo Observatory Synthesis Telescope (MOST) and the day-to-day observing programme of SUMSS. 
The SUMSS survey is dedicated to Michael Large whose expertise and vision made the project possible. 
The MOST is operated by the School of Physics with the support of the Australian Research Council and the Science Foundation for Physics within the University of Sydney.
This publication makes use of data products from the Wide-field Infrared Survey Explorer, 
which is a joint project of the University of California, Los Angeles, and 
the Jet Propulsion Laboratory/California Institute of Technology, 
funded by the National Aeronautics and Space Administration.
This publication makes use of data products from the Two Micron All Sky Survey, which is a joint project of the University of 
Massachusetts and the Infrared Processing and Analysis Center/California Institute of Technology, funded by the National Aeronautics 
and Space Administration and the National Science Foundation.
Funding for the SDSS and SDSS-II has been provided by the Alfred P. Sloan Foundation, 
the Participating Institutions, the National Science Foundation, the U.S. Department of Energy, 
the National Aeronautics and Space Administration, the Japanese Monbukagakusho, 
the Max Planck Society, and the Higher Education Funding Council for England. 
The SDSS Web Site is http://www.sdss.org/.
The SDSS is managed by the Astrophysical Research Consortium for the Participating Institutions. 
The Participating Institutions are the American Museum of Natural History, 
Astrophysical Institute Potsdam, University of Basel, University of Cambridge, 
Case Western Reserve University, University of Chicago, Drexel University, 
Fermilab, the Institute for Advanced Study, the Japan Participation Group, 
Johns Hopkins University, the Joint Institute for Nuclear Astrophysics, 
the Kavli Institute for Particle Astrophysics and Cosmology, the Korean Scientist Group, 
the Chinese Academy of Sciences (LAMOST), Los Alamos National Laboratory, 
the Max-Planck-Institute for Astronomy (MPIA), the Max-Planck-Institute for Astrophysics (MPA), 
New Mexico State University, Ohio State University, University of Pittsburgh, 
University of Portsmouth, Princeton University, the United States Naval Observatory, 
and the University of Washington.
This research has made use of the USNOFS Image and Catalogue Archive
operated by the United States Naval Observatory, Flagstaff Station
(http://www.nofs.navy.mil/data/fchpix/).
The WENSS project was a collaboration between the Netherlands Foundation 
for Research in Astronomy and the Leiden Observatory. 
We acknowledge the WENSS team consisted of Ger de Bruyn, Yuan Tang, 
Roeland Rengelink, George Miley, Huub Rottgering, Malcolm Bremer, 
Martin Bremer, Wim Brouw, Ernst Raimond and David Fullagar 
for the extensive work aimed at producing the WENSS catalog.
TOPCAT\footnote{\underline{http://www.star.bris.ac.uk/$\sim$mbt/topcat/}} 
\citep{taylor05} for the preparation and manipulation of the tabular data and the images.
The Aladin Java applet\footnote{\underline{http://aladin.u-strasbg.fr/aladin.gml}}
was used to create the finding charts reported in this paper \citep{bonnarel00}. 
It can be started from the CDS (Strasbourg - France), from the CFA (Harvard - USA), from the ADAC (Tokyo - Japan), 
from the IUCAA (Pune - India), from the UKADC (Cambridge - UK), or from the CADC (Victoria - Canada).

{}

\appendix
Here we present the spectra of 2FGL J1719.3+1744 and 2FGL J1801.7+4405,
both with uncertain redshift, as reported in the \bzcat\ together with that of 1FGL J1942.7+1033.
We confirmed the redshift for 2FGL J1801.7+4405 and the BL Lac classification of 1FGL J1942.7+1033 \citep{tsarevsky05,masetti13}
while 2FGL J1719.3+1744 (alias BZBJ1719+1745) has a completely featureless spectrum.
\begin{figure}[]
\includegraphics[height=5.7cm,width=6.5cm,angle=0]{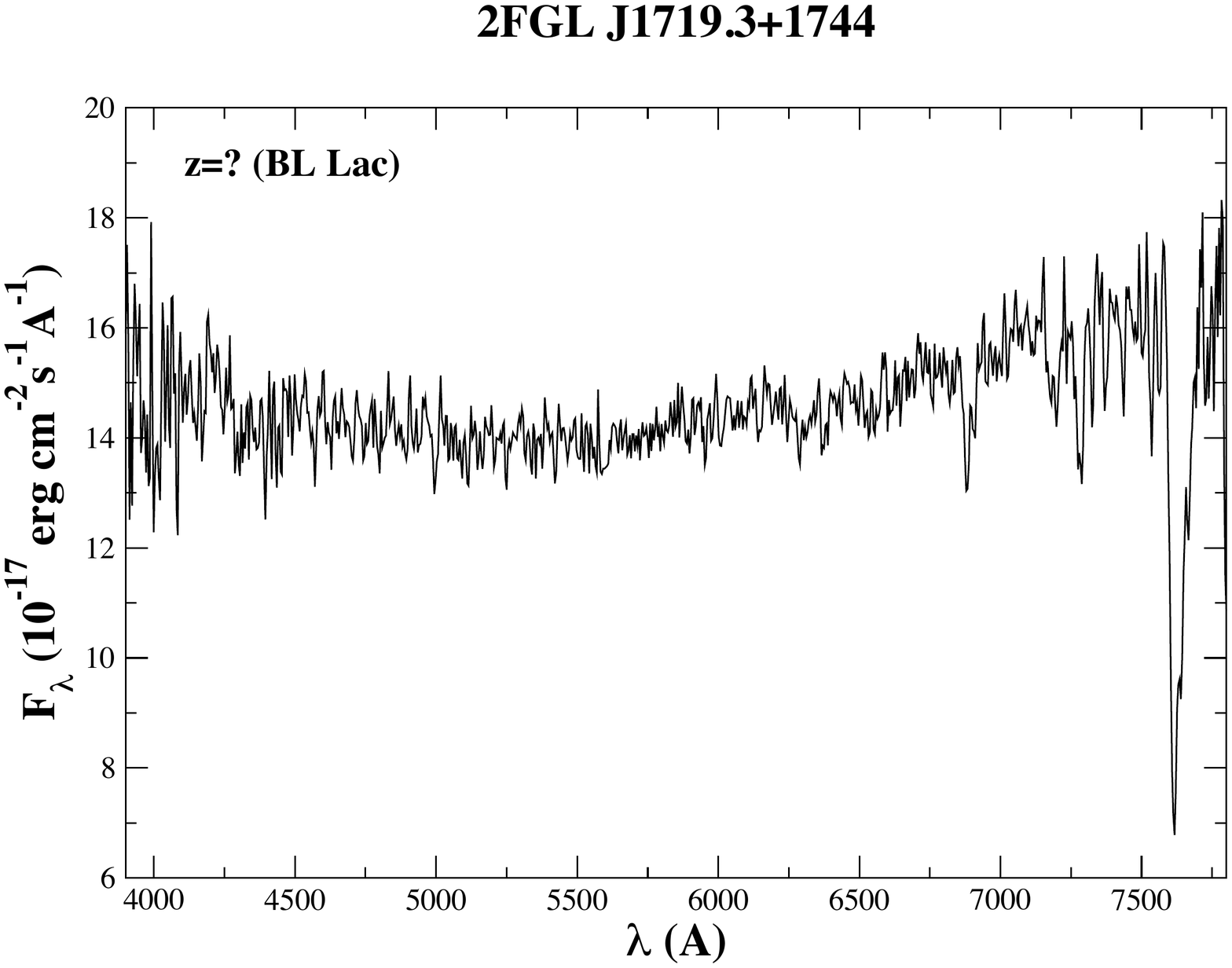}
\includegraphics[height=5.7cm,width=6.5cm,angle=0]{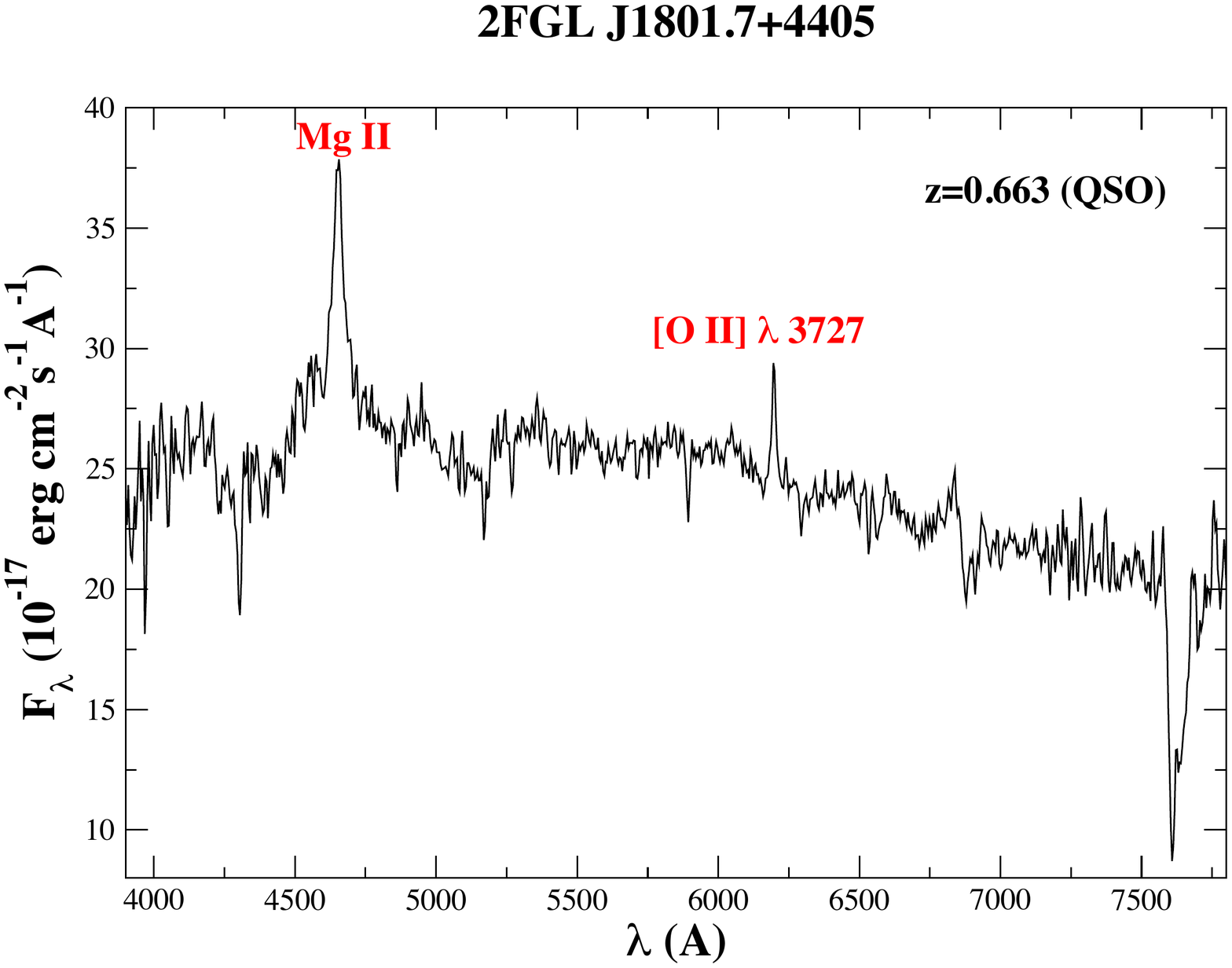}
\includegraphics[height=5.7cm,width=6.5cm,angle=0]{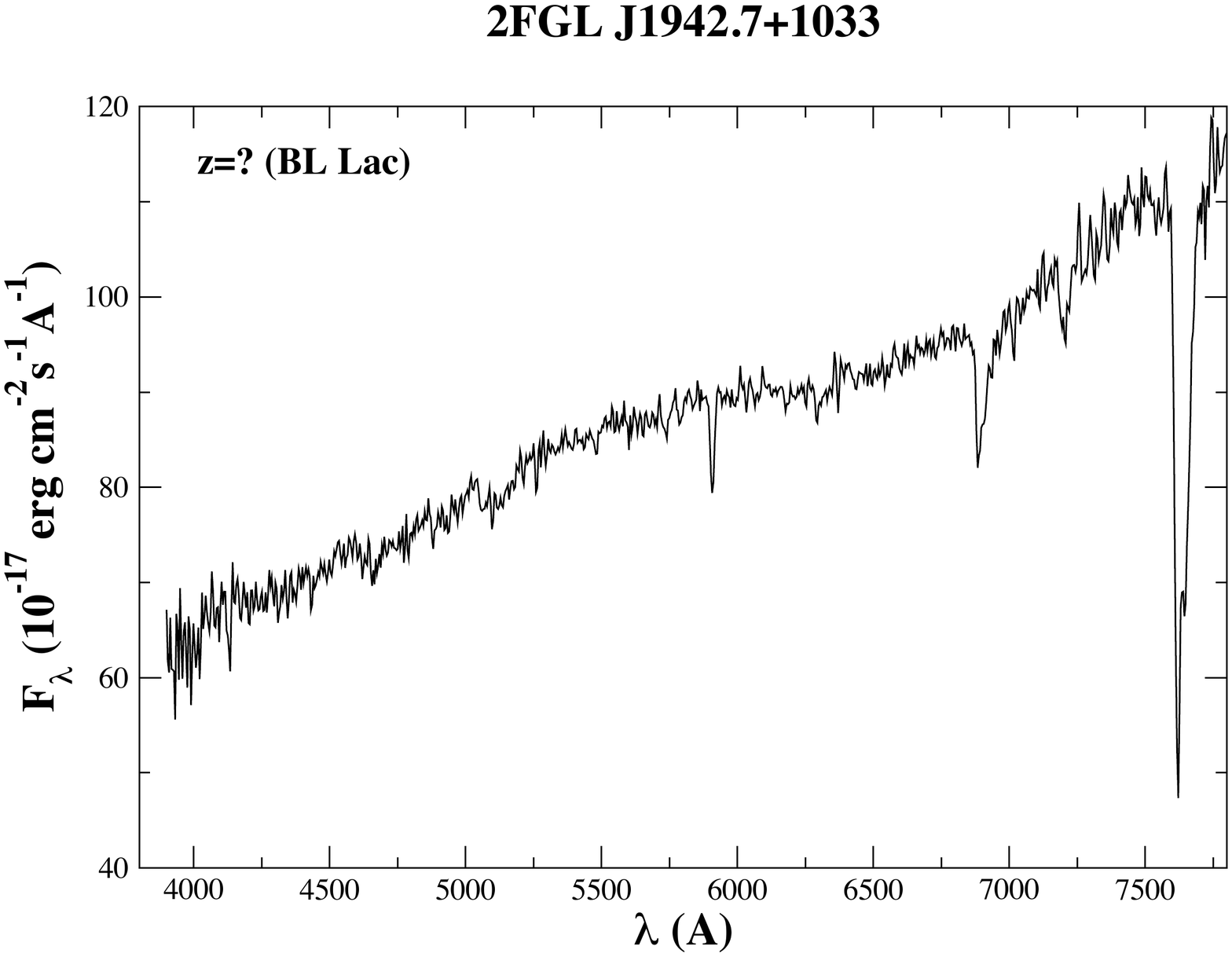}
\end{figure}
\begin{figure}[]
\includegraphics[height=5.cm,width=6.4cm,angle=0]{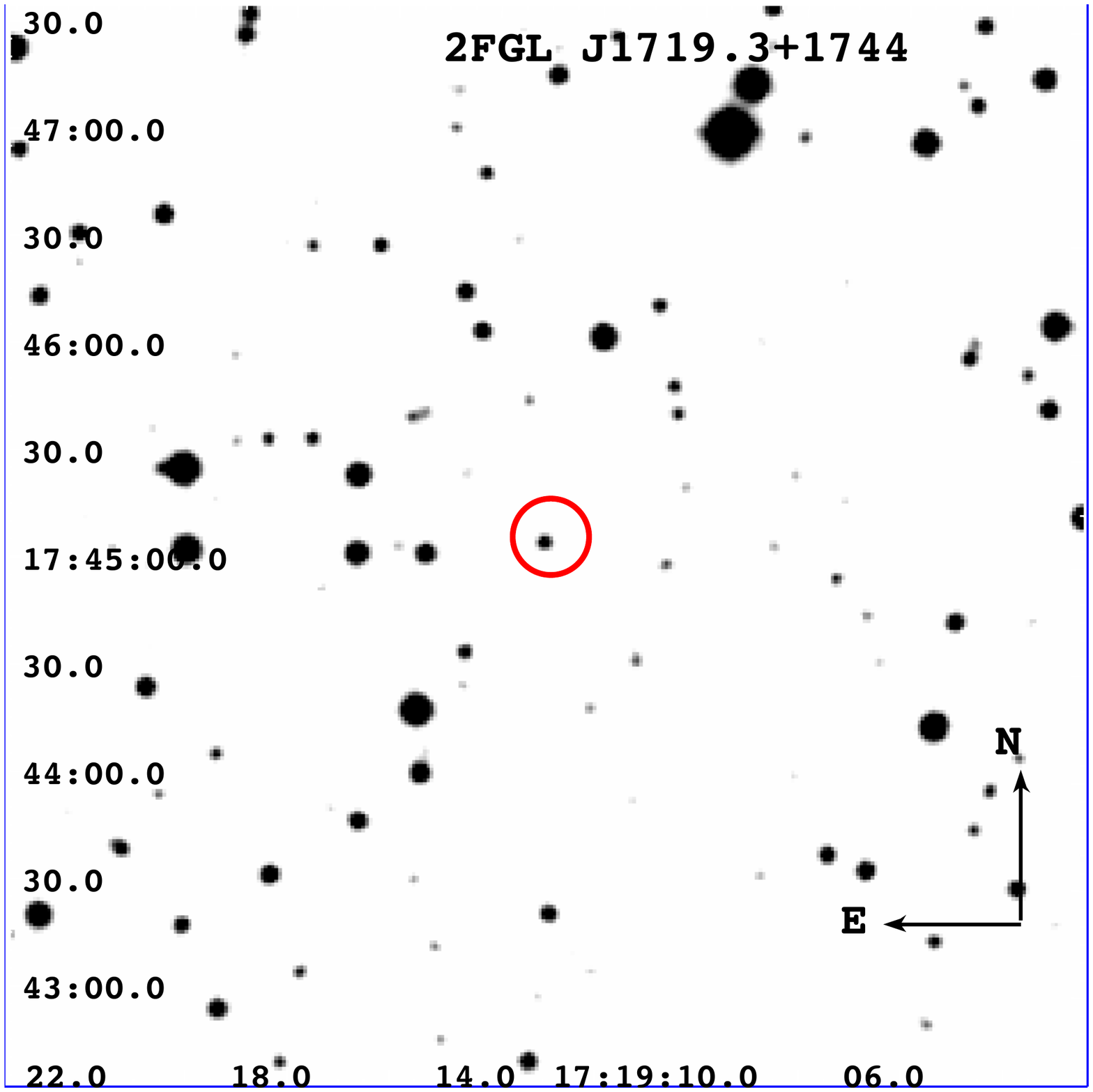}
\includegraphics[height=5.cm,width=6.4cm,angle=-0]{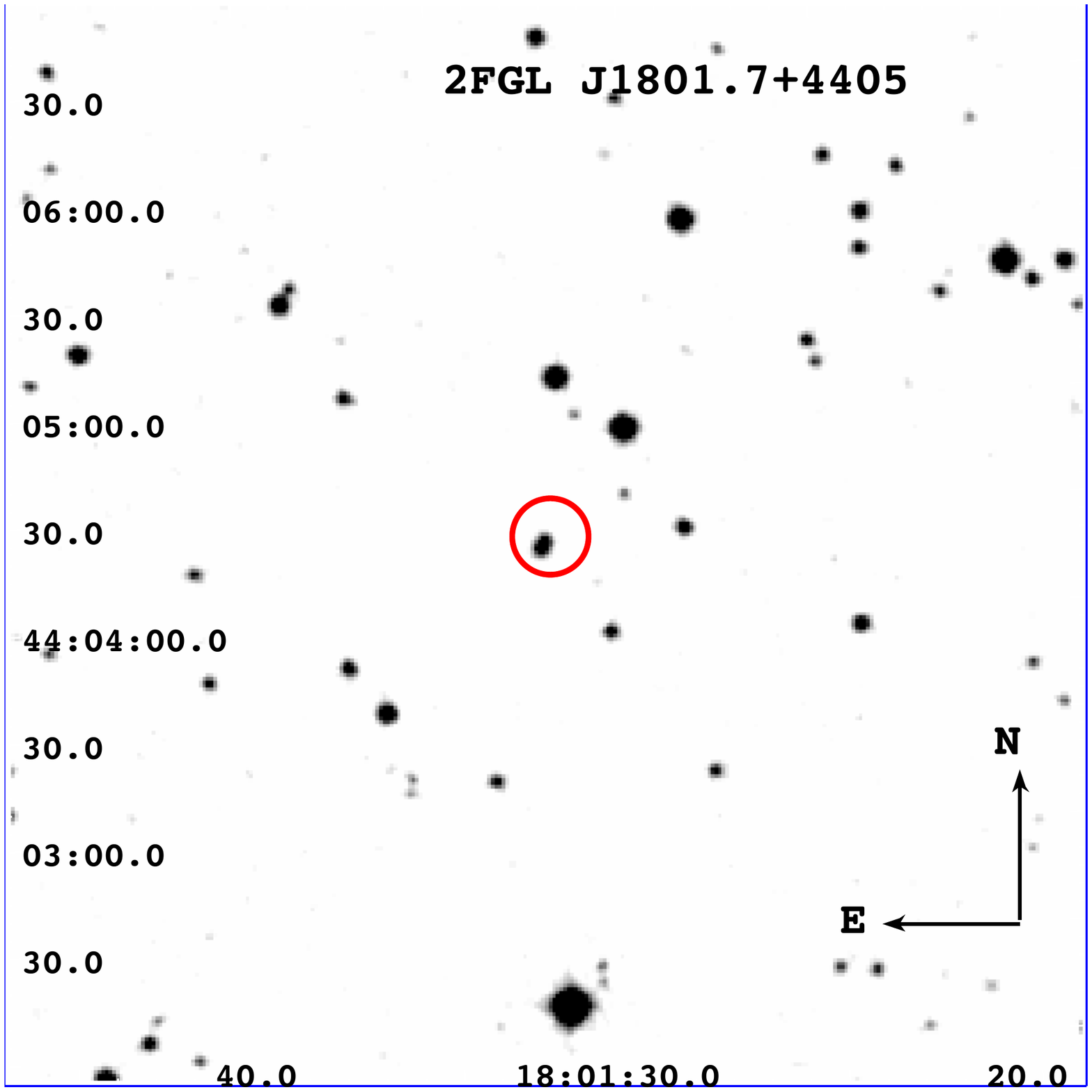}
\includegraphics[height=5.cm,width=6.4cm,angle=0]{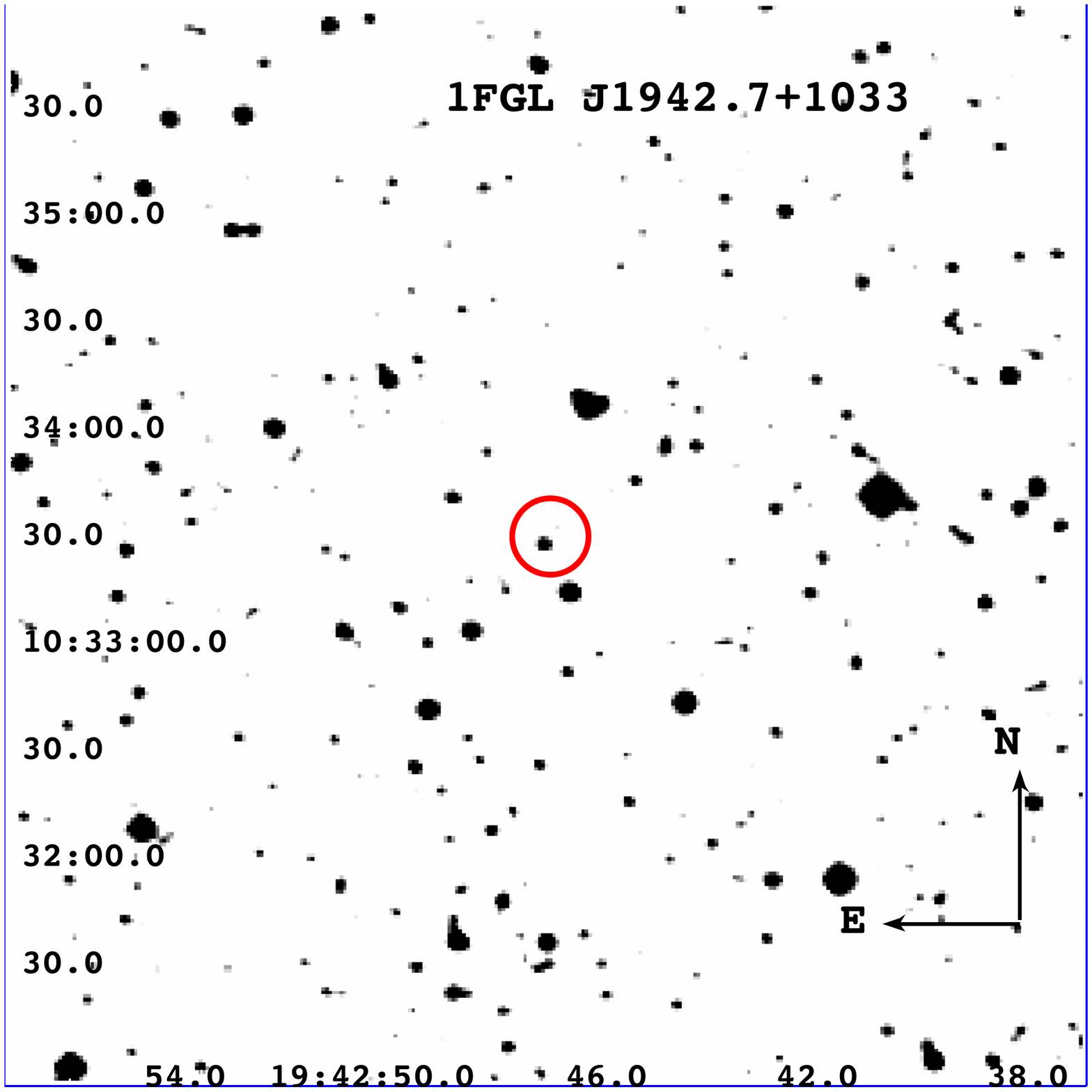}
\caption{Upper panel) The optical spectra of the counterparts associated with 
2FGL J1719.3+1744, 2FGL J1801.7+4405 and 1FGL J1942.7+1033 already available in literature and 
re-observed at OAN in San Pedro M\'artir (M\'exico) on 29 June and 30 June 2014.
The Mg\,{\sc ii} and the [OII] $\lambda$3727 emission lines superimposed to the optical continuum are visible in 2FGL J1801.7+4405
and allowed us to classify the source as a quasar, while the lack of features in the other two spectra led toward a BL Lac classification.
(Lower panel) The 5\arcmin\, $\times$ \,5\arcmin\ finding chart from the Digitized Sky Survey (red filter). 
The potential counterparts pointed during our observations is indicated by the red circle. }
\end{figure}

\end{document}